\documentclass[twocolumn,letterpaper,aps,prc,longbibliography,superscriptaddress,showpacs,nofootinbib,floatfix]{revtex4-1}
\usepackage{graphicx}	% Include figure filed
\usepackage{xcolor}	% temporarily added
\usepackage{xspace}	% Include xspace
\usepackage{amsmath}	% to allow use of \align

\newcommand{\pt}{\mbox{$p_T$}\xspace}

\newcommand{\iaa}{\mbox{$I_{AA}$}\xspace}

\newcommand{\Gsqsn}{\mbox{$\sqrt{s_{_{NN}}}=200~{\rm GeV}$}\xspace}

\newcommand{\ptt}{\mbox{$p_T^t$}\xspace}
\newcommand{\pta}{\mbox{$p_T^a$}\xspace}

\begin{document}

\title{Measurement of two-particle correlations with respect to 
second- and third-order event planes in Au$+$Au collisions 
at $\sqrt{s_{_{NN}}}=200$ GeV}

\newcommand{\abilene}{Abilene Christian University, Abilene, Texas 79699, USA}
\newcommand{\augie}{Department of Physics, Augustana University, Sioux Falls, South Dakota 57197, USA}
\newcommand{\banaras}{Department of Physics, Banaras Hindu University, Varanasi 221005, India}
\newcommand{\barc}{Bhabha Atomic Research Centre, Bombay 400 085, India}
\newcommand{\baruch}{Baruch College, City University of New York, New York, New York, 10010 USA}
\newcommand{\bnlcoll}{Collider-Accelerator Department, Brookhaven National Laboratory, Upton, New York 11973-5000, USA}
\newcommand{\bnlphys}{Physics Department, Brookhaven National Laboratory, Upton, New York 11973-5000, USA}
\newcommand{\caucr}{University of California-Riverside, Riverside, California 92521, USA}
\newcommand{\charlesczech}{Charles University, Ovocn\'{y} trh 5, Praha 1, 116 36, Prague, Czech Republic}
\newcommand{\chonbuk}{Chonbuk National University, Jeonju, 561-756, Korea}
\newcommand{\ciae}{Science and Technology on Nuclear Data Laboratory, China Institute of Atomic Energy, Beijing 102413, People's Republic of China}
\newcommand{\cns}{Center for Nuclear Study, Graduate School of Science, University of Tokyo, 7-3-1 Hongo, Bunkyo, Tokyo 113-0033, Japan}
\newcommand{\colorado}{University of Colorado, Boulder, Colorado 80309, USA}
\newcommand{\columbia}{Columbia University, New York, New York 10027 and Nevis Laboratories, Irvington, New York 10533, USA}
\newcommand{\czechtech}{Czech Technical University, Zikova 4, 166 36 Prague 6, Czech Republic}
\newcommand{\dapnia}{Dapnia, CEA Saclay, F-91191, Gif-sur-Yvette, France}
\newcommand{\debrecen}{Debrecen University, H-4010 Debrecen, Egyetem t{\'e}r 1, Hungary}
\newcommand{\elte}{ELTE, E{\"o}tv{\"o}s Lor{\'a}nd University, H-1117 Budapest, P{\'a}zm{\'a}ny P.~s.~1/A, Hungary}
\newcommand{\eszterhazy}{Eszterh\'azy K\'aroly University, K\'aroly R\'obert Campus, H-3200 Gy\"ongy\"os, M\'atrai \'ut 36, Hungary}
\newcommand{\ewha}{Ewha Womans University, Seoul 120-750, Korea}
\newcommand{\fit}{Florida Institute of Technology, Melbourne, Florida 32901, USA}
\newcommand{\fsu}{Florida State University, Tallahassee, Florida 32306, USA}
\newcommand{\gsu}{Georgia State University, Atlanta, Georgia 30303, USA}
\newcommand{\hiroshima}{Hiroshima University, Kagamiyama, Higashi-Hiroshima 739-8526, Japan}
\newcommand{\howard}{Department of Physics and Astronomy, Howard University, Washington, DC 20059, USA}
\newcommand{\ihepprot}{IHEP Protvino, State Research Center of Russian Federation, Institute for High Energy Physics, Protvino, 142281, Russia}
\newcommand{\illuiuc}{University of Illinois at Urbana-Champaign, Urbana, Illinois 61801, USA}
\newcommand{\inrras}{Institute for Nuclear Research of the Russian Academy of Sciences, prospekt 60-letiya Oktyabrya 7a, Moscow 117312, Russia}
\newcommand{\instpasczech}{Institute of Physics, Academy of Sciences of the Czech Republic, Na Slovance 2, 182 21 Prague 8, Czech Republic}
\newcommand{\isu}{Iowa State University, Ames, Iowa 50011, USA}
\newcommand{\jaea}{Advanced Science Research Center, Japan Atomic Energy Agency, 2-4 Shirakata Shirane, Tokai-mura, Naka-gun, Ibaraki-ken 319-1195, Japan}
\newcommand{\jinrdubna}{Joint Institute for Nuclear Research, 141980 Dubna, Moscow Region, Russia}
\newcommand{\jyvaskyla}{Helsinki Institute of Physics and University of Jyv{\"a}skyl{\"a}, P.O.Box 35, FI-40014 Jyv{\"a}skyl{\"a}, Finland}
\newcommand{\kek}{KEK, High Energy Accelerator Research Organization, Tsukuba, Ibaraki 305-0801, Japan}
\newcommand{\korea}{Korea University, Seoul, 02841, Korea}
\newcommand{\kurchatov}{National Research Center ``Kurchatov Institute", Moscow, 123098 Russia}
\newcommand{\kyoto}{Kyoto University, Kyoto 606-8502, Japan}
\newcommand{\labllr}{Laboratoire Leprince-Ringuet, Ecole Polytechnique, CNRS-IN2P3, Route de Saclay, F-91128, Palaiseau, France}
\newcommand{\lahorelums}{Physics Department, Lahore University of Management Sciences, Lahore 54792, Pakistan}
\newcommand{\lawllnl}{Lawrence Livermore National Laboratory, Livermore, California 94550, USA}
\newcommand{\losalamos}{Los Alamos National Laboratory, Los Alamos, New Mexico 87545, USA}
\newcommand{\lpc}{LPC, Universit{\'e} Blaise Pascal, CNRS-IN2P3, Clermont-Fd, 63177 Aubiere Cedex, France}
\newcommand{\lund}{Department of Physics, Lund University, Box 118, SE-221 00 Lund, Sweden}
\newcommand{\lyon}{IPNL, CNRS/IN2P3, Univ Lyon, Université Lyon 1, F-69622, Villeurbanne, France}
\newcommand{\maryland}{University of Maryland, College Park, Maryland 20742, USA}
\newcommand{\mass}{Department of Physics, University of Massachusetts, Amherst, Massachusetts 01003-9337, USA}
\newcommand{\michigan}{Department of Physics, University of Michigan, Ann Arbor, Michigan 48109-1040, USA}
\newcommand{\muenster}{Institut f\"ur Kernphysik, University of M\"unster, D-48149 M\"unster, Germany}
\newcommand{\muhlenberg}{Muhlenberg College, Allentown, Pennsylvania 18104-5586, USA}
\newcommand{\myongji}{Myongji University, Yongin, Kyonggido 449-728, Korea}
\newcommand{\nagasaki}{Nagasaki Institute of Applied Science, Nagasaki-shi, Nagasaki 851-0193, Japan}
\newcommand{\nara}{Nara Women's University, Kita-uoya Nishi-machi Nara 630-8506, Japan}
\newcommand{\natmephi}{National Research Nuclear University, MEPhI, Moscow Engineering Physics Institute, Moscow, 115409, Russia}
\newcommand{\newmex}{University of New Mexico, Albuquerque, New Mexico 87131, USA}
\newcommand{\nmsu}{New Mexico State University, Las Cruces, New Mexico 88003, USA}
\newcommand{\ohio}{Department of Physics and Astronomy, Ohio University, Athens, Ohio 45701, USA}
\newcommand{\ornl}{Oak Ridge National Laboratory, Oak Ridge, Tennessee 37831, USA}
\newcommand{\orsay}{IPN-Orsay, Univ.~Paris-Sud, CNRS/IN2P3, Universit\'e Paris-Saclay, BP1, F-91406, Orsay, France}
\newcommand{\peking}{Peking University, Beijing 100871, People's Republic of China}
\newcommand{\pnpi}{PNPI, Petersburg Nuclear Physics Institute, Gatchina, Leningrad region, 188300, Russia}
\newcommand{\riken}{RIKEN Nishina Center for Accelerator-Based Science, Wako, Saitama 351-0198, Japan}
\newcommand{\rikjrbrc}{RIKEN BNL Research Center, Brookhaven National Laboratory, Upton, New York 11973-5000, USA}
\newcommand{\rikkyo}{Physics Department, Rikkyo University, 3-34-1 Nishi-Ikebukuro, Toshima, Tokyo 171-8501, Japan}
\newcommand{\saispbstu}{Saint Petersburg State Polytechnic University, St.~Petersburg, 195251 Russia}
\newcommand{\saopaulo}{Universidade de S{\~a}o Paulo, Instituto de F\'{\i}sica, Caixa Postal 66318, S{\~a}o Paulo CEP05315-970, Brazil}
\newcommand{\seoulnat}{Department of Physics and Astronomy, Seoul National University, Seoul 151-742, Korea}
\newcommand{\stonybrkc}{Chemistry Department, Stony Brook University, SUNY, Stony Brook, New York 11794-3400, USA}
\newcommand{\stonycrkp}{Department of Physics and Astronomy, Stony Brook University, SUNY, Stony Brook, New York 11794-3800, USA}
\newcommand{\sungskku}{Sungkyunkwan University, Suwon, 440-746, Korea}
\newcommand{\tenn}{University of Tennessee, Knoxville, Tennessee 37996, USA}
\newcommand{\titech}{Department of Physics, Tokyo Institute of Technology, Oh-okayama, Meguro, Tokyo 152-8551, Japan}
\newcommand{\tsukuba}{Tomonaga Center for the History of the Universe, University of Tsukuba, Tsukuba, Ibaraki 305, Japan}
\newcommand{\vandy}{Vanderbilt University, Nashville, Tennessee 37235, USA}
\newcommand{\waseda}{Waseda University, Advanced Research Institute for Science and Engineering, 17  Kikui-cho, Shinjuku-ku, Tokyo 162-0044, Japan}
\newcommand{\weizmann}{Weizmann Institute, Rehovot 76100, Israel}
\newcommand{\wigner}{Institute for Particle and Nuclear Physics, Wigner Research Centre for Physics, Hungarian Academy of Sciences (Wigner RCP, RMKI) H-1525 Budapest 114, POBox 49, Budapest, Hungary}
\newcommand{\yonsei}{Yonsei University, IPAP, Seoul 120-749, Korea}
\affiliation{\abilene}
\affiliation{\augie}
\affiliation{\banaras}
\affiliation{\barc}
\affiliation{\baruch}
\affiliation{\bnlcoll}
\affiliation{\bnlphys}
\affiliation{\caucr}
\affiliation{\charlesczech}
\affiliation{\chonbuk}
\affiliation{\ciae}
\affiliation{\cns}
\affiliation{\colorado}
\affiliation{\columbia}
\affiliation{\czechtech}
\affiliation{\dapnia}
\affiliation{\debrecen}
\affiliation{\elte}
\affiliation{\eszterhazy}
\affiliation{\ewha}
\affiliation{\fit}
\affiliation{\fsu}
\affiliation{\gsu}
\affiliation{\hiroshima}
\affiliation{\howard}
\affiliation{\ihepprot}
\affiliation{\illuiuc}
\affiliation{\inrras}
\affiliation{\instpasczech}
\affiliation{\isu}
\affiliation{\jaea}
\affiliation{\jinrdubna}
\affiliation{\jyvaskyla}
\affiliation{\kek}
\affiliation{\korea}
\affiliation{\kurchatov}
\affiliation{\kyoto}
\affiliation{\labllr}
\affiliation{\lahorelums}
\affiliation{\lawllnl}
\affiliation{\losalamos}
\affiliation{\lpc}
\affiliation{\lund}
\affiliation{\lyon}
\affiliation{\maryland}
\affiliation{\mass}
\affiliation{\michigan}
\affiliation{\muenster}
\affiliation{\muhlenberg}
\affiliation{\myongji}
\affiliation{\nagasaki}
\affiliation{\nara}
\affiliation{\natmephi}
\affiliation{\newmex}
\affiliation{\nmsu}
\affiliation{\ohio}
\affiliation{\ornl}
\affiliation{\orsay}
\affiliation{\peking}
\affiliation{\pnpi}
\affiliation{\riken}
\affiliation{\rikjrbrc}
\affiliation{\rikkyo}
\affiliation{\saispbstu}
\affiliation{\saopaulo}
\affiliation{\seoulnat}
\affiliation{\stonybrkc}
\affiliation{\stonycrkp}
\affiliation{\sungskku}
\affiliation{\tenn}
\affiliation{\titech}
\affiliation{\tsukuba}
\affiliation{\vandy}
\affiliation{\waseda}
\affiliation{\weizmann}
\affiliation{\wigner}
\affiliation{\yonsei}
\author{A.~Adare} \affiliation{\colorado} 
\author{S.~Afanasiev} \affiliation{\jinrdubna} 
\author{C.~Aidala} \affiliation{\mass} \affiliation{\michigan} 
\author{N.N.~Ajitanand} \altaffiliation{Deceased} \affiliation{\stonybrkc} 
\author{Y.~Akiba} \email[PHENIX Spokesperson: ]{akiba@rcf.rhic.bnl.gov} \affiliation{\riken} \affiliation{\rikjrbrc} 
\author{H.~Al-Bataineh} \affiliation{\nmsu} 
\author{J.~Alexander} \affiliation{\stonybrkc} 
\author{M.~Alfred} \affiliation{\howard} 
\author{K.~Aoki} \affiliation{\kek} \affiliation{\kyoto} \affiliation{\riken} 
\author{Y.~Aramaki} \affiliation{\cns} 
% ...
\collaboration{PHENIX Collaboration} \noaffiliation

\date{\today}

%------------------------------------------------------------------------------|

\begin{abstract}

%\linenumbers 

We present measurements of azimuthal correlations of charged hadron 
pairs in $\sqrt{s_{_{NN}}}=200$ GeV Au$+$Au collisions for the 
trigger and associated particle transverse-momentum ranges of 
$1<p_T^t<10$~GeV/$c$ and $0.5<p_T^a<10$~GeV/$c$. After subtraction 
of an underlying event using a model that includes higher-order 
azimuthal anisotropy $v_2$, $v_3$, and $v_4$, the away-side yield 
of the highest trigger-\pt ($p_T^t>4$~GeV/$c$) correlations is 
suppressed compared to that of correlations measured in $p$$+$$p$ 
collisions. At the lowest associated particle $p_T$ ($0.5<p_T^a<1$ 
GeV/$c$), the away-side shape and yield are modified relative to 
those in $p$$+$$p$ collisions. These observations are consistent 
with the scenario of radiative-jet energy loss. For the low-$p_T$ 
trigger correlations ($2<p_T^t<4$ GeV/$c$), a finite away-side 
yield exists and we explore the dependence of the shape of the 
away-side within the context of an underlying-event model.  
Correlations are also studied differentially versus event-plane 
angle $\Psi_2$ and $\Psi_3$. The angular correlations show an 
asymmetry when selecting the sign of the difference between the 
trigger-particle azimuthal angle and the $\Psi_2$ event plane. This 
asymmetry and the measured suppression of the pair yield out of 
plane is consistent with a path-length-dependent energy loss. No 
$\Psi_3$ dependence can be resolved within experimental 
uncertainties.

\end{abstract}

\maketitle

\section{Introduction}

Energy loss of hard-scattered partons (jet quenching \cite{Wang:1991xy}) 
resulting from the interaction of a colored parton in the quark gluon 
plasma~(QGP) formed in relativistic heavy ion collisions at the 
Relativistic Heavy Ion Collider 
(RHIC)~\cite{Arsene:2004fa,Back:2004je,Adams:2005dq,Adcox:2004mh} has 
been observed in several different ways.  Suppression of single-particle 
and single-jet invariant yields in central A+A 
collisions~\cite{Adler:2003qi,Adler:2003au,Adams:2003im,CMS:2012aa,Aad:2014bxa,Aad:2012vca,Adam:2015ewa} 
provide a baseline measurement of jet quenching.  Measurements of 
correlations between two particles and/or jets give more detailed 
information of the jet quenching process inside the medium 
\cite{Renk:2012hz,Qin:2015srf}. The first jet suppression effect 
observed in azimuthal correlations was an attenuation of the away-side 
yields in high-transverse momenta (\pt) correlations in the most central 
Au$+$Au collisions at $\sqrt{s_{_{NN}}}=$~200~GeV~\cite{Adler:2002tq}.  
The centrality dependence of high-\pt $\pi^0$-hadron 
correlations~\cite{Adare:2010mq} shows a monotonic attenuation of the 
away-side yields with increasing propagation length of partons through 
the medium.  In addition to away-side yield suppression, direct 
photon-hadron 
correlations~\cite{Abelev:2009gu,Chatrchyan:2012gt,STAR:2016jdz}, 
two-particle 
correlations~\cite{Adare:2012qi,Agakishiev:2011st,Nattrass:2016cln}, and 
jet-hadron 
correlations~\cite{Adamczyk:2013jei,Aad:2014wha,Chatrchyan:2014ava,Khachatryan:2016erx} 
show that low-momentum particles correlated with high-\pt jets are 
enhanced in yield, especially at large angles with respect to the jet 
axis.  This may be attributable to the radiation from the parent parton 
or other lost energy absorbed by the surrounding medium. Thus, 
two-particle angular correlations have provided much of the experimental 
information we have about jet energy 
loss~\cite{Aamodt:2011vg,Adare:2010ry,Adare:2010mq,Adams:2006yt,Abelev:2009ah,Agakishiev:2011nb}.

It is important to understand the interactions of partons with the QGP 
at all scales from the hard-scattering scale to the thermal scale. Below 
$E_{\rm jet}\approx $10 GeV, full jet reconstruction is much more 
difficult due the underlying event subtraction. Two-particle 
correlations are important because they can probe lower jet (or parton) 
energies. However, observations of energy loss effects mentioned above, 
especially for lower jet and particle momenta in two-particle 
correlations, have been obscured due to the much larger contribution 
from the underlying event at these momenta.

The underlying event modulations are attributed to hydrodynamic 
collective flow patterns where the importance of higher-order flow 
harmonics was established more 
recently~\cite{Alver:2010gr,Xu:2010du,ATLAS:2012at,Adare:2011tg,ALICE:2011ab,Chatrchyan:2013kba}.  
These patterns are thought to result from the hydrodynamic response of 
the QGP to fluctuating initial geometrical shapes of the overlap region 
of the colliding nuclei. Many hydrodynamic models have been developed 
which capture these effects~\cite{Shen:2014vra,Lin:2004en}, but to date, 
important details of these models are still under development, and their 
full implementation requires involved calculations.  This motivates the 
use of a simpler data-driven model, which will be explored in this work.

The shape of the collective flow in the transverse plane is 
parameterized~\cite{Voloshin:1994mz,Ollitrault:1992bk,Poskanzer:1998yz} 
by a Fourier expansion with
\begin{eqnarray}
v_{n}\left\{\Psi_m\right\}=\left<\cos{n(\phi-\Psi_m)}\right>
\end{eqnarray}
where $v_n$ is the $n^{th}$-order anisotropic flow coefficient, $\phi$ 
is the azimuthal angle of emitted particles, and $\Psi_m$ is the event 
plane angle defined by the $m^{th}$-harmonic number.  For the first 
decade of RHIC operations, only the even harmonics and frequently only 
the $n=2$ term, were considered. The shapes of two-particle correlations 
after subtraction of the $n = 2$-only background motivated the 
introduction of the other harmonics, most importantly $n = 3$ 
\cite{Sorensen:2010zq,Alver:2010gr,Adare:2011tg,ATLAS:2012at,ALICE:2011ab}. 
Under the two-source (flow + jet) model assumption~\cite{Adler:2005ee}, 
this underlying event is directly subtracted to obtain the jet 
contributions. In our previous measurements and most RHIC $A$$+$$A$ 
results, the subtracted flow modulations of the underlying event were 
limited to contributions of $v_2$ and the fourth-order harmonic 
component with respect to the second-order event plane 
$v_4\{\Psi_2\}$~\cite{Adler:2002tq,Abelev:2009af,Adare:2007vu,Adler:2005ee,Adare:2008ae,Adamczyk:2013jei,Adare:2010mq, 
Adare:2010ry,Adare:2006nr,Adare:2012qi}. Only the recent STAR 
measurement~\cite{Agakishiev:2014ada} took into account contributions 
from $v_3$ and the fourth-order harmonic component uncorrelated to the 
second-order event-plane in addition to $v_4\{\Psi_2\}$. 

At low to intermediate \pt in two-particle correlations, intricate 
features appear such as the near-side long-range rapidity correlations 
called the ``ridge''~\cite{Abelev:2009af,Alver:2009id} and the away-side 
``double-humped'' 
structures~\cite{Adler:2005ee,Adare:2006nr,Adare:2008ae,Abelev:2008ac,Adare:2007vu,Aggarwal:2010rf,Adare:2010ry,Agakishiev:2014ada}. 
Across the large rapidity ranges available at the Large Hadron Collider, 
the rapidity-independence and hence the likely geometrical origin of 
most of these structures have been established.  Experiments have shown 
that the ridge and the double-hump structures in the two-particle 
azimuthal correlations for $|\Delta\eta|>1$ for ALICE and 
$|\Delta\eta|>2$ for ATLAS and CMS measured in $p$$+$$p$, $p$$+$Pb, and 
Pb$+$Pb collisions at $\sqrt{s_{_{NN}}}=2.76$ and 
$5$~TeV~\cite{ALICE:2011ab,ATLAS:2012at,Chatrchyan:2012wg} are the same 
in shape and size at much larger rapidity differences. Both the ridge 
and double-hump are successfully explained by the higher-order 
harmonics.  However the mechanism for how the jet correlations combine 
with the flow correlations, especially at small $\Delta\eta$, to yield 
the total two-particle correlation has not been clarified. In 
particular, the correlations left after subtracting a flow-based model 
at small $\Delta\eta$ have not been analyzed in detail.

In this work, we assume a two-source model where the total pair yield is 
a sum of a jet-like component and an underlying-event component.  The 
underlying-event components is modeled using the flow harmonics 
$v_n(n=2,3,4)$, event plane resolutions, and the most important event 
plane correlations between $\Psi_2$ and $\Psi_4$. We assume that the 
$v_n$ measured through the event plane method are the the same as those 
in the correlation functions. Event-by-event $v_n$ 
fluctuations~\cite{Aad:2013xma}, $v_n-v_m$ correlations between 
different orders~\cite{Aad:2015lwa}, normalized symmetric 
cumulants ($v^2_n-v^2_m$ correlations)~\cite{Acharya:2017gsw}, and 
rapidity dependent event plane decorrelations~\cite{Khachatryan:2015oea} 
are not included in this background model. To take into account the 
$v_n-v_m$ and $v^2_n-v^2_m$ correlations in this background model, 
measurements of their original two-dimensional probability distributions 
are necessitated for fine \pt selections. To evaluate a possible effect 
from the $v_n-v_m$ and $v^2_n-v^2_m$ correlations, we performed a toy 
Monte Carlo simulation with the same framework reported in this article, 
assuming a two-dimensional Gaussian with a correlation term between 
$v_n-v_m$. The changes expected are less than the systematic uncertainty 
for $v_n$. Measurements of the rapidity-odd component of the directed 
flow $v_1^{\rm odd}$ using the event plane 
method~\cite{Adams:2004bi,Back:2005pc,Abelev:2008jga,Adamczyk:2014ipa} 
generally yield $v_1^{\rm odd}\sim0$ at $\eta=0$ integrated over all 
\pt. Finite values of \pt differential measurements of $v_1^{\rm 
odd}(\pt)$~\cite{Abelev:2008jga} include momentum conservation and jet 
(mini-jet) effects which are considered signal in this two-particle 
correlation analysis. The rapidity-even component of the directed flow 
$v_1^{\rm even}$ is considered to result from collective expansion of 
the medium. Measurements of $v_1^{\rm even}$ with respect to the 
spectator event plane with the scalar product 
method~\cite{Abelev:2013cva} show its magnitude about 40 times smaller 
than that with respect to the participant event plane obtained with the 
Fourier fits to the two-particle 
correlations~\cite{ATLAS:2012at,Chatrchyan:2012wg}. These observations 
indicate that $v_1^{\rm even}$ has different sensitivity to the 
spectator and participant event planes and warrant further validation of 
the momentum conservation model in the Fourier fits to the two-particle 
correlations. There is currently no concrete $v_1^{\rm even}$ to 
subtract as background. With this reason we do not include contributions 
from $v_1\{\Psi_1\}$ and event plane correlations involving $\Psi_1$ in 
the background model. For the inclusive trigger correlations, we 
estimated a potential impact of $v_5$ modulation using an empirical 
relation $v_5 \sim 0.5 \times v_4$ found in ATLAS $v_n$ 
measurements~\cite{ATLAS:2012at}.

After subtracting the underlying event with the model, we study the 
structures observed at high \pt where the flow backgrounds are 
negligible. Because the jet signal-to-flow background is significantly 
reduced in the low to intermediate \pt region, studying the correlations 
there provides a more stringent test of such a background model. Any 
features left in the residuals can be used to reveal jet energy-loss 
effects at low and intermediate \pt. However, because of our simple 
model, only substantially significant correlations can be attributable 
to the medium effect on jets (i.e.~broadening or suppression) or the 
medium response (i.e.~yields at large angles from the jet).

%=================================================== Fig_1
\begin{figure}[htb]
\includegraphics[width=0.75\linewidth]{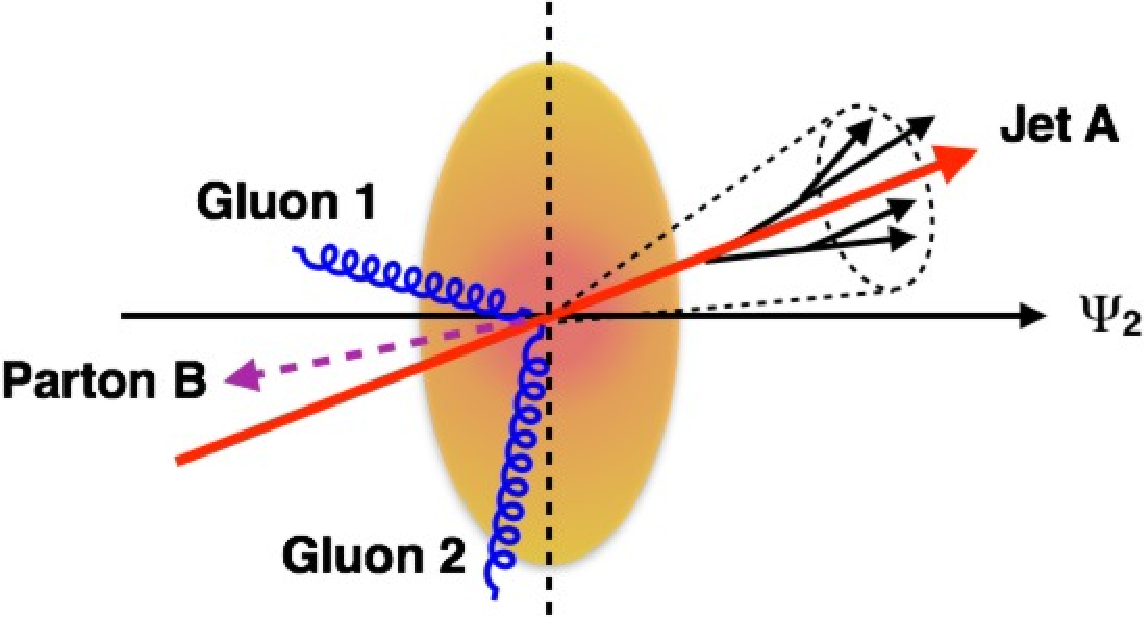} %%01
\caption{
 Two possible gluons (Gluon 1 and Gluon 2) radiated opposite to 
particles detected from Jet A with different medium path lengths. The 
difference in energy loss could lead to asymmetric correlated particle 
yields in the hemisphere to the left of Jet A compared to the right of Jet A.
}
\label{fig:InplaneJetAndGluons}
\end{figure}

%=================================================== Fig_2
\begin{figure}[htb]
\includegraphics[width=0.9\linewidth]{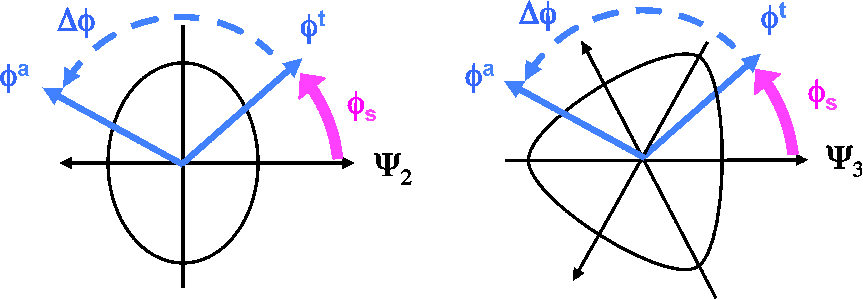}  %%02
\caption{
Schematic picture of a trigger particle selection with respect to 
event-planes and pairing a trigger particle with an associate particle.
}
\label{fig:TrigPsi2Psi3}
\end{figure}

An important goal of jet quenching studies has been to determine the 
density and path-length dependence of energy loss~\cite{Adare:2010mq}. 
Perturbative models of radiative jet quenching and strongly 
coupled jet quenching models predict a different path-length dependence 
for the quenching~\cite{Marquet:2009eq}. Varying the path length by 
selecting azimuthal orientations relative to the second-order event 
plane has been explored for single-particle or single-jet observables at 
high-\pt~\cite{Afanasiev:2009aa,Aad:2012vca}.  Potentially more 
differential information can be obtained from two-particle observables 
coupled with the event-plane.
Figure~\ref{fig:InplaneJetAndGluons} illustrates the trigger (Jet A) being 
emitted to one side of the in-plane direction and the away-side jet 
(Parton B) radiating two gluons (Gluon 1) and (Gluon 2). We, therefore, 
also study two-particle correlations measured differentially with 
respect to $\Psi_2$ and $\Psi_3$ event planes as depicted in 
Fig.~\ref{fig:TrigPsi2Psi3}. Such differential correlations probe the 
path-length and geometrical dependence of energy loss with more 
event-by-event sensitivity and also extend similar studies of high-\pt 
correlations~\cite{Adare:2010mq} down to lower-\pt. We use a new 
method of distinguishing ``left/right'' asymmetry in the $\Psi_n$ 
correlations, which provides more information on the 
background dominated low and intermediate \pt regions by probing 
possible asymmetric parton energy loss because of medium geometry.

In this article: Section~\ref{sec:PHENIXDETECTOR} describes the detector 
set-up of the PHENIX~Experiment. 
Sections~\ref{ssec:ParticleSelections},~\ref{ssec:higher-order-flow-harmonics}, 
and \ref{ssec:two-particle-correlations} describe the analysis 
methodology of particle selections, higher-order flow harmonics, and 
two-particle correlations, respectively. 
Section~\ref{sec:ResultsandDiscussion} presents analysis results and 
discusses their interpretations.  This section first starts with the 
highest \pt trigger selections, \pt $\gtrsim 4$ GeV/$c$, and makes 
connections to known energy-loss effects. Next, lower trigger 
correlations down to 1 GeV/$c$ are presented. Finally the event-plane 
dependence of the intermediate \pt selections are investigated. 
Section~\ref{Summary} summarizes this article.

\section{PHENIX DETECTOR}
\label{sec:PHENIXDETECTOR}

%%% FIGURE : 2007 PHENIX Detector Configuration %%%
%=================================================== Fig_3
\begin{figure}[htb]
\includegraphics[width=1.0\linewidth]{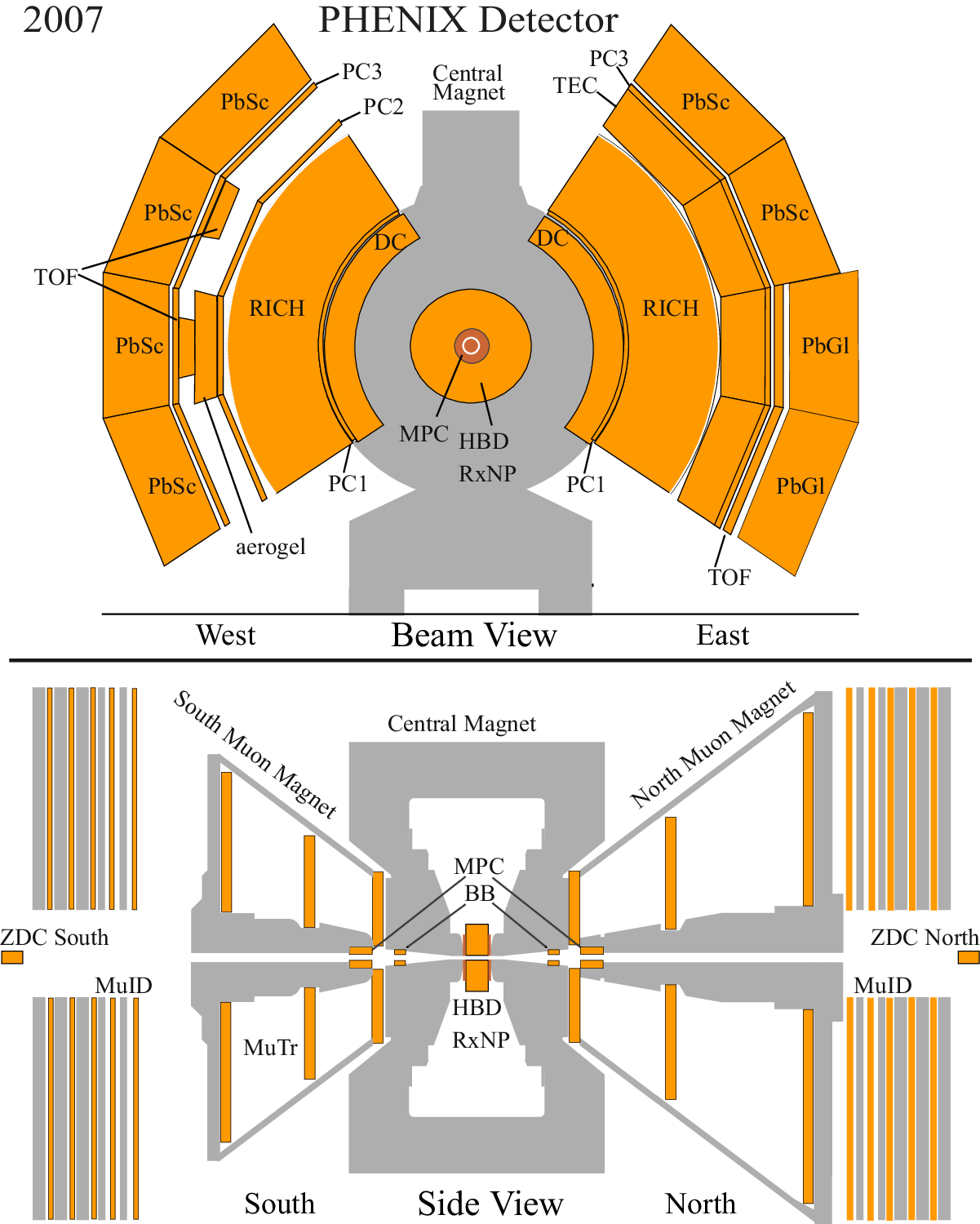}  %%03
\caption{
The PHENIX detector configuration in the 2007 experimental run period. 
The top panel shows the central arm detectors viewed from the beam 
direction. The bottom panel shows the global detectors and muon arm 
viewed from the side perpendicular to the beam direction.
}
\label{fig:phenix-detector}
\end{figure}

The PHENIX detector~\cite{Adcox:2003zp} was designed to measure charged 
hadrons, leptons, and photons to study the nature of the QGP formed in 
ultra-relativistic heavy ion collisions. 
Figure~\ref{fig:phenix-detector} shows the beam view and side view of 
the PHENIX detector including all subsystems for this data taking 
period.

The global detectors, which include the beam-beam counters (BBC), the 
zero-degree calorimeters (ZDC), and the reaction-plane detector (RXN), 
were used to determine event characterizing parameters such as the 
collision vertex, collision centrality, and event-plane orientation. They 
are located on both the south and north side of the PHENIX detectors. 
The BBC is located at $\pm$144 cm ($3<|\eta|<3.9$) from the beam 
interaction point and surrounds the beam pipe with full 
$\Delta\phi=2\pi$ azimuthal acceptance. Each BBC module comprises 64 
quartz \v{C}erenkov radiators equipped with a photomultiplier tube (PMT) 
and measures the total charge (which is proportional to the number of 
particles) deposited in its acceptance. The BBC determines the beam 
collision time, beam collision position along the beam axis direction, 
and collision centrality. The ZDCs~\cite{Adler:2000bd}, located at 18~m 
away from the nominal interaction point, detect the energy deposited by 
spectator neutrons of the two colliding nuclei. The PHENIX minimum-bias 
trigger is provided by the combination of hit information in the ZDC and 
BBC, which requires at least one hit in both the ZDC modules and two 
hits in the BBC modules.

The orientation of higher-order event planes is determined by the BBC 
and the RXN~\cite{Richardson:2010hm}, which have different $\eta$ 
acceptance. The RXNs are located at $\pm$38 cm from the beam interaction 
point and have two rings in each module; RXN-inner and RXN-outer are 
installed to cover $1.5<|\eta|<2.8$ and $1<|\eta|<1.5$, respectively. 
Each ring has 12 scintillators in its azimuthal angle acceptance 
$\Delta\phi=2\pi$.

Charged hadron tracks are reconstructed in the PHENIX central arm 
spectrometer (CNT), which is comprised of two separate arms, east and 
west.  Each arm covers $\left|\eta\right|<0.35$ and $\Delta\phi = 
\pi/2$.

The PHENIX tracking system is composed of the drift chamber (DC) in 
addition to two layers of pad chambers (PC1 and PC3) in the east arm and 
three layers of pad chambers (PC1, PC2, and PC3) in the west arm.  
Momentum is determined by measuring the track curvature through the 
magnetic field by means of a Hough transform with hit information from 
the DC and PC1 with a momentum resolution of $\delta p/p = 1.3\% \oplus 
1.2\%p$~\cite{Adare:2012vq}. Additional track position information is 
provided by the outer layers of the pad chambers and the electromagnetic 
calorimeter (EMCal), which are Lead Glass (PbGl) and Lead Scintillator (PbSc).

The ring imaging \v{C}erenkov counter (RICH) and the EMCal identify and 
exclude electron tracks from the analysis.  The RICH produces a light 
yield for electrons with \pt$>$30~MeV and for pions with \pt$>$5~GeV, 
meaning that a signal in the RICH can be used to separate electrons and 
pions below 5~GeV. Above 5~GeV where this is no longer possible, the 
energy deposited in the EMCal can be used for this separation.  Electrons 
will deposit much more of their total energy than pions will, so that 
the ratio of deposited energy to track momentum is significantly higher 
for electrons than for pions.

\section{Analysis Methodology}
The results presented are based on an analysis of 4.38 billion 
minimum-bias events for Au$+$Au collisions at $\sqrt{s_{_{NN}}}=$~200~GeV 
recorded by the PHENIX detector at RHIC in 2007.

\subsection{Particle Selection}
\label{ssec:ParticleSelections}

Charged hadrons are selected from candidate tracks using cuts similar to 
previous correlation analyses~\cite{Adare:2012qi}. One important cut to 
reject fake tracks, especially decays in the central magnetic field 
before the drift chamber, is an association cut to outer CNT detectors. 
The track trajectories are projected onto outer CNT detectors. The 
nearest hits in the PC3 and the EMCal from the projections are identified 
as hits for the track. The distributions of the distance in the 
azimuthal ($\phi$) and beam ($z_{beam}$) directions between the hits in 
the PC3 and the EMCal and the extrapolated line are fitted with a 
double-Gaussian.  One Gaussian arises from background and the other from 
the signal. Hadron tracks are required to be within $\pm2\sigma$ of the 
signal Gaussian mean in both the $\phi$ and $z_{beam}$ directions in 
both the PC3 and the EMCal. To veto conversion electrons, tracks with 
\pt$<$5 GeV/$c$ having one or more \v{C}erenkov photons in the RICH are 
excluded from this analysis. For \pt$>$5~GeV/$c$, we require $E_{\rm EMCal} > 
0.3 + 0.2$c$ \times \pt$~GeV~\cite{Adler:2005ad,Adare:2008ae}, where 
$E_{\rm EMCal}$ is the cluster energy associated with the track.

\subsection{Higher-Order Flow Harmonics $v_n$}
\label{ssec:higher-order-flow-harmonics}

\subsubsection{Event-plane and Resolution}
\label{sssec:Event-planeandResolution}

Each event plane $\Psi_n$ is determined event-by-event for different 
harmonic numbers $n$ using the RXN and BBC detectors. The RXN detectors 
are used to measure the nominal values of $v_n$ while the BBC detectors 
provide systematic checks to the extracted $v_n$ values. The observed 
event-plane $\Psi_n^{\rm obs}$ is reconstructed as
\begin{eqnarray}
\Psi_n^{\rm obs}&=&\frac{1}{n}\tan^{-1}{\left(\frac{Q_{n,y}}{Q_{n,x}}\right)}.
\end{eqnarray}
Here $Q_{n,x}$ and $Q_{n,y}$ are the flow vector components
\begin{eqnarray}
Q_{n,x}&=&\sum_i w_i \cos(n\phi_i)/\sum_i w_i\\
Q_{n,y}&=&\sum_i w_i \sin(n\phi_i)/\sum_i w_i
\end{eqnarray}
where $\phi_i$ is the azimuthal angle of the $i$-th segment in the 
event-plane detector and $w_i$ is the weight proportional to 
multiplicity in the $i$-th segment. We apply the re-centering and the 
flattening corrections~\cite{Barrette:1997pt,Poskanzer:1998yz} 
separately for each sub-event event-plane.

The $k\times n$~th-order resolution of $n$~th-order event plane is 
defined as ${\rm 
Res}\{kn,\Psi_n\}=\left<\cos{kn(\Psi_n^{\rm obs}-\Psi_n)}\right>$ and can be 
expressed as \cite{Poskanzer:1998yz}
\begin{eqnarray}
{\rm Res}\{kn,\Psi_n\}&=&\frac{\sqrt\pi}{2\sqrt{2}}\chi_n e^{-\frac{\chi_{n}^2}{4}} \nonumber \\
&\times&\left[I_{\frac{k-1}{2}}\left(\frac{\chi^2_n}{4}\right)	+I_{\frac{k+1}{2}}\left(\frac{\chi^2_n}{4}\right)\right] \label{eq:BesselReso}
\end{eqnarray}
where $\chi_n=v_n\sqrt{2M}$, $M$ is the multiplicity used to determine 
the event-plane $\Psi_n$, and $I_{k}$ is the modified Bessel function of 
the first kind.

\begin{figure}[htb]
\includegraphics[width=1.0\linewidth]{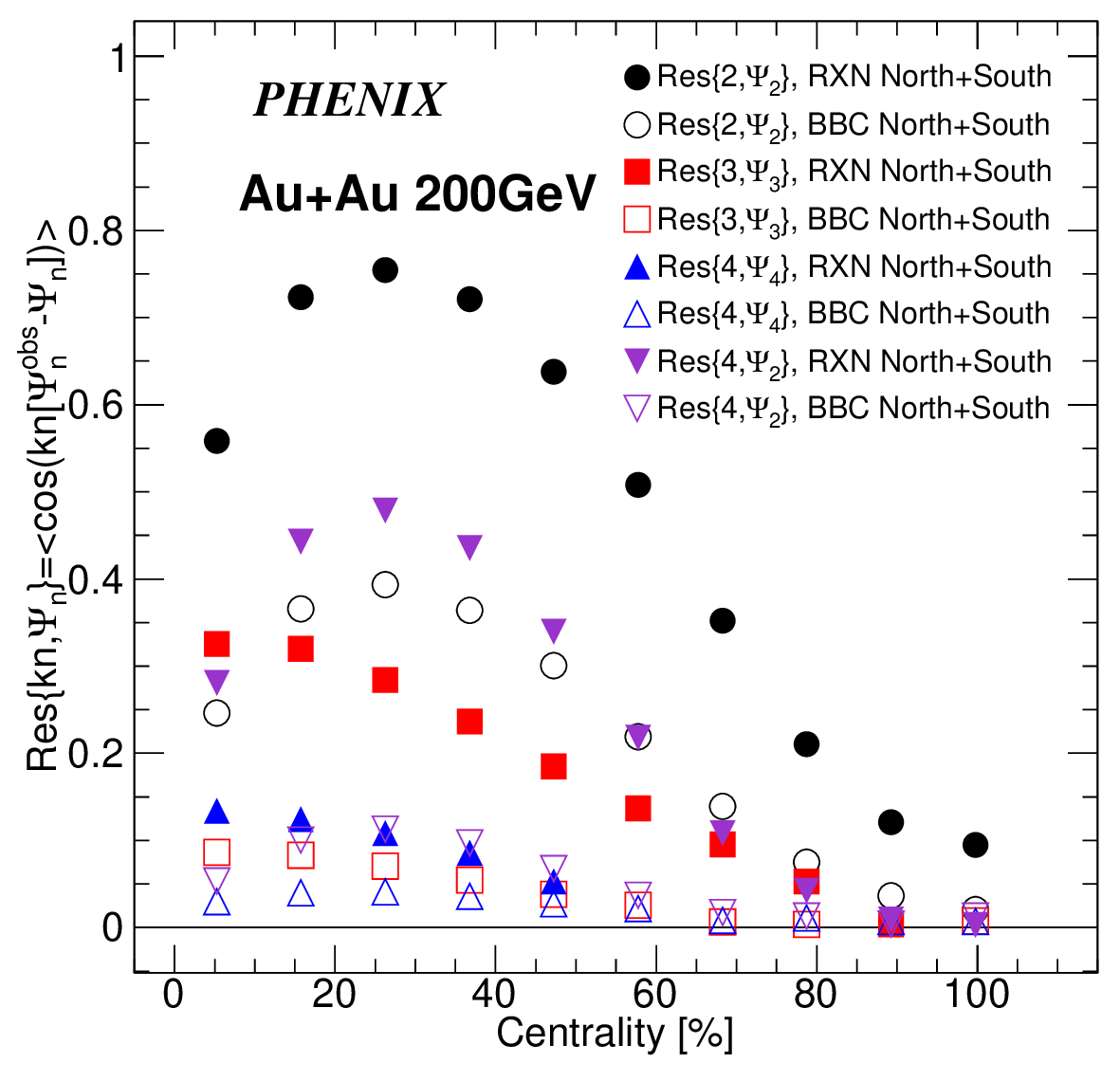}   %%04
\caption{Event-plane resolutions
\mbox{${\rm Res}\{2,\Psi_2\}$},
\mbox{${\rm Res}\{3,\Psi_3\}$},
\mbox{${\rm Res}\{4,\Psi_4\}$}, and
\mbox{${\rm Res}\{4,\Psi_2\}$}
obtained by the combination of the north and south modules of RXN and BBC.
}
\label{fig:EPResolution}
\end{figure}

Because the north (N) and south (S) modules of a given event-plane 
detector have the same pseudorapidity coverage and see the same 
multiplicity and energy for symmetric nucleus-nucleus collisions, the 
north and south modules should have identical resolution in case of no detector biases. We obtain the 
event-plane resolution of an event-plane detector using the two subevent 
method \cite{Poskanzer:1998yz}.
\begin{eqnarray}
{\rm Res}\{kn,\Psi_n\}&=&\left<\cos{kn(\Psi_n^{\rm obs}-\Psi_n)}\right> \nonumber \\
&=&\sqrt{\left<\cos{kn(\Psi_n^{N,obs}-\Psi_n^{S,obs})}\right>}.
\end{eqnarray}
The north+south combined event-plane resolution is determined from 
Eq.~(\ref{eq:BesselReso}) with $\chi_n$ = $\sqrt{2}\chi_{n}^{N,S}$. The 
factor of $\sqrt{2}$ accounts for twice the multiplicity in north+south 
compared to north or south.  Figure~\ref{fig:EPResolution} shows the the 
north+south combined event-plane resolution for both RXN and BBC.

%=================================================== Fig_4

\subsubsection{$v_n$ measurements}

Higher-order flow harmonics 
$v_n$~\cite{Voloshin:1994mz,Poskanzer:1998yz,Alver:2010gr} are measured 
by the event-plane method~\cite{Poskanzer:1998yz}. Charged hadron tracks 
with azimuthal angle $\phi$ are measured with respect to the event plane 
angle $\Psi_n^{\rm obs}$. The flow coefficients $v_{kn}$ are measured as an 
event-average and track-average and correcting by the event plane 
resolution.
\begin{eqnarray}
v_{kn}\{\Psi_n\}=\left<\cos{kn(\phi-\Psi_n^{\rm obs})}\right>/{\rm Res}\{kn,\Psi_n\}.
\end{eqnarray}
Four different observables are studied: $v_2\{\Psi_2\}$, 
$v_3\{\Psi_3\}$, $v_4\{\Psi_4\}$, and $v_4\{\Psi_2\}$. The flow 
harmonics are measured by the nine possible combinations of RXN modules: 
south-inner, south-outer, south-inner+outer, north-inner, north-outer, 
north-inner+outer, south+north-inner, south+north-outer, and 
south+north-inner+outer. The $v_n$ reported is an average over the nine 
different possible RXN combinations, $v_n=\sum_i^{9} v_n^{(i)}/9$, where 
$v_n^{(i)}$ is the flow harmonic in one of the nine RXN module 
combinations.

\subsubsection{Systematic uncertainties and $v_n$ results}
The systematic uncertainties in $v_n$ measurements are from the 
following sources:
\begin{itemize}
\item differences among RXN modules,
\item matching cut width for CNT hadron tracks,
\item rapidity-separation dependence between event-planes and CNT tracks.
\end{itemize}

\begin{figure*}[htb]
\includegraphics[width=0.99\linewidth]{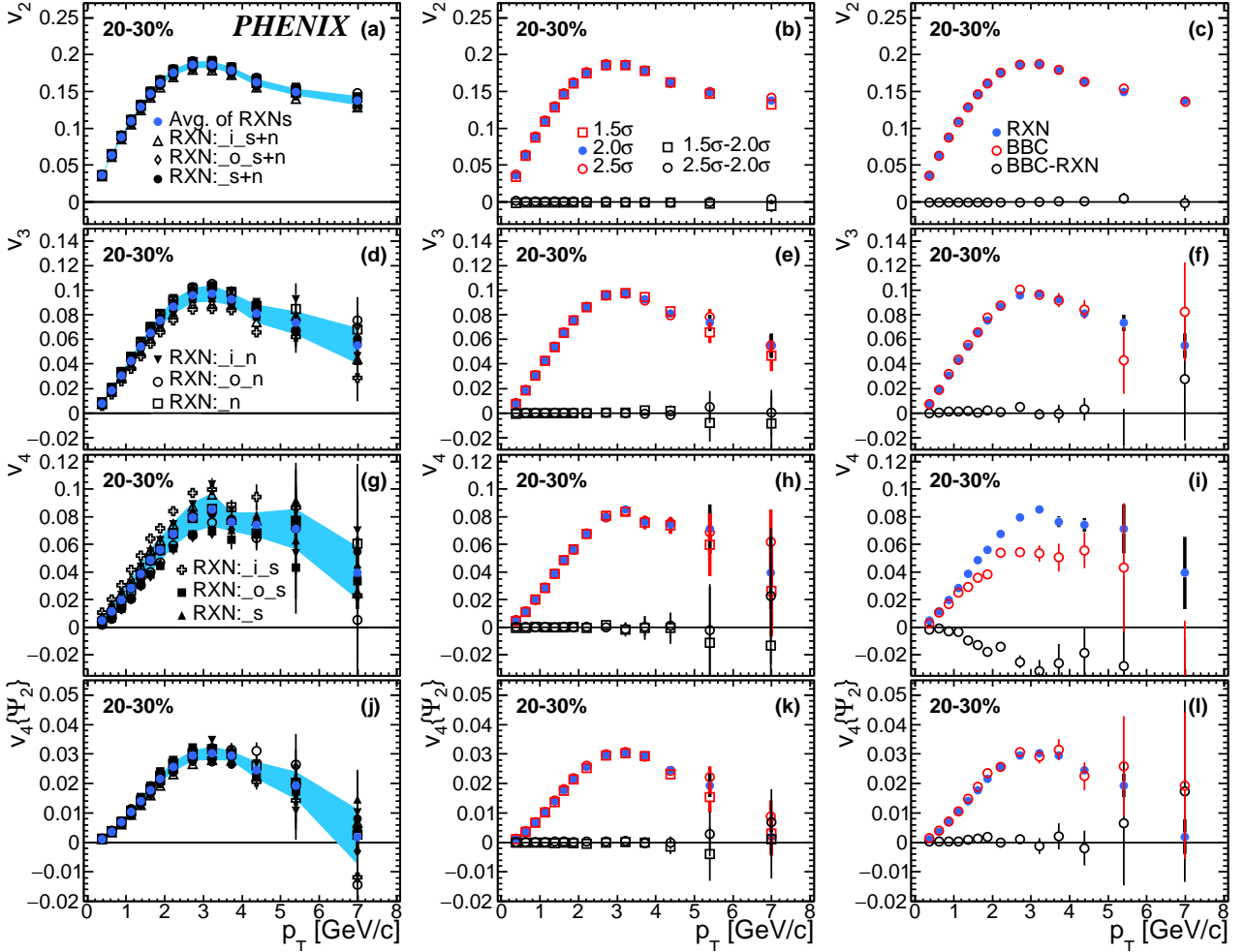}  %%05
\caption{
Higher-order flow harmonics for charged hadrons at midrapidity in Au$+$Au 
collisions at \Gsqsn and their systematics: $v_2$ 
(a-c), $v_3$ (d-f), $v_4$ (g-i), and $v_4\{\Psi_2\}$ (j-l). The source 
of systematic uncertainties are difference among RXN event-planes (a, d, 
g, j), matching cut width for CNT hadron tracks (b, e, h, k), and 
difference between $v_n$ measured with RXN and BBC event planes (c, f, 
i, l). Systematic uncertainties are shown as a shaded band in (a, d, g, 
j) and as an open marker in (b, e, h, k) and (c, f, i, l).
}
\label{fig:v2sys}
\end{figure*}

The systematic uncertainties in the RXN detector $\sigma_{\rm RXN}$ are 
defined by the standard deviation of $v_n$
\begin{eqnarray}
\sigma_{\rm RXN}=\sqrt{\sum_i^9 (v_n^{(i)}-v_n)^2/9}.
\end{eqnarray}
As an example, $v_n$ in 20\%--30\% central collisions measured by different 
RXN event-planes are shown in Fig.~\ref{fig:v2sys} (a,~d,~g,~j). The 
(blue) band indicates $\sigma_{\rm RXN}$.

To evaluate the systematic uncertainty due to track matching, the 
matching cut was varied by $\pm0.5\sigma$ from the nominal $2\sigma$ 
window. We calculated the uncertainty $\sigma_{\rm mat}$ as the average 
deviation between the $v_n$ with the nominal cut and the varied cut
\begin{eqnarray}
\sigma_{\rm mat}=(\left|v_n^{2.5\sigma}-v_n^{2\sigma}\right|+\left|v_n^{1.5\sigma}-v_n^{2\sigma}\right|)/2.
\end{eqnarray}
The variation due to the track matching cut is illustrated in 
Fig.~\ref{fig:v2sys} (b,~e,~h,~k) by showing the $v_n$ in 20\%--30\% 
central collisions measured with tracks having a matching cut of 
1.5$\sigma$, 2$\sigma$ and 2.5$\sigma$. The differences between the 
nominal $2\sigma$ and both $1.5\sigma$ and $2.5\sigma$ are also shown 
and scatter around zero indicating the size of $\sigma_{\rm mat}$.

The systematic uncertainties associated with the rapidity gap between particles and the event-plane $\sigma_{rap}$ are defined by the absolute difference between $v_n$ determined by the RXN average and $v_n$ determined by the BBC.
\begin{eqnarray}
\sigma_{rap}=\left|v_n^{\rm BBC}-v_n^{\rm RXN}\right|.
\end{eqnarray}
The $v_n$ measured with the RXN, the BBC, and their difference are shown in 
Fig.~\ref{fig:v2sys} (c,~f,~i,~l). Except in the case of $v_4$, this systematic 
uncertainty is much less than the uncertainty due to the RXN module variation. The 
small variation in the rapidity gap indicates that the contamination from nonflow correlations
does not dominate the uncertainty on the extraction of $v_n$.

The total systematic uncertainties $\sigma_{v_n}$ are the quadrature sum 
of these individual systematic uncertainties
\begin{eqnarray}
\sigma_{v_n}=\sqrt{\sigma_{\rm RXN}^2+\sigma_{\rm mat}^2+\sigma_{rap}^2}.
\end{eqnarray}
These total systematic uncertainties are conservatively assigned symmetrically.
In nearly all $p_T$ and centrality classes, the RXN 
systematic uncertainty dominates the total uncertainty.

\begin{figure*}[htb]
\includegraphics[width=0.99\linewidth]{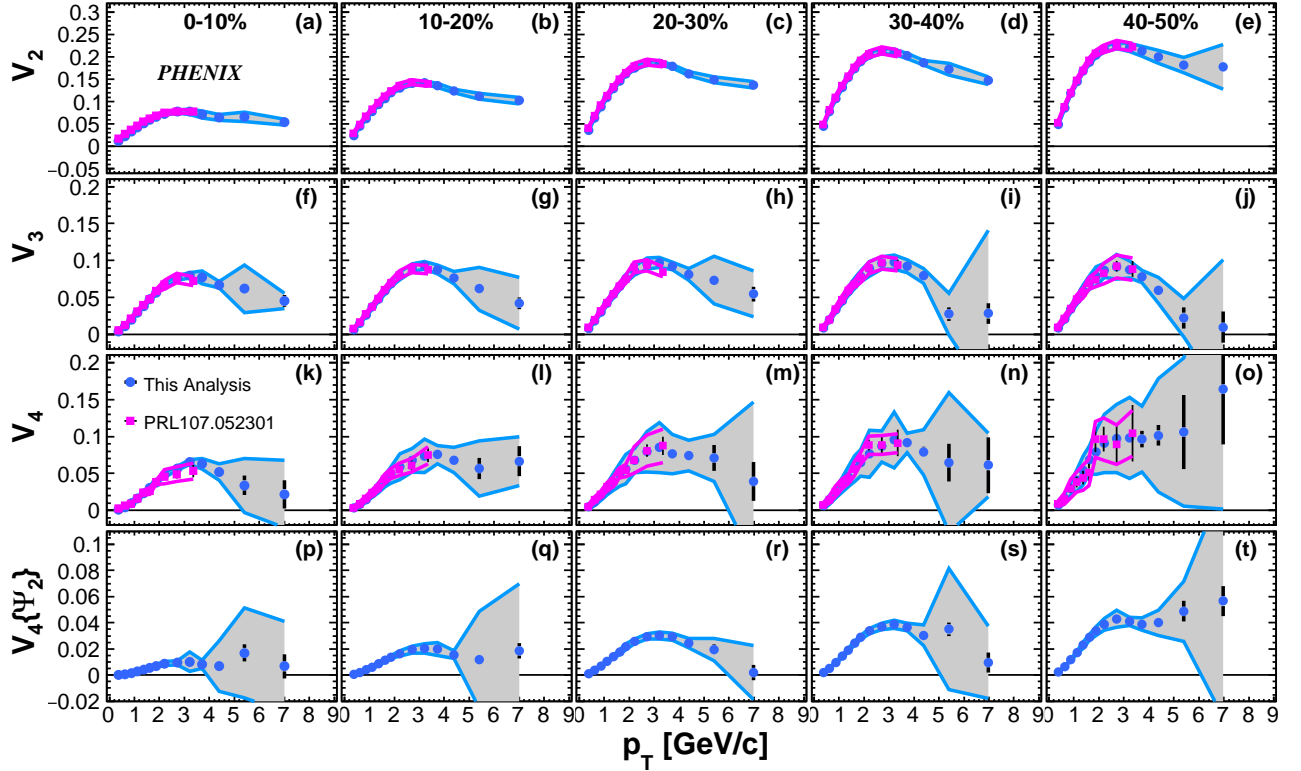}  %%06
\caption{
Higher-order flow harmonics for charged hadrons at midrapidity in Au$+$Au 
collisions at $\sqrt{s_{_{NN}}}=200$~GeV. Coefficients are determined 
using the event-plane method for $v_2$ ((a)-(e)), $v_3$ ((f)-(j)), $v_4$ 
((k)-(o)), and $v_4\{\Psi_2\}$ ((p)-(t)). The columns represent 
centrality bins 0\%--10\% (a,f,k,p), 10\%--20\% (b,g,l,q), 20\%--30\% 
(c,h,m,r), 30\%--40\% (d,i,n,s), and 40\%--50\% (e,j,o,t). Coefficients 
obtained in this analysis are shown by blue points and those measured in 
Ref.~\cite{Adare:2011tg} are shown by magenta points. Shaded bands and 
magenta lines indicate systematic uncertainties of those measurements.
}

\label{fig:vn-all}
\end{figure*}

The $v_n$ results are shown in Fig.~\ref{fig:vn-all} and compared with 
previous PHENIX $v_n$ measurements~Ref.~\cite{Adare:2011tg}. They are 
consistent within uncertainties where they overlap. For the two-particle 
correlations, we calculate $v_n$ in four large $p_T$ bins as indicated in Table~\ref{tab:vntable}.

%=================================================== Fig_5

%-------------------------------------------------- Table I
\begin{table*}[htb]
  \caption{
Data table for $v_2$, $v_3$, $v_4$ and $v_4\left\{\Psi_2\right\}$ 
(\%) in Au$+$Au collisions at \Gsqsn. 
The first uncertainties are statistical while the second uncertainties 
are total systematic. In all instances the statistical error is not 
identically zero but it is much smaller than the systematic uncertainty.
  }
  \begin{ruledtabular}\begin{tabular}{cccccc}
    Centrality & $p_T$ GeV/$c$ & $v_2 (\%)$ & $v_3 (\%)$ & $v_4  (\%)$ & $v_{4}\left\{\Psi_{2}\right\}  (\%)$ \\
    \hline
    0\%--10\% & 0.5--1.0 & $2.67\pm0.00\pm0.14$ & $1.45\pm0.00\pm0.09$ & $0.64\pm0.01\pm0.16$ & $0.109\pm0.004\pm0.052$ \\
    & 1.0--2.0 & $4.92\pm0.00\pm0.21$ & $3.67\pm0.00\pm0.18$ & $2.19\pm0.01\pm022$ & $0.396\pm0.005\pm0.058$ \\
    & 2.0--4.0 & $7.39\pm0.00\pm0.34$ & $6.96\pm0.01\pm0.29$ & $5.12\pm0.01\pm0.23$ & $0.90\pm0.01\pm0.10$ \\
    & 4.0--10 & $6.46\pm0.00\pm0.67$ & $6.61\pm0.01\pm0.41$ & $5.0\pm0.0\pm1.1$ & $0.7\pm0.0\pm1.4$ \\
\\
    10\%--20\% & 0.5--1.0 & $5.09\pm0.00\pm0.19$ & $1.94\pm0.00\pm0.16$ & $1.04\pm0.01\pm0.30$ & $0.270\pm0.003\pm0.033$ \\
    & 1.0--2.0 & $9.03\pm0.00\pm0.26$ & $4.59\pm0.00\pm0.30$ & $2.95\pm0.01\pm0.60$ & $0.82\pm0.00\pm0.10$ \\
    & 2.0--4.0 & $13.4\pm0.0\pm0.04$ & $8.28\pm0.01\pm0.51$ & $6.2\pm0.0\pm1.7$ & $1.74\pm0.00\pm0.24$ \\
    & 4.0--10 & $12.2\pm0.0\pm0.04$ & $7.5\pm0.0\pm1.0$ & $6.7\pm0.0\pm1.3$ & $1.54\pm0.01\pm0.50$ \\
\\
    20\%--30\% & 0.5--1.0 & $7.26\pm0.00\pm0.20$ & $2.29\pm0.00\pm0.22$ & $1.44\pm0.01\pm0.51$ & $0.481\pm0.003\pm0.046$ \\
    & 1.0--2.0 & $12.5\pm0.0\pm0.3$ & $5.17\pm0.01\pm0.40$ & $3.6\pm0.0\pm1.0$ & $1.33\pm0.00\pm0.11$ \\
    & 2.0--4.0 & $17.9\pm0.0\pm0.4$ & $8.93\pm0.01\pm0.61$ & $7.2\pm0.0\pm2.0$ & $2.69\pm0.00\pm0.18$ \\
    & 4.0--10 & $16.1\pm0.0\pm0.5$ & $7.98\pm0.01\pm0.76$ & $7.3\pm0.0\pm2.4$ & $2.34\pm0.01\pm0.28$ \\
\\
    30\%--40\% & 0.5--1.0 & $8.83\pm0.00\pm0.22$ & $2.49\pm0.01\pm0.31$ & $1.79\pm0.02\pm0.64$ & $0.682\pm0.004\pm0.050$\\
    & 1.0--2.0 & $14.9\pm0.0\pm0.4$ & $5.52\pm0.01\pm0.53$ & $4.3\pm0.0\pm1.3$ & $1.84\pm0.00\pm0.12$ \\
    & 2.0--4.0 & $20.7\pm0.0\pm0.5$ & $9.13\pm0.01\pm0.88$ & $8.1\pm0.0\pm3.0$ & $3.49\pm0.01\pm0.21$ \\
    & 4.0--10 & $18.4\pm0.0\pm0.5$ & $7.4\pm0.0\pm1.1$ & $7.7\pm0.0\pm3.4$ & $3.01\pm0.01\pm0.96$ \\
\\
    40\%--50\% & 0.5--1.0 & $9.76\pm0.00\pm0.25$ & $2.56\pm0.01\pm0.35$ & $2.11\pm0.04\pm0.70$ & $0.823\pm0.006\pm0.052$ \\
    & 1.0--2.0 & $16.3\pm0.0\pm0.4$ & $5.62\pm0.01\pm0.68$ & $5.2\pm0.0\pm1.8$ & $2.19\pm0.01\pm0.14$ \\
    & 2.0--4.0 & $21.9\pm0.0\pm0.7$ & $8.8\pm0.0\pm1.0$ & $9.3\pm0.1\pm4.2$ & $3.99\pm0.01\pm0.43$ \\
    & 4.0--10 & $19.8\pm0.0\pm1.6$ & $5.6\pm0.0\pm1.9$ & $10.4\pm0.1\pm8.0$ & $4.1\pm0.0\pm1.2$ \\
  \end{tabular}\end{ruledtabular}
  \label{tab:vntable}
\end{table*}

%=================================================== Fig_6

\subsection{Two-Particle Correlations}
\label{ssec:two-particle-correlations}

\subsubsection{Pair Selections}
\begin{figure}[htb]
\includegraphics[width=1.0\linewidth]{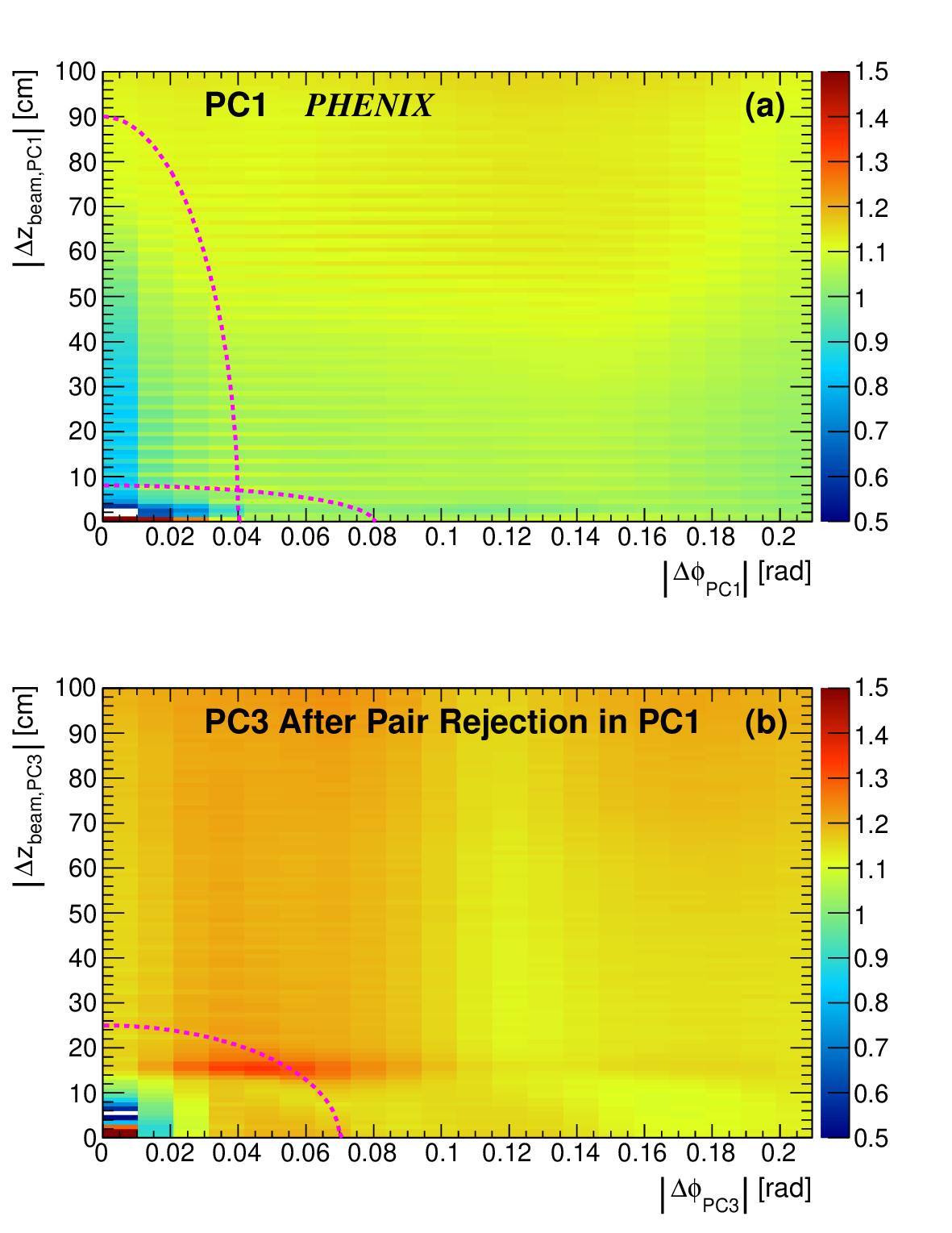}  %%07
\caption{
The ratio of the real-event to mixed-event distributions of distances 
$\Delta z_{beam}$-$\Delta\phi$ between hits in a pair of tracks in (a) 
PC1 and (b) PC3 after the PC1 cut. The region encircled by dashed 
(magenta) curves are excluded from this analysis.
}
\label{fig:pairrejection}
\end{figure}

Selected tracks are paired for correlations. 
%In real events two tracks cannot come arbitrarily close together. The 
Two tracks cannot be reconstructed arbitrarily close together. The 
tracking algorithm would split or merge the tracks. Therefore, there is 
an acceptance difference for pairs in real and mixed events. These effects are 
%estimated by calculating the distances $\Delta\phi$(rad) and $\Delta z_{beam}$(cm) between hits in the PC1 and the PC3, where 
estimated from the distributions of the distances $\Delta\phi$(rad) and $\Delta z_{beam}$(cm) between hits in the PC1 and the PC3, where 
$\Delta\phi$(rad) is the relative azimuthal angle and $\Delta 
z_{beam}$(cm) is the relative length between two track hits in both real 
%and mixed events. The ratios of these distributions are shown in 
and mixed events. The ratios of the real-to-mixed event distributions are shown in 
Fig.~\ref{fig:pairrejection}. The ratio is normalized to arbitrary units.
The dip and spike structures starting from $\Delta\phi=\Delta z_{beam}=0$ indicate
inefficient and over-efficient regions, respectively. The dashed lines indicate
the cuts used to remove these inefficient and over-efficient regions:
\begin{eqnarray}
\sqrt{\left(\Delta\phi_{PC1}/0.04\right)^2 + \left(\Delta z_{beam,PC1}/90\right)^2} <1,\nonumber\\
\sqrt{\left(\Delta\phi_{PC1}/0.08\right)^2 + \left(\Delta z_{beam,PC1}/8.0\right)^2} <1,\\
\sqrt{\left(\Delta\phi_{PC3}/0.07\right)^2 + \left(\Delta z_{beam,PC3}/25\right)^2} <1.\nonumber
\end{eqnarray}

%=================================================== Fig_7

\subsubsection{Inclusive Trigger Correlations}
\label{sssec:inccor}
Two-particle correlations are calculated as
\begin{eqnarray}
C(\Delta\phi)=\frac{N^{\rm real}(\Delta\phi)}{N^{\rm mixed}(\Delta\phi)}\frac{\int d\Delta\phi^{\prime} N^{\rm mixed}(\Delta\phi^{\prime})}{\int d\Delta\phi^{\prime} N^{\rm real}(\Delta\phi^{\prime})}
\end{eqnarray}
where \mbox{$\Delta\phi=\phi^a-\phi^t$} is the relative azimuthal angle 
between trigger and associated hadrons and $N^{\rm real}(\Delta\phi)$ and 
$N^{\rm mixed}(\Delta\phi)$ are pair distributions in the real and mixed 
events, respectively. $N^{\rm real}(\Delta\phi)$ reflects the physical 
correlation among trigger and associated hadrons from jets and from the 
underlying event as well as the dihadron detector acceptance effects. 
$N^{\rm mix}(\Delta\phi)$ is obtained by pairing trigger and associated 
hadrons from randomly selected pairs of events that have similar 
collision vertices and centralities so that it reflects only the 
dihadron acceptance effects. The collision centrality is divided into 10\% 
steps and the collision vertex in the range of $\pm30$ cm is divided into 10 bins
for this event-mixing. Taking the ratio between the real and 
mixed distributions corrects for the nonuniform azimuthal acceptance 
for dihadrons so that $C(\Delta\phi)$ contains only physical effects. 
%The correlation function is normalized so that the underlying event 
%modulates approximately around unity.
% The correlation function is normalized to unity.

\begin{figure}[htb]
\includegraphics[width=1.0\linewidth]{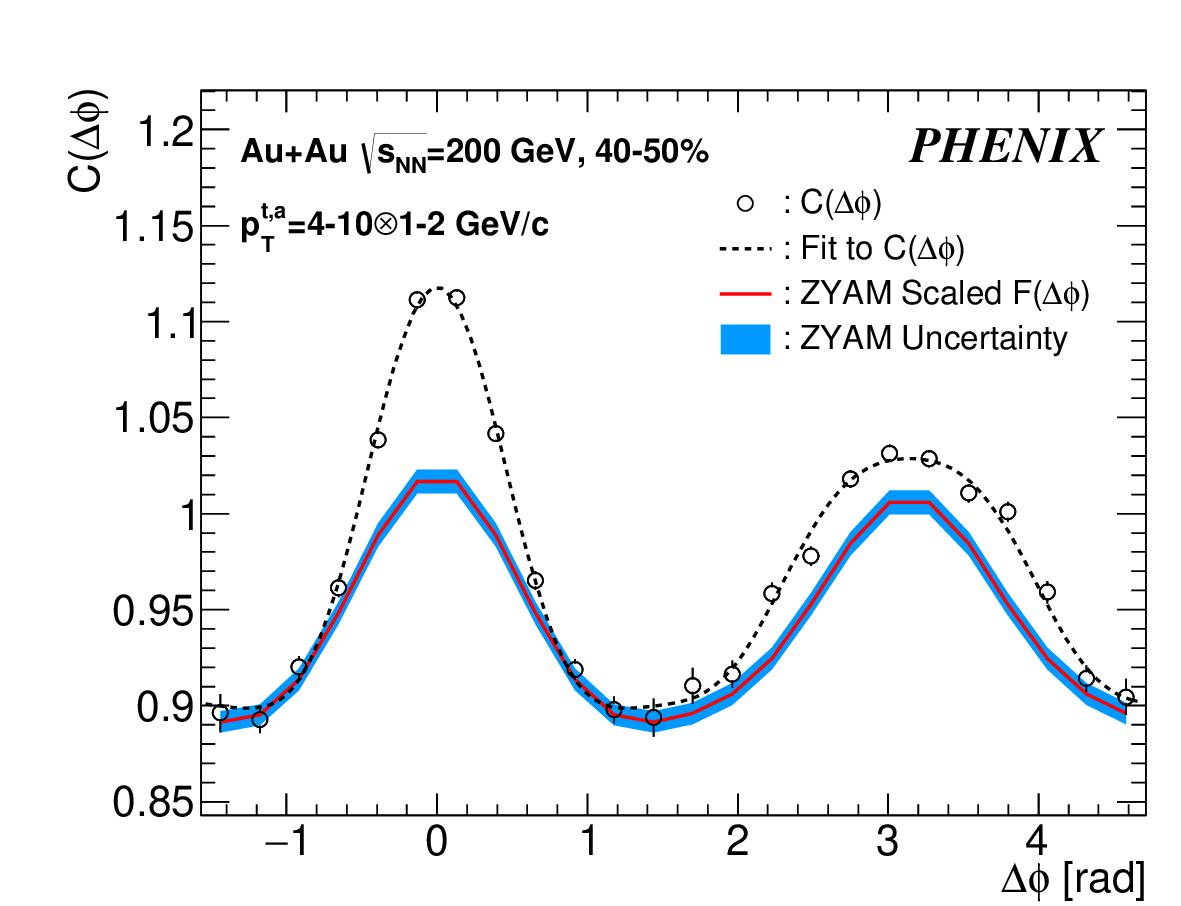}  %%08
\caption{
Example of ZYAM extraction where the correlation function 
$C(\Delta\phi)$ (open circle) is fitted (dashed line). The normalization 
of the underlying event model (red solid line) is adjusted to match the 
minimum value of the fit. The (blue) band indicates the uncertainty on 
the ZYAM extraction determined by the statistical uncertainty of 
$C(\Delta\phi)$ near the minimum.
}
\label{fig:zyam}
\end{figure}

Within the two-source model \cite{Adler:2005ee}, the correlation 
function $C(\Delta\phi)$ is composed of a jet-like term $J(\Delta\phi)$ 
and an underlying-event term that includes modulations from flow 
$F(\Delta\phi)$. We use the following model for the underlying 
event~\cite{Poskanzer:1998yz}
\begin{equation}
F(\Delta\phi)=1+\sum_{n=2}^{4}2v_n^tv_n^a\cos{n\Delta\phi}.
\end{equation}
The jet-like correlation is then obtained by subtracting $F(\Delta\phi)$ 
from $C(\Delta\phi)$ as
\begin{eqnarray}
J(\Delta\phi)=C(\Delta\phi)-b_{zyam}F(\Delta\phi).
\end{eqnarray}
The scaling factor $b_{zyam}$ is determined with the zero yield at 
minimum (ZYAM) method~\cite{Ajitanand:2005jj,Adams:2005ph,Adler:2005ee}. 
In the ZYAM assumption, $F(\Delta\phi)$ is scaled such that 
$J(\Delta\phi)$ has a minimum of exactly zero.  This therefore gives the 
lower boundary of possible jet-like correlations. The ZYAM scaling 
factor $b_{zyam}$ is determined by fitting the correlation function 
$C(\Delta\phi)$ with Fourier series for 
$-\frac{\pi}{2}<\Delta\phi<\frac{3\pi}{2}$ and identifying the single 
%point where this fit and $F(\Delta\phi)$ have the minimum contact point. 
point where this fit and $F(\Delta\phi)$ have the contact point and $J(\Delta\phi)$ is zero.
The statistical uncertainty $e_{zyam}$ of the $\Delta\phi$ bin 
containing the ZYAM point is used to scale $F(\Delta\phi)$ to estimate 
the systematic uncertainty due to ZYAM
\begin{eqnarray}
\sigma_{zyam}(\Delta\phi)=e_{zyam}F(\Delta\phi).
\end{eqnarray}
An example of the ZYAM determination is shown in Fig.~\ref{fig:zyam}.

%=================================================== Fig_8

The jet-like correlations $J(\Delta\phi)$ are scaled to the per-trigger 
yield $Y(\Delta\phi)$
\begin{eqnarray}
Y(\Delta\phi)=\frac{1}{N^t}\frac{dN^{ta}}{d\Delta\phi}
=\frac{\int d\Delta\phi N^{\rm real}(\Delta\phi)}{2\pi\epsilon^aN^t}J(\Delta\phi),
\end{eqnarray}
where $N^t$ is the number of trigger hadrons, $N^{ta}$ is the number of
pairs, and $\epsilon^a$ is the single-hadron tracking efficiency in the 
associated hadron \pt range. The efficiency is estimated via detector 
simulations for acceptance and occupancy effects as discussed in 
Ref.~\cite{Adler:2003au,Adare:2008ae,Adare:2010ry,Adare:2010mq}. The 
tracking efficiency of a trigger particle is canceled by the ratio of 
$\int d\Delta\phi N^{\rm real}/N^t$.

\subsubsection{Event Plane-Dependent Correlations}

Event plane-dependent two-particle correlations $C(\Delta\phi, \phi_s)$ 
are defined as
\begin{eqnarray}
C(\Delta\phi,\phi_s)&=&\frac{N^{\rm real}(\Delta\phi,\phi_s)}{N^{\rm mixed}(\Delta\phi,\phi_s)} \nonumber \\*
&\times& \frac{\int\int d\Delta\phi^{\prime} d\phi_s^{\prime} N^{\rm mixed}(\Delta\phi^{\prime},\phi_s^{\prime})}{\int\int d\Delta\phi^{\prime} d\phi_s^{\prime} N^{\rm real}(\Delta\phi^{\prime},\phi_s^{\prime})}, \label{eq:EPC}
\end{eqnarray}
where \mbox{$\phi_s=\phi^t-\Psi_n$} and 
\mbox{$N^{\rm real}(\Delta\phi,\phi_s)$} and 
\mbox{$N^{\rm mix}(\Delta\phi,\phi_s)$} are the event plane-dependent pair 
$\Delta\phi$ distributions in real and mixed events, respectively.
%We use the event-plane determined by the entire RXN acceptance for this
%event-plane dependence study.
We use the event-plane determined by the entire RXN acceptance providing the best event-plane 
resolution among PHENIX subsystems i.e. the best sensitivity for this event-plane dependence study.
Other event planes were not used because those planes have worse resolution.

Similar to inclusive correlations, event plane-dependent jet-like 
correlations $J(\Delta\phi, \phi_s)$ are obtained by subtracting the 
event plane-dependent flow background term $F(\Delta\phi,\phi_s)$ from 
$C(\Delta\phi,\phi_s)$ with a ZYAM scale factor as
\begin{eqnarray}
J(\Delta\phi,\phi_s)=C(\Delta\phi,\phi_s)-b_{zyam}F(\Delta\phi,\phi_s).
\end{eqnarray}
We use the same $b_{zyam}$ as determined from the inclusive correlations 
from the same trigger, associated and centrality selection. An 
analytical formula for $F(\Delta\phi,\phi_s)$ including the $n=2$ event 
plane dependence exists~\cite{Bielcikova:2003ku}, however, it is not 
easily applied with finite correlations between the $n=2$ and $n=4$ 
event planes. For this reason, a Monte Carlo simulation is employed to 
estimate $F(\Delta\phi,\phi_s)$. This is described in 
Sec.~\ref{sssec:EPFlowBKG} below.

The event plane-dependent jet-like correlations are converted into event 
plane-dependent per-trigger yield as
\begin{eqnarray}
Y(\Delta\phi,\phi_s)&=&\frac{1}{N^t_{\phi_s}}\frac{dN^{ta}_{\phi_s}}{d\Delta\phi} \nonumber \\
&=&\frac{\int d\Delta\phi N^{\rm real}(\Delta\phi,\phi_s)}{2\pi\epsilon^aN^t_{\phi_s}}J(\Delta\phi,\phi_s) \label{eq:EPDependentPTY}
\end{eqnarray}
where $N^t_{\phi_s}$ is the number of trigger hadrons and 
$N^{ta}_{\phi_s}$ is the number of pairs in the trigger event-plane bin.

\subsubsection{Flow Background Model Including Event Plane Dependence }
\label{sssec:EPFlowBKG}

With the assumption that the measured $v_n$ from the event plane method 
are purely from collective dynamics of the medium, flow-like azimuthal 
distributions of single hadrons can be generated by performing a Monte 
Carlo simulation inputting the experimentally measured $v_n$, the 
resolution of the event planes, and the strength of correlation among 
different order event planes. The single-hadron azimuthal distributions 
due to collective flow can be described by a superposition of $v_n$ as
\begin{eqnarray}
\frac{dN}{d\phi} = 1 + \sum_{n=2}^{4}2v_n\cos{n(\phi-\Psi_n^{\rm true})},  \label{v2v3v4}
\end{eqnarray}
where $\phi$ is the azimuthal angle of the emitted hadrons and 
$\Psi_n^{\rm true}$ is a true $n$th-order event plane defined over 
$[-\pi/n,\pi/n]$. Separate distributions using $v_n$ for each \pt ranges 
of trigger and associated particles are used in the simulation. The 
trigger and associated distributions in real events share a common 
$\Psi_n$, while those in mixed events do not.

The experimental event plane resolution is introduced through a 
dispersion term $\Delta\Psi_n$ where \mbox{$\Psi_n^{\rm obs} = 
\Psi_n^{\rm true}+\Delta\Psi_n$}. We calculate $\Delta\Psi_n$ as
\begin{eqnarray}
\Delta\Psi_n=\frac{e^{-\frac{\chi_{n}^2}{2}}}{\pi}\left[1+z_n\sqrt{\pi}[1+{\rm erf}(z_n)]e^{z_n^2}\right] \label{eq:Erf}
\end{eqnarray}
where \mbox{$z_n=\chi_{n}/\sqrt{2}\cos(n\Delta\Psi_n)$} and 
${\rm erf}(z_n)$ is the error 
function~\cite{Ollitrault:1997di,Poskanzer:1998yz}. This equation can be 
solved for $\Delta\Psi_n$ by using the experimentally-determined 
$\chi_n$ from the measured event plane resolutions using 
Eq.~(\ref{eq:BesselReso}).

Because a weak correlation between $\Psi_2^{\rm true}$ and $\Psi_3^{\rm true}$ 
exists \cite{Adare:2011tg}, the directions of $\Psi_2^{\rm true}$ and 
$\Psi_3^{\rm true}$ are generated independently. The direction of 
$\Psi_4^{\rm true}$ is generated assuming a correlation with 
$\Psi_2^{\rm true}$, $\Psi_4^{\rm true} = \Psi_2^{\rm true} + \Delta\Psi_{42}$. We 
estimate $\Delta\Psi_{42}$ assuming the correlation between the two 
event planes follow similar functional forms as the dispersion of event 
planes due to the resolution. That is, we assume,
\begin{eqnarray}
\Delta\Psi_{42}=\frac{e^{-\frac{\chi_{42}^2}{2}}}{\pi}\left[1+z_{42}\sqrt{\pi}[1+{\rm erf}(z_{42})]e^{z_{42}^2}\right] \label{eq:Erf42}
\end{eqnarray}
where \mbox{$z_{42}=\chi_{42}/\sqrt{2}\cos{4\Delta\Psi_{42}}$}. The 
parameter $\chi_{42}$ is assumed to be similar to 
Eq.~(\ref{eq:BesselReso})
\begin{eqnarray}
\left<\cos(4\Delta\Psi_{42})\right>&=&\frac{\sqrt{\pi}}{2\sqrt{2}}\chi_{42} e^{-\chi_{42}^2/4} \nonumber \\
	&\times&	\left[I_{0}\left(\frac{\chi^2_{42}}{4}\right)+I_{1}\left(\frac{\chi^2_{42}}{4}\right)\right] \label{eq:-BesselChi42}
\end{eqnarray}
where \mbox{$\left<\cos(4\Delta\Psi_{42})\right>=v_4\{\Psi_2\}/v_4$}~\cite{Yan:2015jma}. 
The functional shape of Eq.~(\ref{eq:Erf42}) is verified by event plane 
correlation studies using the BBCs and the RXNs following the method 
described in Ref.~\cite{Aad:2014fla}. The correlation strength between 
$\Psi_2^{\rm true}$ and $\Psi_3^{\rm true}$, 
$\left<\cos{6(\Psi_2-\Psi_3)}\right>$, is measured to be consistent with 
zero within large statistical uncertainties. Potential impacts of 
$\left<\cos{6(\Psi_2-\Psi_3)}\right>$ to the event plane dependent 
correlations are estimated using the value of 
$\left<\cos{6(\Psi_2-\Psi_3)}\right>$ reported in 
Ref.~\cite{Aad:2014fla} by the ATLAS Experiment. The impact of 
$\left<\cos{6(\Psi_2-\Psi_3)}\right>$ is within the systematic 
uncertainties described later.

We use the averaged $\chi_{42}$ value between \mbox{$2<\pt<4$}~GeV/$c$ 
and $1<\pt<2$~GeV/$c$ for event plane-dependent correlations of 
($2<\ptt<4)\otimes$~($1<\pta<2$), ($2<\ptt<4)\otimes$~($2<\pta<4$), and 
($4<\ptt<10)\otimes$~($2<\pta<4$)~GeV/$c$ because 
\mbox{$\left<\cos(4\Delta\Psi_{42})\right>$} would contain 
auto-correlations from jets at high $p_T$.

%=================================================== Fig_9
\begin{figure*}[htb]
\begin{minipage}{0.60\linewidth}
\includegraphics[width=0.99\linewidth]{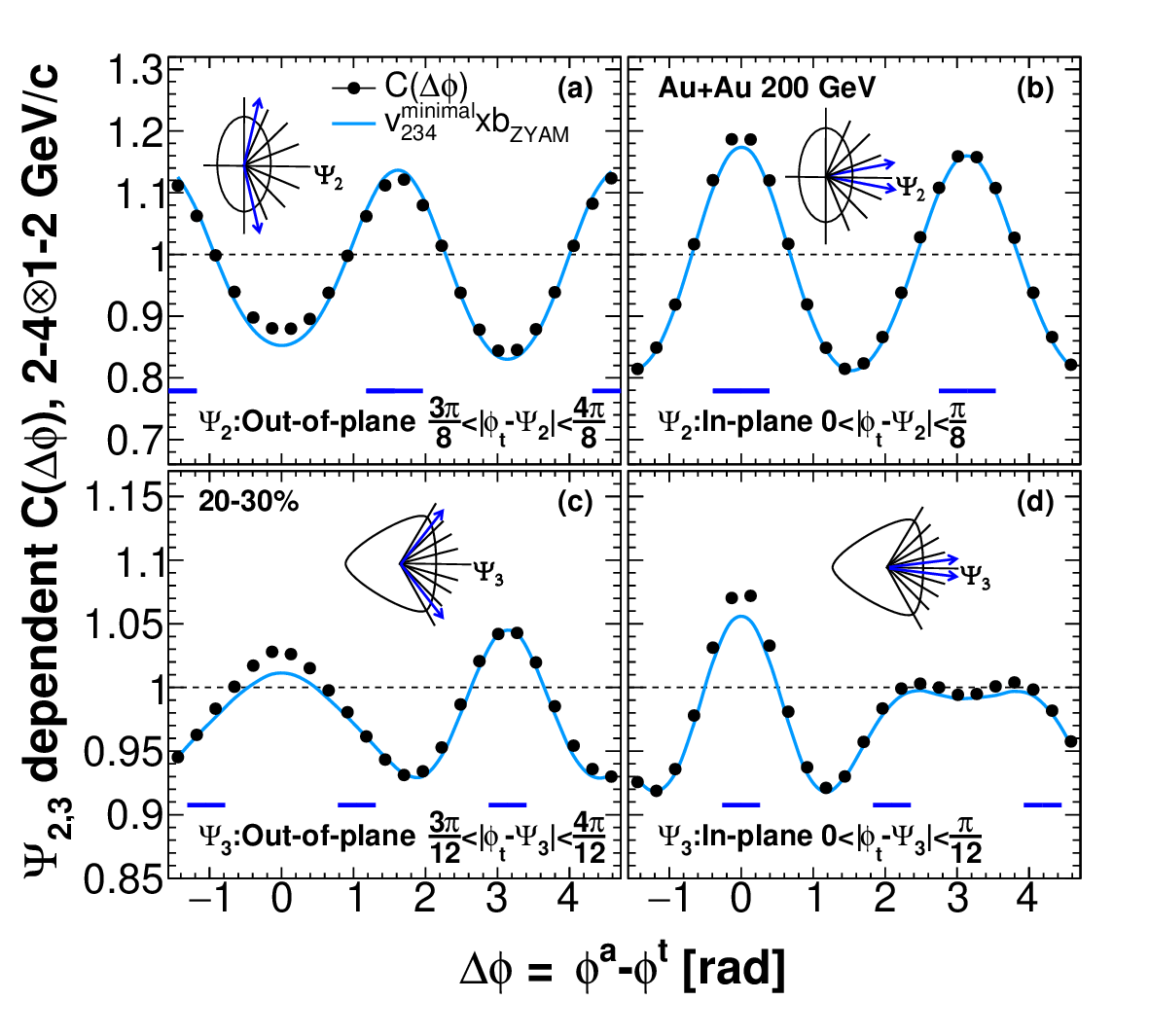}  %%09
\end{minipage}
%\hspace{0.02\linewidth}
\begin{minipage}{0.3\linewidth}
\caption{
Event plane-dependent $C(\Delta\phi)$ (black circles) and event 
plane-dependent model flow background (blue lines)of ($2<\ptt<4) 
\otimes$~($1<\pta<2$)~GeV/$c$. Trigger particles are selected in (a) 
out-of-plane $3\pi/8<|\phi^t-\Psi_2|<4\pi/8$ of $\Psi_2$, (b) in-plane 
$0<|\phi^t-\Psi_2|<\pi/8$ of $\Psi_2$, (c) out-of-plane 
$3\pi/12<|\phi^t-\Psi_3|<4\pi/12$ of $\Psi_3$, (d) in-plane 
$0<|\phi^t-\Psi_3|<\pi/12$ of $\Psi_3$. Schematic pictures in each panel 
also depict these ranges of the trigger particle selections with respect 
to event plane $\Psi_n$.
}
\label{fig:EPFlowCorFitAV}
\end{minipage}
%\end{figure*}
%=================================================== Fig_10
%\begin{figure*}[htb]
\begin{minipage}{0.60\linewidth}
\includegraphics[width=0.99\linewidth]{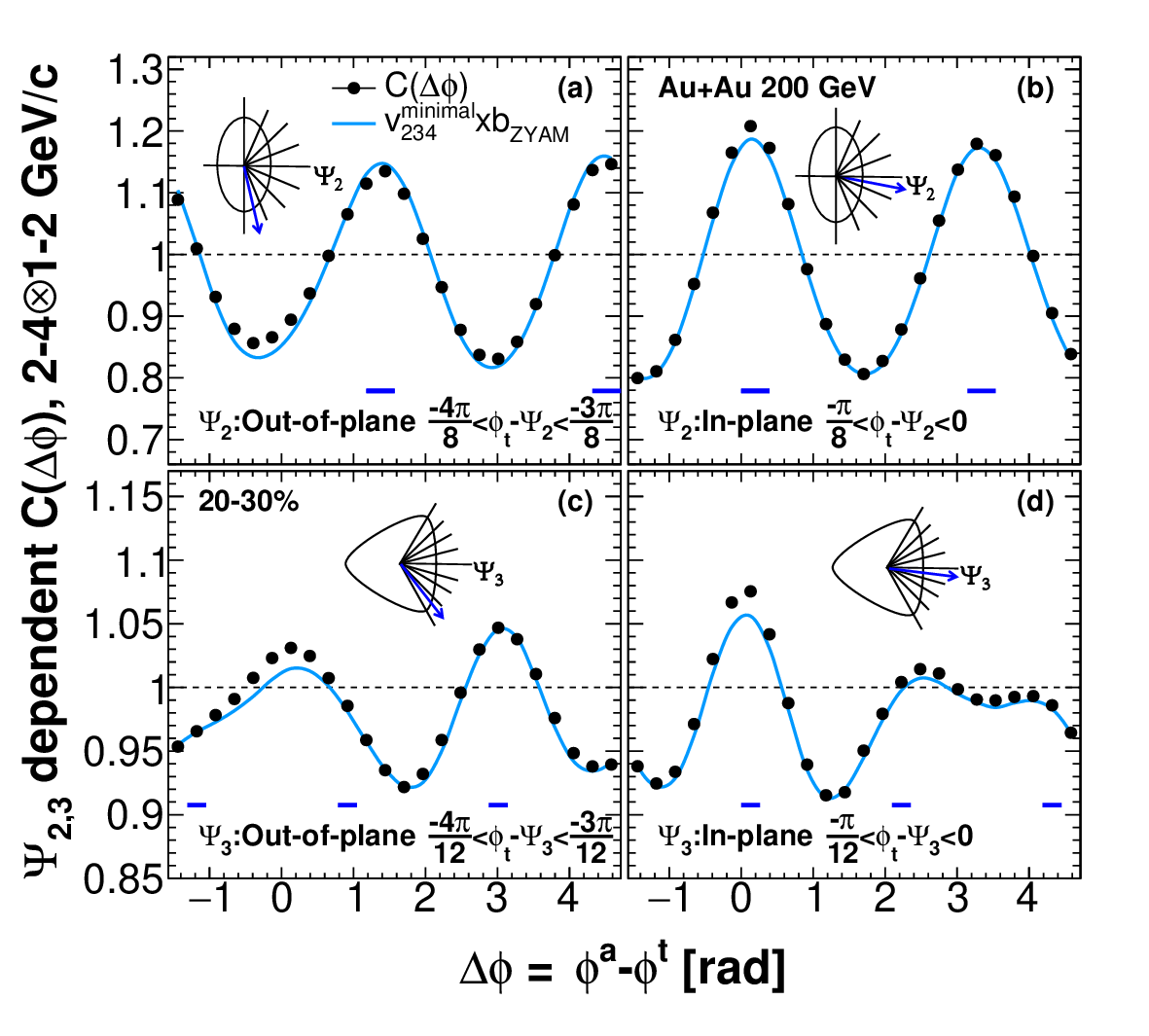}  %%10
\end{minipage}
%\hspace{0.02\linewidth}
\begin{minipage}{0.3\linewidth}
\caption{
Event plane-dependent $C(\Delta\phi)$ (black circles) and event 
plane-dependent model flow background (blue lines) of ($2<\pt<4) 
\otimes$~($1<\pta<2$)~GeV/$c$. Trigger particles are selected in (a) 
out-of-plane $-4\pi/8<\phi^t-\Psi_2<-3\pi/8$ of $\Psi_2$, (b) in-plane 
$-\pi/8<\phi^t-\Psi_2<0$ of $\Psi_2$, (c) out-of-plane 
$-4\pi/12<\phi^t-\Psi_3<-3\pi/12$ of $\Psi_3$, (d) in-plane 
$-\pi/12<\phi^t-\Psi_3<0$ of $\Psi_3$. Schematic pictures in each panel 
also depict these ranges of the trigger particle selections with respect 
to event-plane $\Psi_n$.
}
\label{fig:EPFlowCorFitNP}
\end{minipage}
\end{figure*}

The event plane-dependent background shapes are determined by generated 
particles in this simulation using Eq.~(\ref{eq:EPC}). 
Figure~\ref{fig:EPFlowCorFitAV} shows event plane-dependent correlations 
and backgrounds with a selection of the absolute trigger azimuthal angle 
relative to the event-planes \mbox{$|\phi^t-\Psi_n|$}. The backgrounds 
agree with the experimental correlations except at 
\mbox{$\Delta\phi=0,\pi$} where contributions from jets are expected. 
Figure~\ref{fig:EPFlowCorFitNP} shows event plane-dependent correlations 
and backgrounds with a selection of trigger azimuthal angle relative to 
event planes \mbox{$\phi^t-\Psi_n < 0$}. Agreement between the 
experimental correlations and the background except at 
\mbox{$\Delta\phi=0,\pi$} is also observed here.  Other event-plane 
dependent correlations and backgrounds with a selection of trigger 
azimuthal angle relative to event-planes $\phi^t-\Psi_n<0$ for different 
collision centralities and $p_T^{t,a}$ selections are shown in 
the Appendix.

\subsection{Unfolding of Event Plane-Dependent Correlations} 
\label{ssec:IterUnfold} 

In this analysis, $\phi_s$ is divided into 8 bins. The width of the 
$\phi_s$ bins is $\pi/8$ and $\pi/12$ when correlating with $\Psi_2$ and 
$\Psi_3$, respectively. The event plane-dependent per-trigger yields 
$Y(\Delta\phi,\phi_s)$ are smeared across neighboring event plane bins 
due to limited experimental resolution of the event planes. We unfold 
the smearing to obtain the true event-plane dependence of the 
correlations. Two different methods are used to check the unfolding
procedure: (I) iterative Bayesian unfolding, $Y^{\rm itr}_{\rm unf}$, 
and (II) correcting the event plane-dependence of the per-trigger yield 
based on a Fourier analysis, $Y^{\rm fit}_{\rm unf}$.

 \subsubsection{Iterative Bayesian Unfolding}

The Iterative Bayesian Unfolding Method presented in 
Ref.~\cite{D'Agostini:1994zf,Adye:2011gm} is applied to this analysis 
with the following formulation
 \begin{eqnarray}
 \hat{n}(\phi_{s,i},\Delta\phi)&=&\sum_j M_{ij}n^{\rm obs}(\phi_{s,j},\Delta\phi) \\
 M_{ij}&=&\frac{P(\phi_{s,j}|\phi_{s,i})n(\phi_{s,i},\Delta\phi)}{\varepsilon_i\sum_l P(\phi_{s,j}|\phi_{s,l})n(\phi_{s,l},\Delta\phi)},
 \end{eqnarray}
where $\hat{n}(\phi_{s},\Delta\phi)$ is the unfolded distribution, 
$n^{\rm obs}(\phi_{s},\Delta\phi)$ is the experimentally observed 
distribution, $n(\phi_{s},\Delta\phi)$ is the prior distribution,
$P(\phi_{s,j}|\phi_{s,i})$ is the conditional probability matrix where 
$\phi_{s,i}$ is measured to be $\phi_{s,j}$, and 
$\varepsilon_i=\sum_jP(\phi_{s,j}|\phi_{s,i})$ is the efficiency. In the 
iterative calculation, $\hat{n}(\phi_{s},\Delta\phi)$ also serves as the 
prior distribution of the next loop. We perform this unfolding 
separately for every $\Delta\phi$ bin.
 
We define the experimentally observed distribution as 
$n^{\rm obs}(\phi_{s},\Delta\phi)=1+Y(\phi_s,\Delta\phi)$ using the measured 
event plane-dependent per-trigger yield. The offset is to prevent a 
divergence in the iteration due to small yields near the ZYAM point. In 
the initial loop of the iteration, we define the prior distribution as 
$n(\phi_{s},\Delta\phi)=n^{\rm obs}(\phi_{s},\Delta\phi)$.
 
\begin{figure}[tbh]
\includegraphics[width=1.0\linewidth]{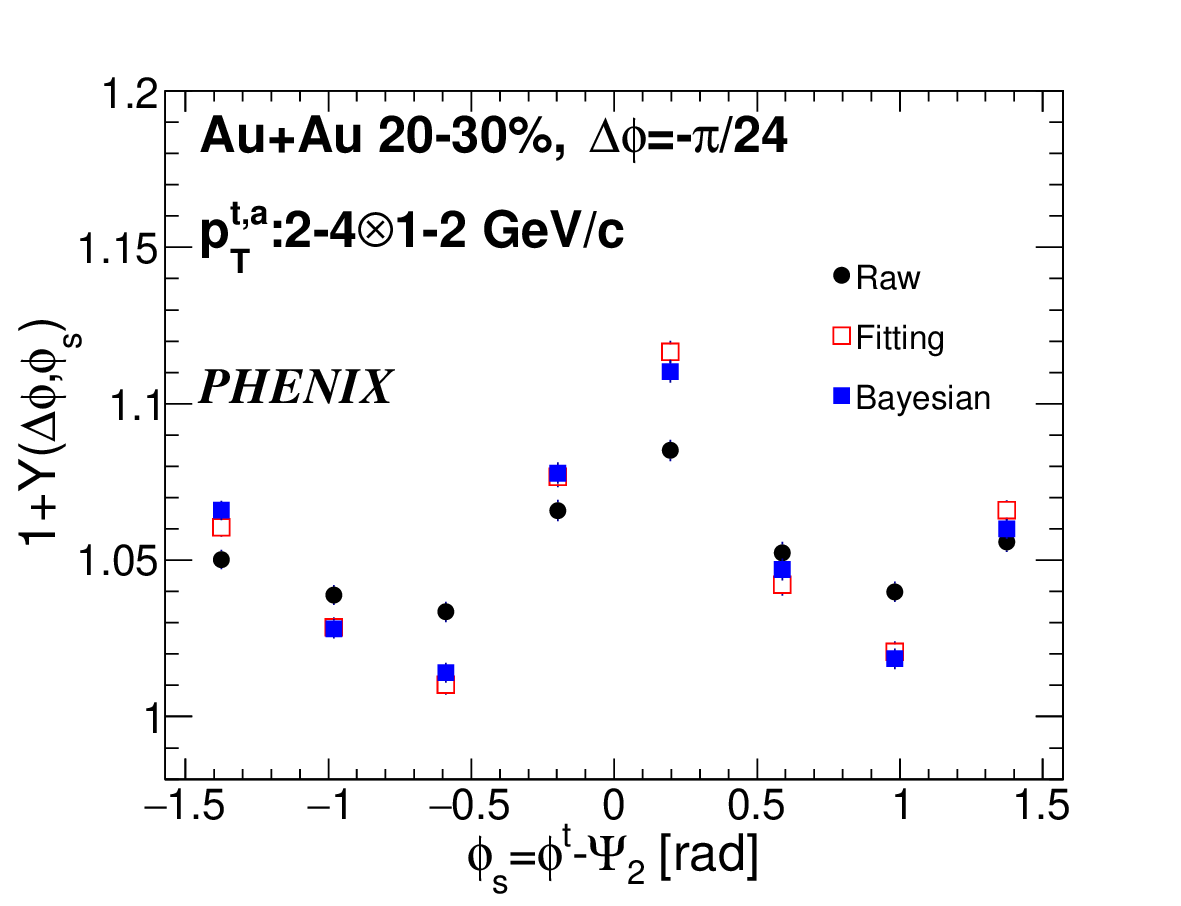} %%11
\caption{
The raw event plane-dependent per-trigger yields offset by 1, 
$1+Y(\phi_s,\Delta\phi=-\pi/24)$ (black circles). The resulting 
corrected per-trigger yields using iterative Bayesian unfolding (blue 
filled squares) and from Fourier fitting (red open squares).
}
\label{fig:FourierUnfold}
\end{figure}

The probability distribution of the relative azimuthal angle between the 
true event plane $\Psi_n$ and the measured event plane $\Psi_n^{\rm obs}$ 
can be translated into the difference between real and observed $\phi_s$ as
 \begin{eqnarray}
 \Psi_n-\Psi_n^{\rm obs}=(\phi^t-\Psi_n^{\rm obs})-(\phi^t-\Psi_n)=\phi_s^{\rm obs}-\phi_s. \quad
 \end{eqnarray}
With this probability distribution of \mbox{$\phi_s^{\rm obs}-\phi_s$}, the 
probability matrix $P(\phi_{s,j}|\phi_{s,i})$ is determined by the 
degree of the contamination by neighboring $\phi_s$ bin as
 \begin{align}
P(\phi_{s,j}|\phi_{s,i})=\left(
\begin{array}{cccccccc}
 		s_0 & s_1 & s_2 & s_3 & s_4 & s_5 & s_6 & s_7\\
 		s_7 & s_0 & s_1 & s_2 & s_3 & s_4 & s_5 & s_6\\
 		s_6 & s_7 & s_0 & s_1 & s_2 & s_3 & s_4 & s_5\\
 		s_5 & s_6 & s_7 & s_0 & s_1 & s_2 & s_3 & s_4\\
 		s_4 & s_5 & s_6 & s_7 & s_0 & s_1 & s_2 & s_3\\
 		s_3 & s_4 & s_5 & s_6 & s_7 & s_0 & s_1 & s_2\\
 		s_2 & s_3 & s_4 & s_5 & s_6 & s_7 & s_0 & s_1\\
 		s_1 & s_2 & s_3 & s_4 & s_5 & s_6 & s_7 & s_0\\
\end{array}
 		\right)
 \end{align}
where \mbox{$s_n$~$(n\neq0)$} is the contamination fraction from the 
$n$-th $\phi_s$ bin away from a selected $\phi_s$ bin, and $s_0$ is the 
fraction of the true signal in the selected $\phi_s$ bin. A study in 
previous identified particle $v_2$ measurements of the PHENIX 
experiment~\cite{Adare:2012vq} using the same data sample as this 
analysis showed that the tracking efficiency is independent of $\phi_s$. 
Thus, we normalize the probability as $\sum s_n=1$, 
i.e.~$\varepsilon=1$. Due to the cyclic boundary condition in the 
azimuthal angle direction, symmetric elements of 
$P(\phi_{s,j}|\phi_{s,i})$ are identical i.e.~\mbox{$s_5=s_3$}, 
\mbox{$s_6=s_2$}, and \mbox{$s_7=s_1$}. The matrix 
$P(\phi_{s,j}|\phi_{s,i})$ depends only on the order of event-planes and 
centrality. An example of corrections based on this iterative method at 
$-\frac{\pi}{24}<\Delta\phi<0$ for 
\mbox{($2<\ptt<4$)~$\otimes$~($1<\pta<2$)}~GeV/$c$ in 20\%--30\% central 
collisions is shown in Fig.~\ref{fig:FourierUnfold} together with an 
example of the Fourier analysis method introduced in 
Section~\ref{sssec:FourierCorrection}.

\subsubsection{Fourier Oscillation Correction of the Event Plane-Dependence of Correlations}
\label{sssec:FourierCorrection}

The second method to correct the event plane-dependence of the 
per-trigger yield is a Fourier analysis. $Y(\Delta\phi,\phi_s)$ is 
offset by 1 to prevent divergences in the correction due to small values 
due to the ZYAM subtraction of the background. A Fourier series should 
be able to fit the event plane-dependence of 
\mbox{$1+Y(\Delta\phi,\phi_s)$}, and the fit function to the 
$\Psi_2$-dependent case at a given $\Delta\phi$ can be written as
\begin{eqnarray}
F^Y(\Delta\phi,\phi_s)&=&a_0\left[1+\sum_{n=2,4}2a_n\cos{n(\phi_s+\Delta\phi)}\right] \quad 
\end{eqnarray}
and similarly the $\Psi_3$-dependent case can be written as
\begin{eqnarray}
F^Y(\Delta\phi,\phi_s)&=&a_0\left[1+2a_3\cos{3(\phi_s+\Delta\phi)}\right]
\end{eqnarray}
where $a_0$ is a normalization and $a_{2}$, $a_{3}$, and $a_{4}$ are the 
azimuthal anisotropies of \mbox{$1+Y(\phi_s,\Delta\phi)$}. In the 
fitting functions $F^Y(\Delta\phi,\phi_s)$, the phase shift $\Delta\phi$ 
is necessary in \mbox{$1+Y(\phi_s,\Delta\phi)$} because the associated 
yields are at \mbox{$\phi^a-\Psi_n=\phi_s+\Delta\phi$} (see 
Figure~\ref{fig:TrigPsi2Psi3}).

With the assumption that the coefficients determined from the fits are 
diluted by the event plane resolutions, the effects can be corrected in 
a manner analogous to the single-particle azimuthal anisotropy $v_n$ as 
performed in Ref.~\cite{Adare:2010mq}. For the $\Psi_2$-dependent case, 
the correction is given as
\begin{eqnarray}
F^{Y,cor}(\Delta\phi,\phi_s)&=&a_0\left[1+\sum_{n=2,4}\frac{2a_n\cos{n(\phi_s+\Delta\phi)}}{{\rm Res}\{n,\Psi_2\}}\right] \nonumber \\
\end{eqnarray}
and for the $\Psi_3$-dependent case it is given as
\begin{eqnarray}
F^{Y,cor}(\Delta\phi,\phi_s)&=&a_0\left[1+\frac{2a_3\cos{3(\phi_s+\Delta\phi)}}{{\rm Res}\{3,\Psi_3\}}\right]. \quad
\end{eqnarray}
The correction coefficient to $1+Y^{\rm cor}$ is then given by the ratio 
$F^{Y,cor}(\phi_s)/F^Y(\phi_s)$, which then fixes the corrected 
per-trigger yield as
\begin{eqnarray}
1+Y^{\rm cor}(\phi_s,\Delta\phi)&=&\frac{F^{Y,cor}(\phi_s,\Delta\phi)}{F^{Y}(\phi_s,\Delta\phi)}
\times(1+Y(\phi_s,\Delta\phi)). \nonumber \\
\end{eqnarray}

%=================================================== Fig_11

\subsection{Systematic Uncertainties for Per-Trigger Yields}
\subsubsection{Efficiency}

Systematic uncertainties in tracking efficiency are estimated to be 
approximately 10\% for $\pt<4$~GeV/$c$ and 13\% for $\pt>4$~GeV/$c$ 
independent of 
centrality~\cite{Adler:2003au,Adare:2008ae,Adare:2010ry,Adare:2010mq}.

\subsubsection{Inclusive Per-Trigger Yields}

Systematic uncertainties on the yields from the matching cut and from 
the $v_n$ measurements are determined by the variations on the 
parameters discussed below. The systematic uncertainty from the matching 
cut $\sigma_{\rm mat}$ after flow subtraction is derived in similar manner 
as in previous publications \cite{Adare:2008ae} as

\begin{eqnarray}
\sigma_{\rm mat}=|Y^{mat=2.5\sigma}(\Delta\phi)-Y^{mat=1.5\sigma}(\Delta\phi)|/2.
\end{eqnarray}
The systematic uncertainties from $v_n$ are evaluated by taking the 
quadrature sum of residuals from the 1-$\sigma$ uncertainties on the 
$v_n$ for all orders of $n$ used in the subtraction.  Formally the 
calculation is given by
\begin{eqnarray} \sigma_{v}=\sqrt{\sum_{k=2,3,4}\sum_{l=\pm
1}\frac{|Y^{v_k^{l\sigma}}(\Delta\phi)-Y^{v_k}(\Delta\phi)|^2}{2}}.
\end{eqnarray}
where the second $Y(\Delta\phi)$ refers to the yields resulting from the 
default set of measured $v_n$ values. The total systematic uncertainties 
$\sigma_{\rm in}$ in the inclusive trigger yields are given by
\begin{eqnarray}
\sigma_{\rm in}=\sqrt{\sigma_{v}^2+\sigma_{\rm mat}^2}.
\end{eqnarray}
We studied the inclusion of a $v_5$ term assuming $v_5 = 0.5 \times v_4$ 
consistent with the ATLAS measurements~\cite{ATLAS:2012at}. The results 
were completely consistent with the quoted uncertainties. Uncertainties 
due to ZYAM will be discussed later.

\subsubsection{Event Plane-Dependent Per-Trigger Yields}
\label{sssec:epdependentpty}

%=================================================== Fig_12
\begin{figure*}[htb]
\includegraphics[width=0.99\linewidth]{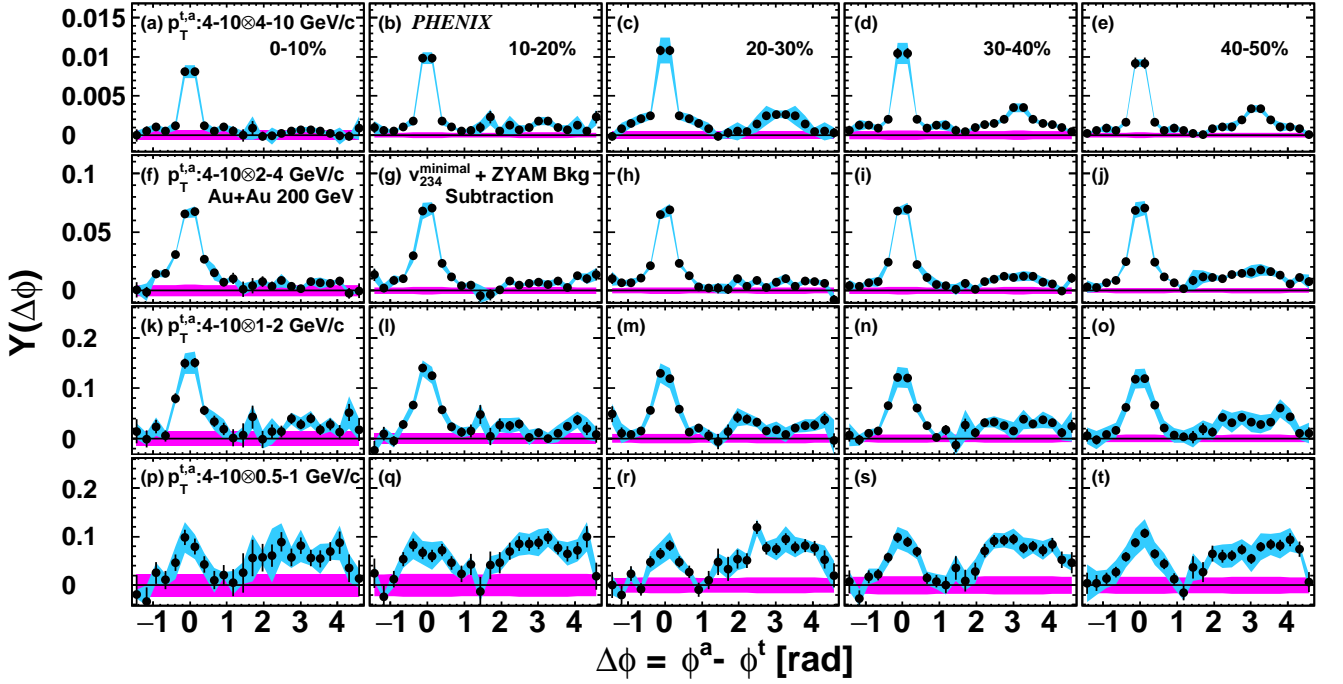}  %%12
\caption{
Per-trigger yields $Y(\Delta\phi)$ of dihadrons pairs measured in Au$+$Au 
collisions at \Gsqsn after subtracting the underlying event model 
with several $p_T$ selections:
(a)--(e) ($4<\ptt<10$)~$\otimes$~($4<\pta<10$)~GeV/$c$,
(f)--(j) ($4<\ptt<10$)~$\otimes$~($2<\pta<4$)~GeV/$c$,
(k)--(o) ($4<\ptt<10$)~$\otimes$~($1<\pta<2$)~GeV/$c$, and
(p)--(t) ($4<\ptt<10$)~$\otimes$~($0.5<\pta<1$)~GeV/$c$.
The columns represent centrality bins 0\%--10\% (a,f,k,p), 10\%--20\% 
(b,g,l,q), 20\%--30\% (c,h,m,r), 30\%--40\% (d,i,n,s), 40\%--50\% (e,j,o,t). 
Systematic uncertainties due to track matching and the $v_n$ are shown 
by blue bands around the points. Uncertainties from ZYAM are shown by 
the purple bands around zero yield.
}
\label{fig:InclusiveTrig4-10}
\end{figure*}

In addition to systematic uncertainties considered in the inclusive 
per-trigger yields, systematic uncertainties due to the 
$\cos{4(\Psi_4-\Psi_2)}$ correlation strength are also taken into 
account before the unfolding of event plane resolution effects. The 
value of $\chi_{42}$ is determined with $v_4\left\{\Psi_4\right\}$ and 
$v_4\left\{\Psi_2\right\}$, and the systematic uncertainties of 
$\chi_{42}$ are propagated from those of $v_4\left\{\Psi_2\right\}$. 
Systematic uncertainties on the yield due to $\chi_{42}$ are given by
\begin{eqnarray}
\sigma_{\chi_{42}}=\sqrt{\sum_{l=\pm 1}\frac{|Y^{\chi_{42}^{l\sigma}}(\Delta\phi,\phi_s)-Y^{\chi_{42}}(\Delta\phi,\phi_s)|^2}{2}}.
\end{eqnarray}
The systematic uncertainties before unfolding are
\begin{eqnarray}
\sigma_{\rm bef}=\sqrt{\sigma_{v}^2+\sigma_{\rm mat}^2+\sigma_{\chi_{42}}^2}.
\end{eqnarray}

For the event plane-dependent per-trigger yields, the systematic 
uncertainty due to the impact of the finite event plane resolution on 
the correlations has contributions from the method and the number of 
iterations in the Bayesian method.  The uncertainty due to the method is 
given by
\begin{eqnarray}
\sigma_{Met}=|Y^{\rm fit}_{\rm unf}-Y^{\rm itr}_{\rm unf}|,
\end{eqnarray}
where $Y^{\rm itr}_{\rm unf}$ is the result using the iterative Baysean method 
and $Y^{\rm fit}_{\rm unf}$ is the result using the Fourier fitting method.  The 
uncertainty due to the number of interactions using for unfolding is 
given by the difference between the number of iterations (Nit) for 
$n=5$ and $n=10$
\begin{eqnarray}
\sigma_{\rm Nit}=|Y^{n=5}_{\rm unf}-Y^{n=10}_{\rm unf}|.
\end{eqnarray}
These unfolding uncertainties are added in quadrature to the 
uncertainties before unfolding
\begin{eqnarray}
\sigma_{\rm tot}=\sqrt{{\sigma_{\rm bef}}^2+{\sigma_{Met}}^2+{\sigma_{\rm Nit}}^2 }.
\end{eqnarray}

\noindent In event plane-dependent per-trigger yields, we also unfolded 
the upper and lower boundaries of ZYAM uncertainties propagated from 
statistical uncertainties from the data. The systematic uncertainties 
associated with ZYAM are not included into the total systematic 
uncertainties $\sigma_{\rm tot}$. 
%How they are taken into account is described later.
These variations are discussed below.

\section{Results and Discussion}
\label{sec:ResultsandDiscussion}

\subsection{Inclusive Per-Trigger Yields I: High-$p_T$ Trigger Particles}

We first present the highest trigger $p_T$ correlations. The jet-like 
correlations should be dominated by $2\rightarrow2$ scattering. Pairs of 
particles with $\Delta\phi = 0$, the near side, are from both particles 
fragmenting from a single jet. Pairs of hadrons around $\Delta\phi = 
\pi$, the away side, occur when each particle fragments from 
back-to-back jets. In high-$p_T$ correlations the jet momentum fraction 
for the associated particle is approximated by
\begin{eqnarray}
z_T = \pta/\ptt. 
\end{eqnarray}

Per-trigger yields with trigger particles from $4<\ptt<10$~GeV/$c$ 
paired with associated particles from $0.5<\pta<10$~GeV/$c$ are shown in 
Fig.~\ref{fig:InclusiveTrig4-10}. The band around zero indicates the 
systematic uncertainty due the ZYAM assumption. The band around the data 
points is the systematic uncertainty from all other sources. Systematic 
uncertainties from the associated tracking efficiency and matching and 
the ZYAM normalization are fully correlated point-to-point. The 
underlying event subtraction is correlated point-to-point and can affect 
the shape. For the highest trigger \pt correlations, the dominant 
systematic is not the underlying event subtraction.

The near-side yield is centrality independent 
(Fig.~\ref{fig:InclusiveTrig4-10}). This is consistent with measurements 
of the two-particle correlation that indicated the near-side yields are 
not modified 
\cite{Agakishiev:2011st,Abelev:2009ah,Abelev:2009af,Aamodt:2011vg}. The 
lack of centrality dependence is also consistent with the picture that 
triggering on high-$p_T$ particles biases the origin of the 
hard-scattering toward the surface of the QGP such that the leading 
parton loses little to no energy.

The away-side peak is evident in several \pta and centrality selections. 
The evolution of the away shape and yield with centrality and \pta is 
similar to previous measurements where only $v_2$ is assumed to contribute 
to the underlying event~\cite{Adare:2008ae,Adare:2007vu,Adams:2006yt}.
The away-side peak becomes sharper and more pronounced as \pta increases or the centrality 
selection becomes more peripheral. In more central collisions and lower 
\pta when the away-side structure is present, it is broader than in the 
highest \pta and peripheral centrality selection. The trends are 
consistent with a picture where the associated parton opposite the 
trigger loses energy and scatters in the medium. At the lowest \pta a 
very wide plateau-like away-side structure is observed with similar 
shape and magnitude in all centralities. Similar low-momentum and 
large-angle yields have been observed in prior measurements 
\cite{Adare:2012qi,Adamczyk:2013jei,Aad:2014wha,Chatrchyan:2014ava}.

\begin{figure}[htb]
\includegraphics[width=0.98\linewidth]{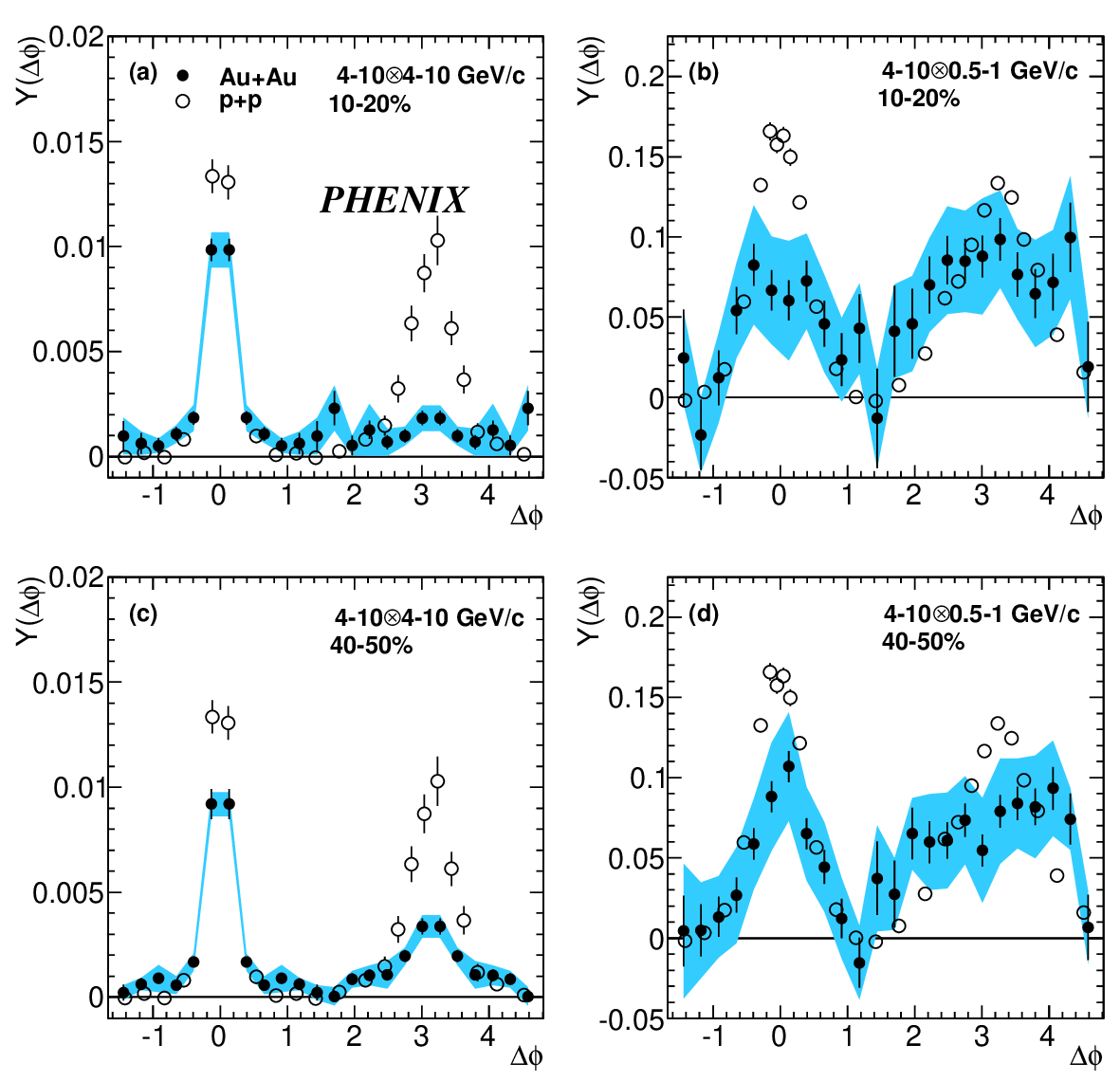}  %%13
\caption{
Comparison per-trigger yields between Au$+$Au and $p$+$p$ collisions at 
\Gsqsn from 4--10 GeV/$c$ triggers correlated with associated 
particles 4--10 GeV/$c$ (a,c) and 0.5--1 GeV/$c$ (b,d) in 10\%--20\% (a,b) 
and 40\%--50\% (c,d) collisions after subtraction of the 
underlying-event model. In the highest associated $p_T$ correlations an 
away-side suppression is observed. In the lowest associated $p_T$ 
correlations an enhanced yield at angles far from $\Delta\phi=\pi$ is 
observed. The background normalization (ZYAM) uncertainty shown in the 
purple band around zero on Figure \ref{fig:InclusiveTrig4-10} is 
included in the blue band around the points in this figure.
}
\label{fig:compJFsAApp}
\end{figure}

Figure \ref{fig:compJFsAApp} shows the comparison between the highest 
$\ptt$ correlations for each centrality with the same distributions 
measured in $p$+$p$ collisions from a previous analysis 
\cite{Adare:2008ae}. In that paper the lowest $\pta$ bin was 0.4-1.0 
GeV/$c$ compared to 0.5--1.0 GeV/$c$ in this analysis. Therefore, the 
lowest $\pta$ bin from $p$+$p$ was modified by a $\Delta\phi$-dependent 
correction determined from {\sc pythia} 6 \cite{Sjostrand:2006za}. The 
correction, which has negligible uncertainties compared to those from 
other sources, was determined from the ratio of fits to the {\sc pythia} 
dihadron $\Delta\phi$ per-trigger yield distributions with 
$0.5<\pta<1.0$ GeV/$c$ and $0.4<\pta<1.0$ GeV/$c$.

Previous correlation analyses that relied on $v_2$-only subtraction 
indicated the near-side yield was enhanced in Au$+$Au compared to $p$+$p$, 
the so-called ``ridge''~\cite{Abelev:2009af,Adare:2008ae}. Our updated 
underlying event model has reduced the near-side yield as 
expected~\cite{Alver:2010gr}. In fact, the yields are slightly suppressed 
relative to $p$+$p$. The integrated away-side yields show modification 
relative to $p$+$p$. Figures~\ref{fig:compJFsAApp}(b) and (d) show the 
comparisons of the per-trigger yields for the lowest $\pta$. The 
away-side shapes of the Au$+$Au distributions are different than $p$+$p$. 
The large-angle enhancement of the per-trigger yield at low associated 
particle momentum is qualitatively consistent with measurements of direct photon-hadron 
and jet-hadron correlations with fully reconstructed 
jets\cite{Adare:2012qi,Adamczyk:2013jei,Aad:2014wha,Chatrchyan:2014ava}.

%=================================================== Fig_13

\begin{figure}[htb]
\includegraphics[width=1.0\linewidth]{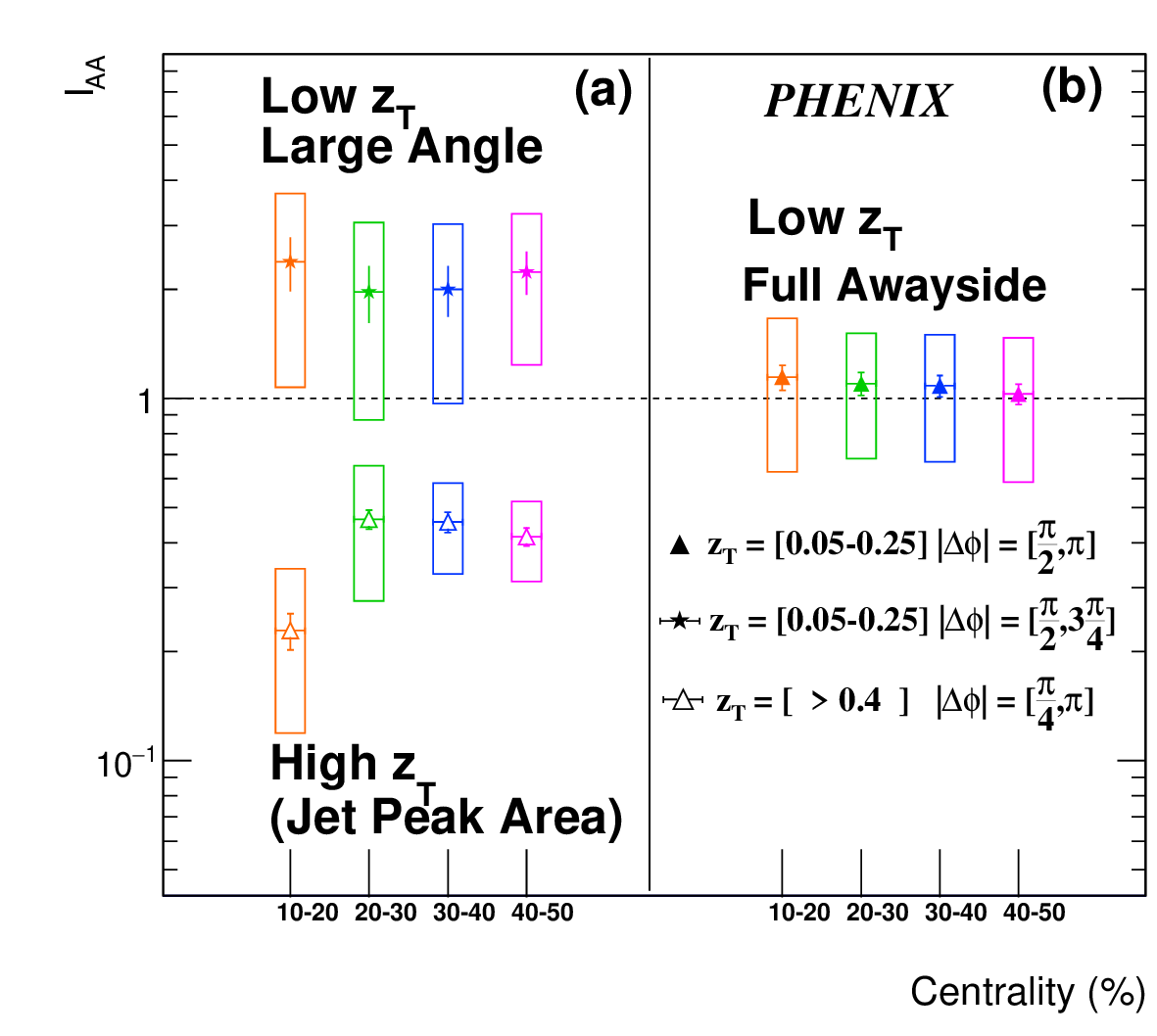}  %%14
\caption{
\iaa : Ratio of away-side yields in Au$+$Au to $p$+$p$ from 
Fig.~\ref{fig:compJFsAApp} in various $\Delta\phi$ integration regions 
for the high and low $z_T=\pta/\ptt$. Away-side yields show the 
well-known suppression at high $z_{T}$, most pronounced for the small 
angle region around the usual away-side peak center $|\Delta\phi- \pi| < 
\pi/4$. At low $z_{T}$, the large angle integration region, 
$|\Delta\phi- \pi| > \pi/4$, show an enhancement in \iaa which is 
significantly higher than the high $z_{T}$ suppressed values, and 
generally enhanced above unity The full away-side integration region at 
low $z_{T}$ is also higher than the suppressed level with at least 
1$\sigma$ significance for most centrality bins.
}
\label{fig:FragmentationFunc}
\end{figure}

To explore these features quantitatively, we calculate the ratio 
($I_{AA}$) of the away-side yields in Au$+$Au to those in $p$+$p$
\begin{equation} 
\iaa = \frac{Y_{{\rm Au}+{\rm Au}}(\Delta\phi)}{Y_{p+p}(\Delta\phi)}.
\end{equation}
Figure~\ref{fig:FragmentationFunc}(b) shows $\iaa$ vs.~centrality when 
integrating the away side $0 < |\Delta\phi-\pi| < \pi/2$ for 4--10 
GeV/$c$ hadrons paired with 0.5--1.0 GeV/$c$ hadrons. $\iaa$ is unity 
within uncertainties indicating that yield suppression is disfavored. 
Figure~\ref{fig:FragmentationFunc}(a) shows $\iaa$ for two different 
angular regions of integration and $\pta$ selections. First, for $\pta$ 
from 4--10 GeV/$c$ (high $z_{T}$) and integrating 
$0<|\Delta\phi-\pi|<\pi/4$, the jet peak region, $\iaa$ is less than 
unity indicating the pair yields are suppressed relative to those in 
$p$$+$$p$. This is consistent with previous measurements of strong 
suppression of high \pta \cite{Adare:2012qi,Adare:2010ry}. When 
integrating $\pi/4<|\Delta\phi-\pi|<\pi/2$ for $\pta$ from 0.5--1.0 
GeV/$c$ (low $z_{T}$), $\iaa\sim1$ within systematic uncertainties. This 
would indicate that the yield in Au$+$Au is similar to $p$+$p$. However, 
it is more instructive to compare the $\iaa$ for a fixed \ptt, which 
approximately fixes the jet energy. 
Figure~\ref{fig:FragmentationFunc}(a) shows that the low-$z_{T}$ 
fragments at large angles from $\Delta\phi = \pi$ are significantly 
enhanced compared to the suppressed level of high-$z_{T}$ fragments 
within the jet region. Both the high-$z_{T}$ suppression relative to 
$p$+$p$ and the enhanced level of low-$z_{T}$ fragments at large angles 
are consistent with a radiative energy loss model where the away-side 
jet traverses the medium, loses energy, and the energy gets 
redistributed to larger angles.

%=================================================== Fig_14

\subsection{Inclusive Per-Trigger Yields II: Intermediate-$p_T$ Trigger Particles}

%=================================================== Fig_15
\begin{figure*}[htb]
\includegraphics[width=0.99\linewidth]{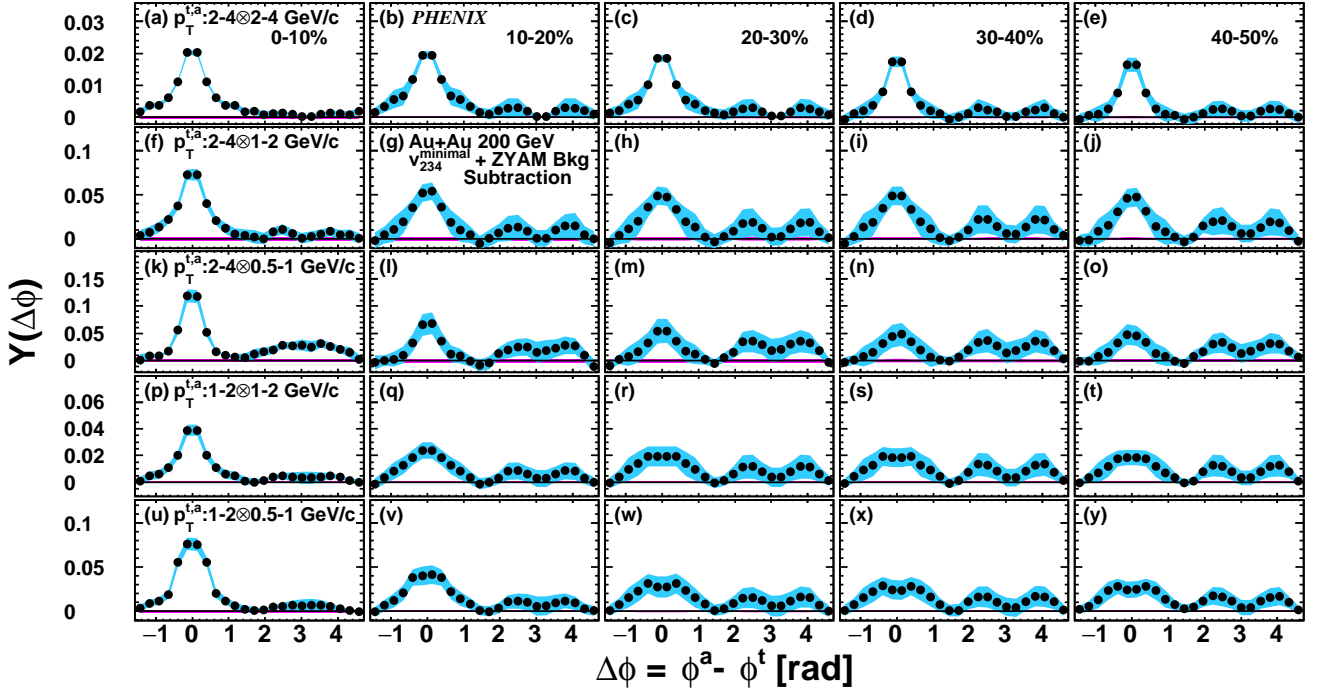}  %%15
\caption{
Per-trigger yields $Y(\Delta\phi)$ of dihadron pairs measured in Au$+$Au 
collisions after subtracting the underlying event-model with several 
$p_T$ selections of the trigger and associated particles ($p_T^{t,a}$):
((a)-(e)) ($2<\ptt<4$)~$\otimes$~($2<\pta<4$)~GeV/$c$,
((f)-(j)) ($2<\ptt<4$)~$\otimes$~($1<\pta<2$)~GeV/$c$,
((k)-(o)) ($2<\ptt<4$)~$\otimes$~($0.5<\pta<1$)~GeV/$c$,
((p)-(t)) ($1<\ptt<2$)~$\otimes$~($1<\pta<2$)~GeV/$c$, and
((u)-(y)) ($1<\ptt<2$)~$\otimes$~($0.5<\pta<1$)~GeV/$c$.
The columns represent centrality bins 0\%--10\% (a,f,k,p,u), 10\%--20\% 
(b,g,l,q,v), 20\%--30\% (c,h,m,r,w), 30\%--40\% (d,i,n,s,x), 40\%--50\% 
(e,j,o,t,y). Systematic uncertainties are shown by (blue) bands around 
the points. Uncertainties from ZYAM are shown by (purple) bands around 
zero yield.
}
\label{fig:InclusiveTrig1-4}
\end{figure*}

%=================================================== Fig_16
\begin{figure}[htb]
\includegraphics[width=1.0\linewidth]{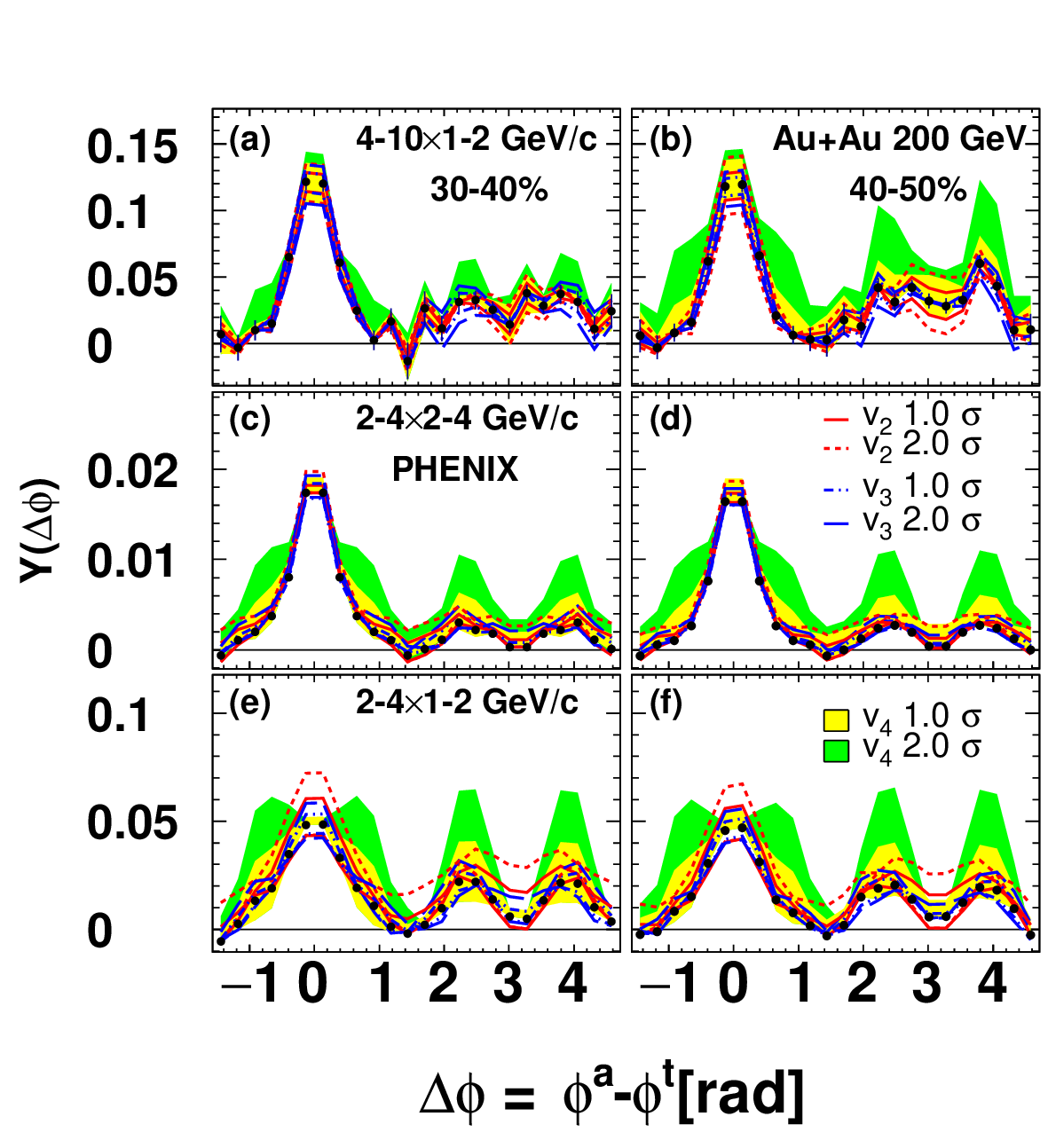} %%16
\caption{
 Per-trigger yields $Y(\Delta\phi)$ of dihadron pairs measured in Au$+$Au 
collisions after subtracting the underlying event-model with several 
$p_T$ selections:
((a)-(b)) ($4<\ptt<10$)~$\otimes$~($1<\pta<2$)~GeV/$c$,
((c)-(d)) ($2<\ptt<4$)~$\otimes$~($2<\pta<4$)~GeV/$c$, and
((e)-(f)) ($2<\ptt<4$)~$\otimes$~($1<\pta<2$)~GeV/$c$.
The columns represent centrality bins 30\%--40\% (a,c,e) and 40\%--50\% (b,d,f). The lines and bands further break down of the uncertainty contributions from each different order of the $v_n$ subtraction. The systematic uncertainties are point-to-point correlated. If the yield at $\Delta\phi=\pi$ is reduced, the away-side yield outside the region $\Delta\phi=\pi$ is increased. 
}
\label{fig:InclusiveTrigBrazil}
\end{figure}

%Given the success of our background model at reproducing prior 
Given the success of reproducing prior 
correlation results at high $\ptt$, we study lower $\ptt$ correlations 
to attempt to measure jet-like correlations at lower momentum 
transfer $Q^2$.  Per-trigger yields with trigger particles of $1<\ptt<2$ 
and $2<\ptt<4$~GeV/$c$ paired with associated particles of $0.5$ GeV/$c 
<\pta<\ptt$ in several centrality selections are shown in 
Fig.~\ref{fig:InclusiveTrig1-4}. As in Fig.~\ref{fig:InclusiveTrig4-10}, 
the ZYAM uncertainties are shown as a band around zero while $v_n$ 
uncertainties are combined as the band around the data points. At these 
$\ptt$ the jet-like signal-to-underlying-event background is reduced 
making the contribution of the $v_n$ uncertainties dominant. Because the 
$v_n$ uncertainties are point-to-point correlated, it is important to 
recognize that the yields and shape change due to that correlation. For 
example, if the $v_2$ subtracted is too large, the effect on the away 
side is a reduced peak and an enhanced large-angle yield. If the $v_2$ 
subtracted is too small, the away-side becomes more peaked. In the 
discussion that follows we only make statements that have a significant
variation over the systematic uncertainties.
%a large systematic significance.

The away-side yield and shape varies with both $\pta$ and centrality. In 
these \pt selections and in the most central collisions, the away-side 
seems to completely disappear. If our background model represents all 
nonjet correlations, the disappearance is presumably due to jet 
quenching. Compared to $v_2$-only subtraction \cite{Adare:2008ae}, the 
very large displaced away-side peaks are reduced primarily due to the 
subtraction of $v_3$ in the underlying event \cite{Alver:2010gr}. Both 
the flat away-side and the near-side peak shape, seems relatively 
centrality independent.

To better assess the systematic significance of the correlation 
features at the lower \pta\ selections, 
Figure~\ref{fig:InclusiveTrigBrazil} shows a further breakdown of the 
uncertainty contributions from different orders of the $v_n$ 
subtraction. The lines and bands show the 1$\sigma$ and 2$\sigma$ 
variation of each of the $v_n$ components independently. Here the ZYAM 
level is recalculated for every $v_n$ variation. These 
uncertainties are point-to-point correlated where a reduction at 
$\Delta\phi = \pi$ leads to an increase at angles away from $\pi$. The 
$v_4$ dominates the uncertainty in the away-side shape. The location of 
any double-peak structure in the away-side is also strongly dependent on 
the underlying event background subtraction. The peak in the 
uncertainties from $v_2$, $v_3$ and $v_4$ are all at slightly different 
places. Therefore, a robust and well-motivated background model is 
necessary to extract detailed shape information at these $\pta$. With 
our quoted systematic uncertainties, we cannot distinguish between a 
single broad away-side peak structure or a double-hump structure.

\subsection{Event Plane-Dependent Correlations}

\begin{figure*}[htb]
\begin{minipage}{0.67\linewidth}
\includegraphics[width=0.49\linewidth]{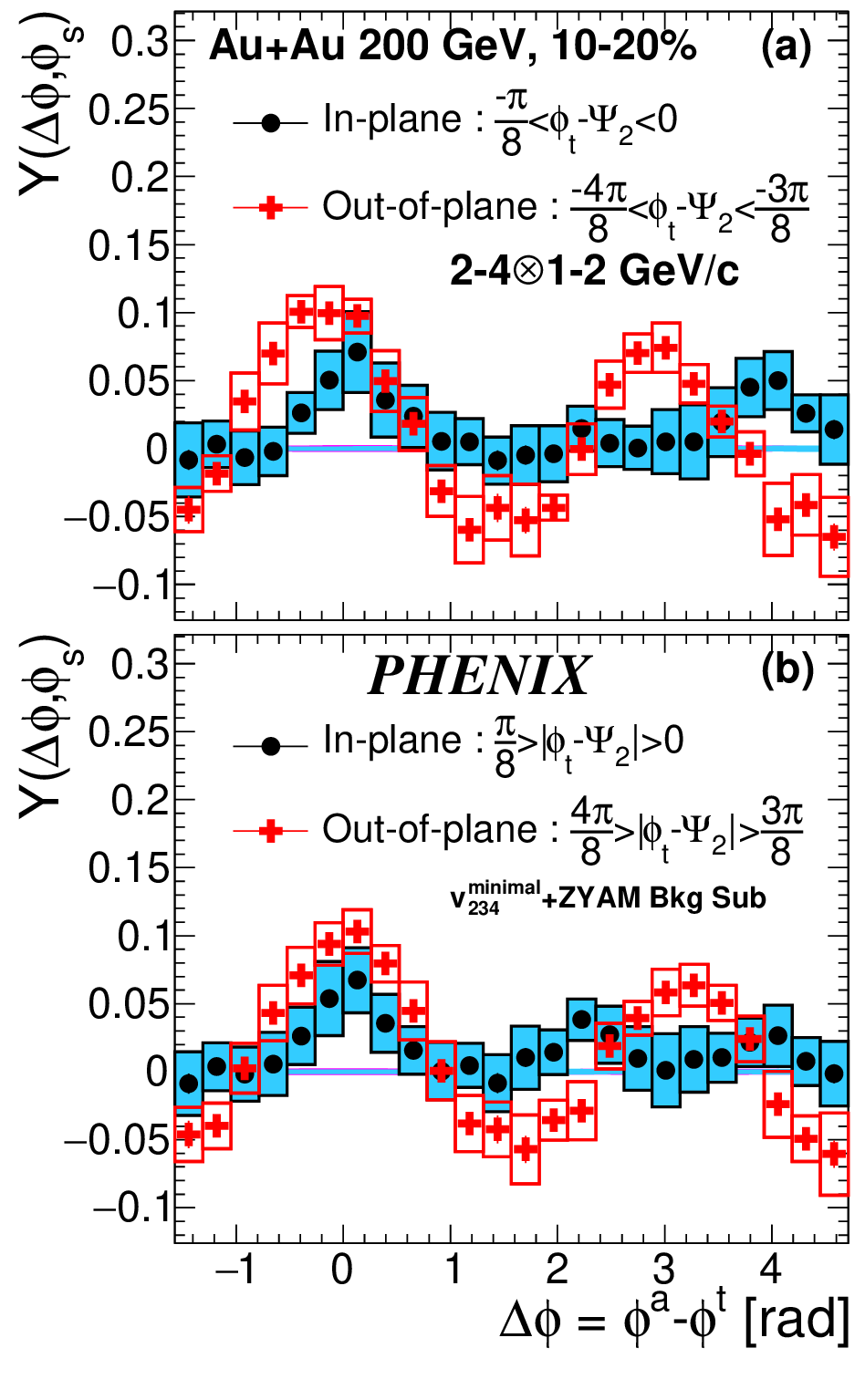} %%17a
\includegraphics[width=0.49\linewidth]{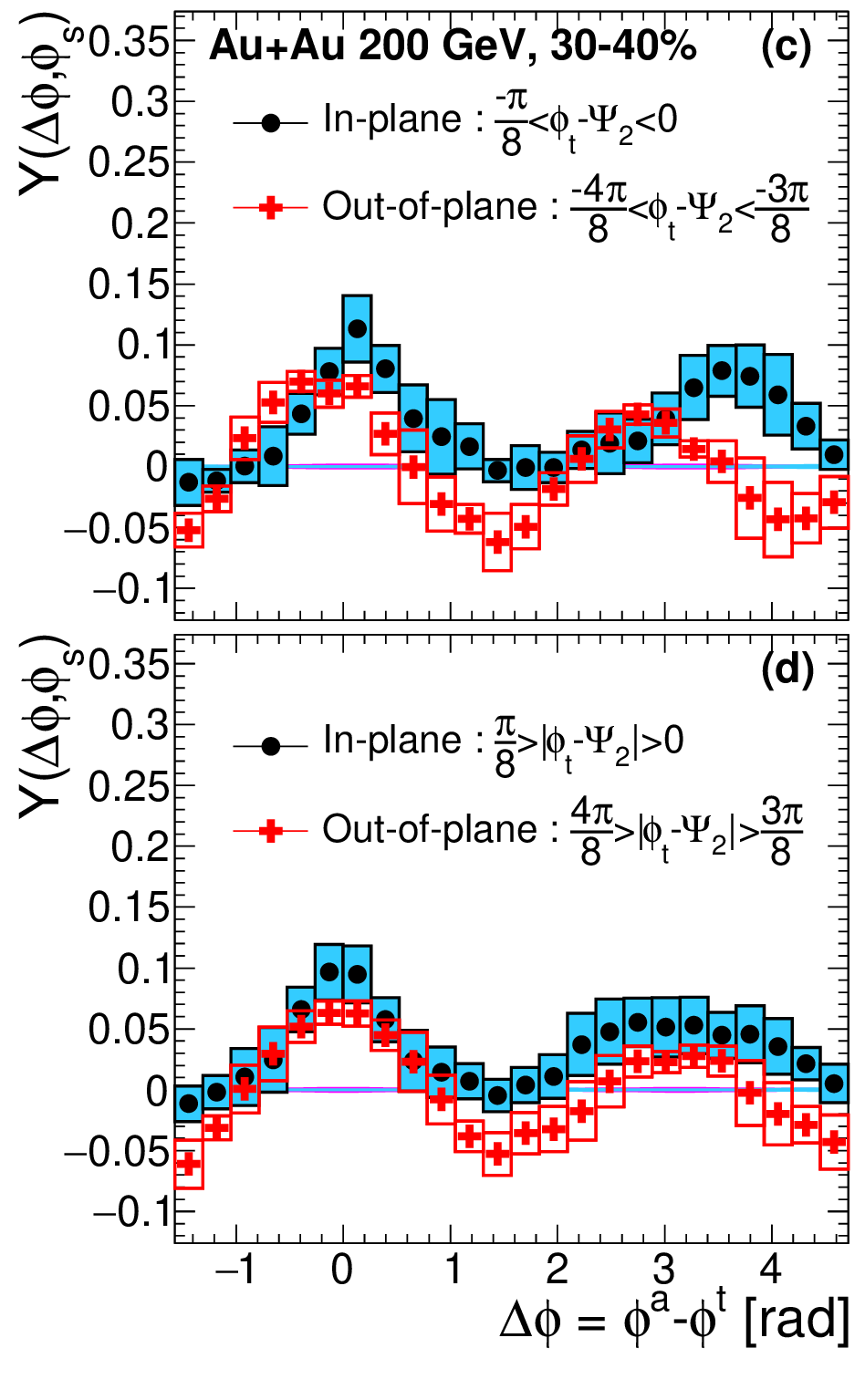} %%17b
\end{minipage}
%\hspace{0.05\linewidth}
\begin{minipage}{0.28\linewidth}
\caption{
 $\Psi_2$ event plane-dependent per-trigger yields $Y(\Delta\phi)$ of 
dihadron pairs measured in Au$+$Au collisions after subtracting the 
underlying event-model for ($2<\ptt<4$)~$\otimes$~($1<\pta<2$)~GeV/$c$ 
and 10\%--20\% (a,b) and 30\%--40\% (c,d). In-plane $0<|\phi^t-\Psi_2|<\pi/8$ 
are (black) circles and Out-of-plane $3\pi/8<|\phi^t-\Psi_2|<4\pi/8$ are 
(red) crosses. The sign of $\phi^t-\Psi_2$ is negative in the panels 
(a, c) and both positive and negative in the panels (b, d). The ZYAM 
systematic uncertainties are shown in the band around zero yield. The 
other systematic uncertainties are shown in the boxes around the points.
}
\label{fig:TwoPanelPTYExaVertPsi2Pt10_13}
\end{minipage}
%\end{figure*}
%=================================================== Fig_18
%\begin{figure*}[htb]
\begin{minipage}{0.67\linewidth}
\includegraphics[width=0.99\linewidth]{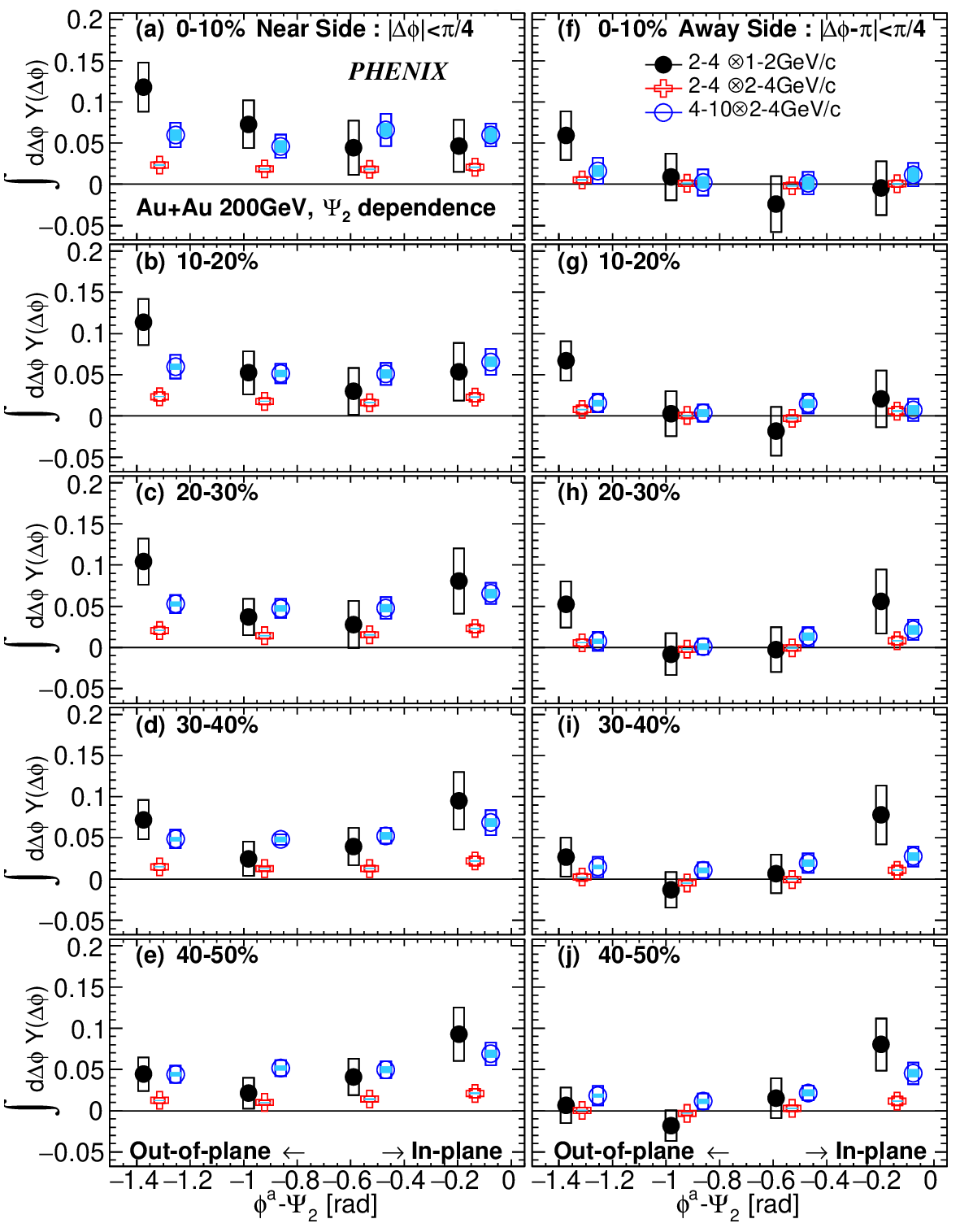} %%18
\end{minipage}
%\hspace{0.05\linewidth}
\begin{minipage}{0.31\linewidth}
\caption{
 Integrated per-trigger yields as a function of associated particle 
angle relative to $\Psi_2$ event plane $\phi^a - \Psi_2$ integrated over 
the near side \mbox{$|\Delta\phi|<\pi/4$} (a--e) 
and over the away side \mbox{$|\Delta\phi-\pi|<\pi/4$} (f--j). The 
columns represent centrality bins (a,f) 0\%--10\%, (b,g) 10\%--20\%, 
(c,h) 20\%--30\%, (d,i) 30\%--40\%, (e,j) 40\%--50\%. The ranges 
of $\ptt\otimes\pta$ are 
(filled [black] circles) ($2<\ptt<4$)~$\otimes$~($1<\pta<2$)~GeV/$c$,
(open [red] squares), $2<\ptt<4$~$\otimes$~($2<\pta<4$)~GeV/$c$, and 
(open [blue] circles) ($4<\ptt<10$)~$\otimes$~($2<\pta<4$)~GeV/$c$.
 The ZYAM systematic uncertainties are shown in the solid light blue boxes around the data points. The other systematic uncertainties are shown in the open boxes around the points.
}
\label{fig:PTYAngPsi2NoFitVertr}
\end{minipage}
\end{figure*}

Figure~\ref{fig:TwoPanelPTYExaVertPsi2Pt10_13} shows $\Psi_2$-dependent 
per-trigger yields of trigger particles from $2<\ptt<4$ GeV/$c$ with 
associated particles from $1<\pta<2$ GeV/$c$ in 10\%--20\% and 30\%--40\% 
central collisions. In Figs.~\ref{fig:TwoPanelPTYExaVertPsi2Pt10_13} (b) 
and (d), the trigger is selected to be either in-plane 
$0<|\phi_s|<\pi/8$ or out-of-plane $3\pi/8<|\phi_s|<\pi/2$, 
respectively. Similar to the inclusive per-trigger yields, we observe a 
broad away-side structure in both cases. The use of a common ZYAM point 
results in a slight over-subtraction in the out-of-plane bins. The over 
subtraction can be corrected by determining a ZYAM point for each 
$\phi_s$ selection, which however makes event-plane dependent correlations integrated 
over the $\phi_s$ bins different from inclusive correlations. This would result in moving all yield points up and 
does not affect the discussion of the shape that follows.

In Figs.~\ref{fig:TwoPanelPTYExaVertPsi2Pt10_13} (a) and (c), we chose the 
trigger to have a particular sign of $\phi_s$. That is, we choose 
$-\pi/8<\phi_s<0$ and $-\pi/2<\phi_s<-3\pi/8$ for the in-plane and 
out-of-plane, respectively. Choosing the sign of $\phi_s$ to be negative 
results in always choosing the trigger to be ``below'' the event plane, 
if the event plane is the horizontal. When sign-selecting $\phi_s$, an 
asymmetry around $\Delta\phi\sim0$ and $\Delta\phi\sim\pi$ is observed. 
Such an asymmetry does not exist when choosing both signs of $\phi_s$. 
In the in-plane trigger case, there is a preference for the associated 
particle to be emitted toward the in-plane direction $i.e.$ the thinner 
side of the overlap region.  Referring back to 
Fig.~\ref{fig:InplaneJetAndGluons}, our data suggests that Gluon 1 is 
more likely to be measured than Gluon 2.  Other $\Psi_2$ dependent 
per-trigger yields with a selection of trigger azimuthal angle relative 
to event-planes $\phi^t-\Psi_n<0$ for different collision centralities 
and $p_T^{t,a}$ selections are shown in the Appendix.

The integrated per-trigger yields are shown in Fig.~\ref{fig:PTYAngPsi2NoFitVertr} as 
a function of associated particle angle with respect to $\Psi_2$. 
Figure~\ref{fig:PTYAngPsi2NoFitVertr} shows data for all centralities and 
all four orientations with respect to the event plane. We note that the 
use of a single ZYAM level for all event plane bins can result in 
negative yields for certain $\Delta\phi$.

In Figs.~\ref{fig:PTYAngPsi2NoFitVertr} (a)-(e), the yields have been 
integrated for the near side $|\Delta\phi| = < \pi/4$. For all 
trigger/associated combinations there is weak to no dependence of the 
yield on event plane orientation. This trend persists for all centrality 
selections.

%=================================================== Fig_17

%=================================================== Fig_19
\begin{figure*}[htb]
\begin{minipage}{0.67\linewidth}
\includegraphics[width=0.49\linewidth]{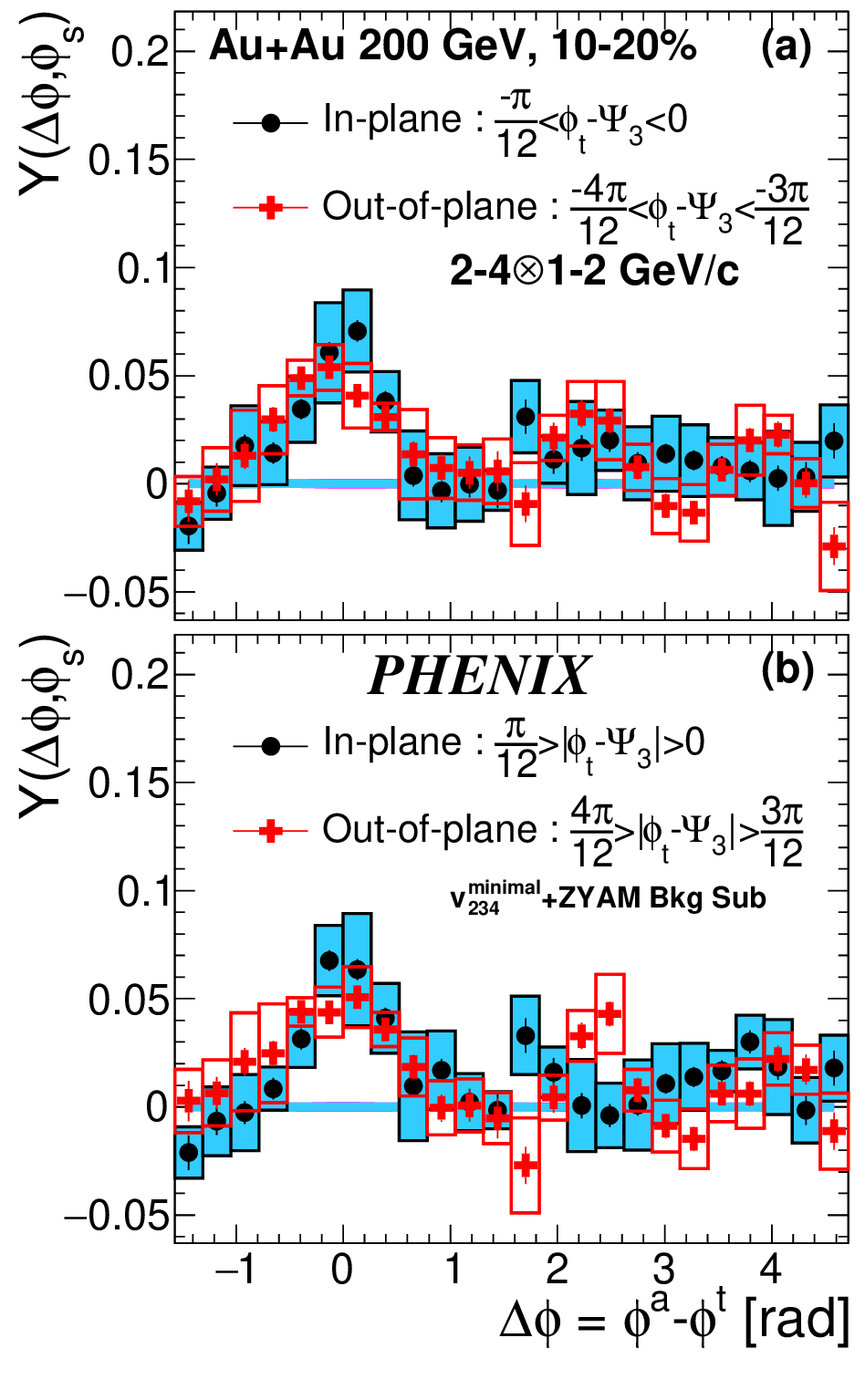}  %%19a
\includegraphics[width=0.49\linewidth]{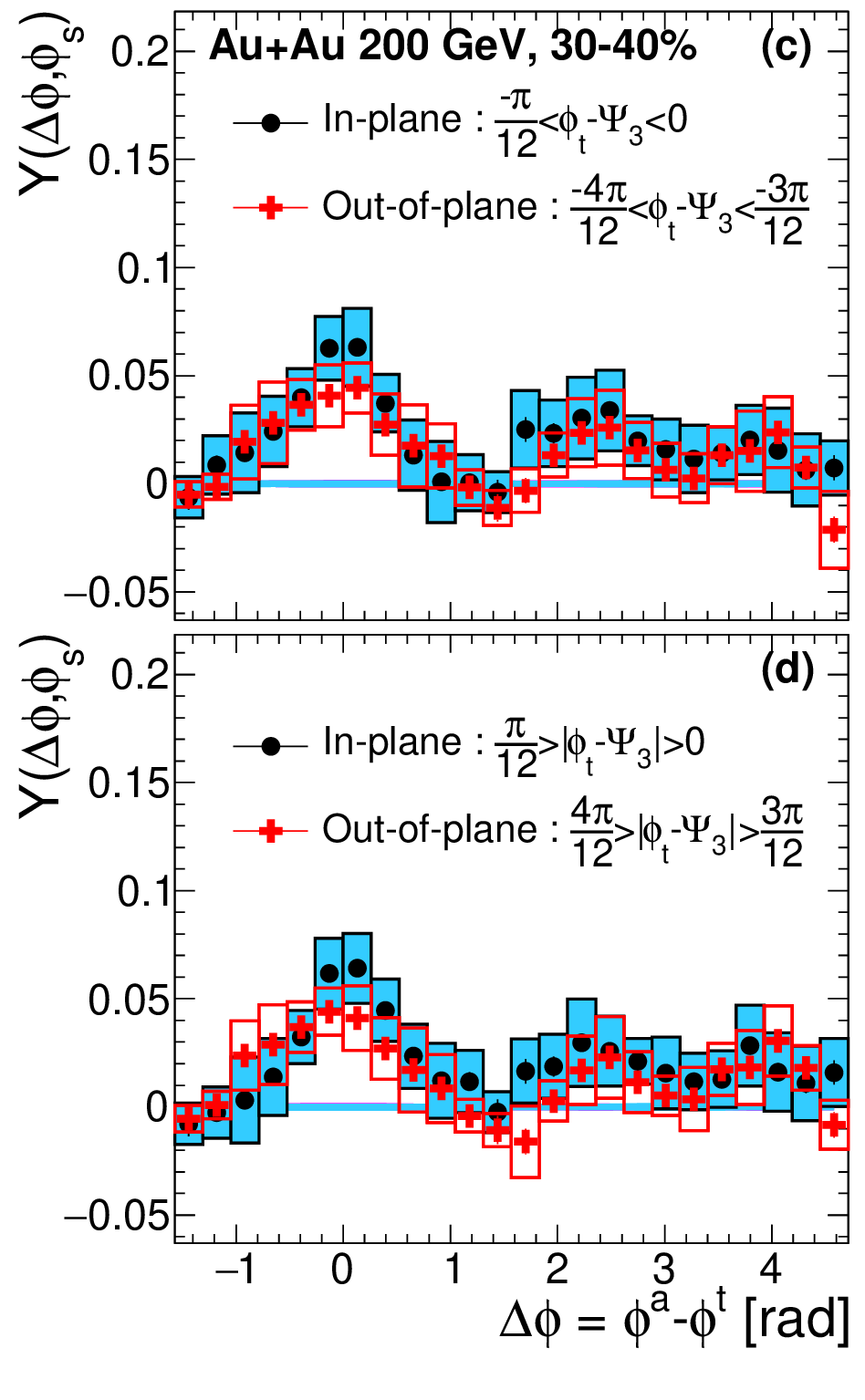}  %%19b
\end{minipage}
%\hspace{0.05\linewidth}
\begin{minipage}{0.28\linewidth}
\caption{
 $\Psi_3$ event plane-dependent per-trigger yields $Y(\Delta\phi)$ of 
dihadron pairs measured in Au$+$Au collisions after subtracting the 
underlying event model fo ($2<\ptt<4$)~$\otimes$~($1<\pta<2$)~GeV/$c$ 
and 10\%--20\% (a,b) and 30\%--40\% (c,d). In-plane 
\mbox{$0<|\phi^t-\Psi_3|<\pi/12$} are (black) circles and out-of-plane 
$3\pi/12<|\phi^t-\Psi_3|<4\pi/12$ are (red) crosses. The sign of 
$\phi^t-\Psi_3$ is negative in the panels (a,c) and both positive and 
negative in the panels (b,d). The ZYAM systematic uncertainties are show 
in the band around zero yield. The other systematic uncertainties are 
shown in the boxes around the points.
}
\label{fig:TwoPanelPTYExaVertPsi3Pt10_13}
\end{minipage}
%\end{figure*}
%=================================================== Fig_20
%\begin{figure*}[htb]
\begin{minipage}{0.67\linewidth}
\includegraphics[width=0.99\linewidth]{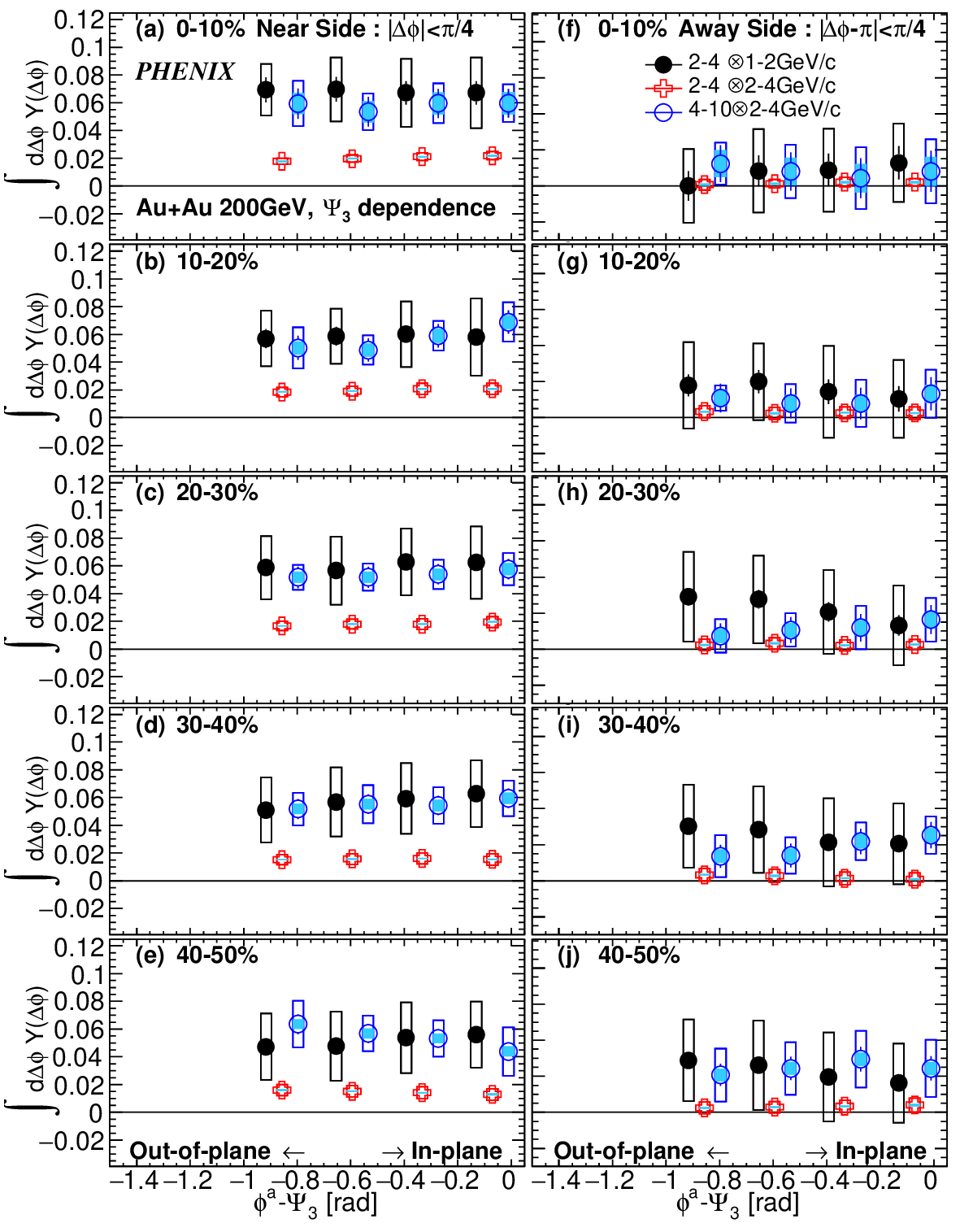}  %%20
\end{minipage}
%\hspace{0.05\linewidth}
\begin{minipage}{0.31\linewidth}
\caption{
 Integrated per-trigger yields as a function of associated particle angle relative to $\Psi_3$ event plane for ((a)-(e)) near-side $|\Delta\phi|<\pi/4$ and for ((f)-(j)) away-side $|\Delta\phi-\pi|<\pi/4$. The columns represent centrality bins 0\%--10\% (top), 10\%--20\%, 20\%--30\%, 30\%--40\%, 40\%--50\% (bottom). The ranges of $\ptt\otimes\pta$ are ($2<\ptt<4$)~$\otimes$~($1<\pta<2$)~GeV/$c$ (filled black circles), $2<\ptt<4$~$\otimes$~$2<\pta<4$ GeV/$c$ (red squares), and ($4<\ptt<10$)~$\otimes$~($2<\pta<4$)~GeV/$c$ (open blue circles). The ZYAM systematic uncertainties are shown in the solid light blue boxes around the data points. The other systematic uncertainties are shown in the open boxes around the points.
}
\label{fig:PTYAngPsi3NoFitVert}
\end{minipage}
\end{figure*}

In Figs.~\ref{fig:PTYAngPsi2NoFitVertr} (f)-(j), the yields have been 
integrated for the away side $|\Delta\phi-\pi|<\pi/4$. For the event plane 
selections in the range $ -1.2<\phi^{a}-\Psi_2<-0.5$, no significant yields 
are generally observed for both the highest and lowest trigger \pt.
In the most central collisions, for ($2<\ptt<4$)~$\otimes$~($1<\pta<2$)~GeV/$c$,
the largest yield is out of plane. The difference between the in-plane and
out-of-plane yields is approximately 1 sigma.
This trend is opposite in the most peripheral collision selection where the largest 
yield is in plane. In 10\%--40\% central collisions similar yields are 
observed in-plane versus out of plane. 
%Because each of the trends with event plane is not affected by 
%the level of the ZYAM uncertainty, 
For ($2<\ptt<4$)~$\otimes$~($1<\pta<2$)~GeV/$c$, there is a possible trend with centrality 
that out-of-plane yield is reduced from central to peripheral collisions whereas in-plane
yield increases from central to peripheral collisions.  The significance of this possible 
trend is approximately 1 sigma.
%The trends with event plane and centrality are observed in 
%both the highest and lowest trigger \pt.

Correlations selecting the trigger within a certain azimuthal angles 
from the $\Psi_3$ plane are shown in 
Figure~\ref{fig:TwoPanelPTYExaVertPsi3Pt10_13}. The triangular shape of 
the third Fourier component restricts the range: $-\pi/3 < \Psi_3 \le 
\pi/3$. The out-of-plane direction is at $\pm\pi/3$ radians relative to $\Psi_3$. In 
Figs.~\ref{fig:TwoPanelPTYExaVertPsi3Pt10_13} (b) and (d), we do not 
select the sign for $\phi_s$. Similar to the inclusive distributions, 
there is a broad away-side structure that has a small yield. In 
Figs.~\ref{fig:TwoPanelPTYExaVertPsi3Pt10_13} (a) and (c), we select for 
$\phi_s<0$. No discernible asymmetry is observed. It is possible that 
unfolding with the smaller $\Psi_3$ event plane resolution could obscure 
any effect.  Other $\Psi_3$ dependent per-trigger yields with a 
selection of trigger azimuthal angle relative to event-planes 
$\phi^t-\Psi_n<0$ for different collision centralities and $p_T^{t,a}$ 
selections are shown in the Appendix.

Similar to the $\Psi_2$-dependent correlations, we also integrate the 
per-trigger yields. This is shown in Fig.~\ref{fig:PTYAngPsi3NoFitVert} 
for each centrality and associated particle azimuthal angle with respect 
to $\Psi_3$. Figures~\ref{fig:PTYAngPsi3NoFitVert} (a)-(e) show the near 
side integral.  Figures~\ref{fig:PTYAngPsi3NoFitVert} (f)-(j) show the 
away side integral. In all cases, no event plane-dependent or centrality 
dependent trends are observed within uncertainties.

\clearpage %

\section{Summary} \label{Summary} 

In summary, we reported the two-particle azimuthal dihadron correlation 
measurements at $|\Delta\eta|<0.7$ in Au$+$Au collisions at 
$\sqrt{s_{_{NN}}}=$~200~GeV with and without subtraction of an underlying event.
The underlying event model includes modulations from higher-order flow 
coefficients $v_n$~$(n=2,3,4)$ that assumes only the expected 
correlation of 2nd and 4th order event planes.

We tested this two-source model by studying high-\pt ($>4$ GeV/$c$) 
triggers. We observe suppression of high-$z_{T}$ jet fragments as well 
as enchantment of low-$z_{T}$ jet fragments off the away-side jet axis. 
These results are consistent with previous dihadron and $\gamma$-hadron 
correlations and jet analyses 
\cite{Adler:2003qi,Adler:2003au,Adams:2003im,CMS:2012aa,Aad:2014bxa,Aad:2012vca}.

At lower trigger \pt $2<\ptt<4$ GeV/$c$, the near-side distribution is 
not enhanced compared to $p$$+$$p$, which traditionally is associated with 
the ridge. When a significant away-side yield exists, the double-hump 
structure that had been observed when subtracting a $v_2$-only 
underlying event is significantly reduced. Given our model assumptions 
and the systematic uncertainties on $v_n$, we cannot precisely determine 
if the away-side distribution is a single broadened peak or has further 
structure that may peak away from $\Delta\phi = \pi$.

We also present dihadron correlations selecting on $\phi_s$, the angle 
of the trigger with respect to the event plane. When 
requiring $\phi_s$ to chose one side of the overlap region, the 
away-side developed an asymmetry in the $\Delta\phi$ distribution 
where the away-side yield is largest on the same side of the event 
plane. Such an asymmetry is qualitatively consistent with path 
length-dependent energy loss where the away-side jet would have less 
medium to traverse when emerging from the same side of the overlap than 
being exactly back-to-back or through the opposite side. The observed 
asymmetry when the sign of the trigger with respect to the event plane 
is selected should set additional constraints on models of parton energy 
loss, and/or models of the underlying event.

%%%%%%%%%%%%%%%%%%%%%%%%%  Acknowledgements

\section*{ACKNOWLEDGMENTS} 

We thank the staff of the Collider-Accelerator and Physics
Departments at Brookhaven National Laboratory and the staff of
the other PHENIX participating institutions for their vital
contributions.  We acknowledge support from the
Office of Nuclear Physics in the
Office of Science of the Department of Energy,
the National Science Foundation,
Abilene Christian University Research Council,
Research Foundation of SUNY, and
Dean of the College of Arts and Sciences, Vanderbilt University
(U.S.A),
Ministry of Education, Culture, Sports, Science, and Technology
and the Japan Society for the Promotion of Science (Japan),
Conselho Nacional de Desenvolvimento Cient\'{\i}fico e
Tecnol{\'o}gico and Funda\c c{\~a}o de Amparo {\`a} Pesquisa do
Estado de S{\~a}o Paulo (Brazil),
Natural Science Foundation of China (People's Republic of China),
Croatian Science Foundation and
Ministry of Science and Education (Croatia),
Ministry of Education, Youth and Sports (Czech Republic),
Centre National de la Recherche Scientifique, Commissariat
{\`a} l'{\'E}nergie Atomique, and Institut National de Physique
Nucl{\'e}aire et de Physique des Particules (France),
Bundesministerium f\"ur Bildung und Forschung, Deutscher
Akademischer Austausch Dienst, and Alexander von Humboldt Stiftung (Germany),
J. Bolyai Research Scholarship, EFOP, the New National Excellence
Program ({\'U}NKP), NKFIH, and OTKA (Hungary),
Department of Atomic Energy and Department of Science and Technology (India),
Israel Science Foundation (Israel),
Basic Science Research and SRC(CENuM) Programs through NRF funded by the 
Ministry of Education and the Ministry of Science and ICT (Korea);
Physics Department, Lahore University of Management Sciences (Pakistan),
Ministry of Education and Science, Russian Academy of Sciences,
Federal Agency of Atomic Energy (Russia),
VR and Wallenberg Foundation (Sweden),
the U.S. Civilian Research and Development Foundation for the
Independent States of the Former Soviet Union,
the Hungarian American Enterprise Scholarship Fund,
the US-Hungarian Fulbright Foundation,
and the US-Israel Binational Science Foundation.

%################################################################## Appendix B

\section*{APPENDIX}
\label{appendix}
\begin{figure*}[htb]
\includegraphics[width=0.99\linewidth]{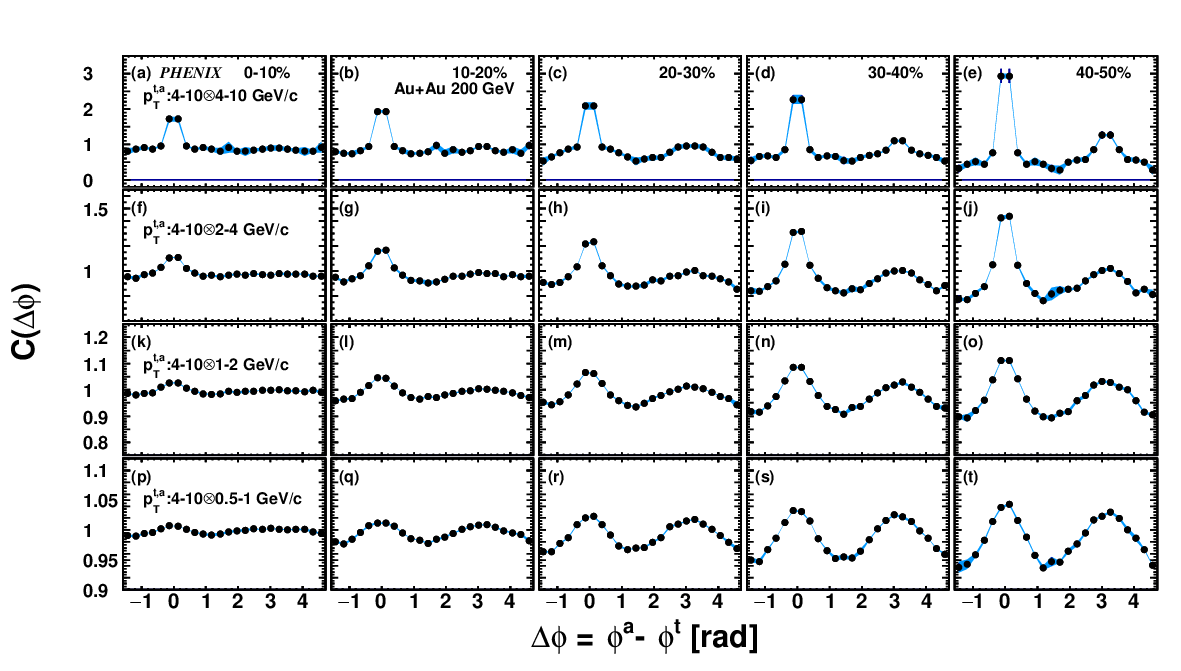}  %%21
\caption{
Correlations $C(\Delta\phi)$ of dihadrons pairs measured in Au$+$Au 
collisions at \Gsqsn before subtracting the underlying event model 
with several $p_T$ selections:
(a)--(e) ($4<\ptt<10$)~$\otimes$~($4<\pta<10$)~GeV/$c$,
(f)--(j) ($4<\ptt<10$)~$\otimes$~($2<\pta<4$)~GeV/$c$,
(k)--(o) ($4<\ptt<10$)~$\otimes$~($1<\pta<2$)~GeV/$c$, and
(p)--(t) ($4<\ptt<10$)~$\otimes$~($0.5<\pta<1$)~GeV/$c$.
The columns represent centrality bins 0\%--10\% (a,f,k,p), 10\%--20\% 
(b,g,l,q), 20\%--30\% (c,h,m,r), 30\%--40\% (d,i,n,s), 40\%--50\% (e,j,o,t). 
Systematic uncertainties due to track matching are shown 
by blue bands around the points.
}
\label{fig:HighPtInclusiveCorWOVnSubtraction}
%\end{figure*}
%=================================================== Fig_22
%\begin{figure*}[htb]
\includegraphics[width=0.99\linewidth]{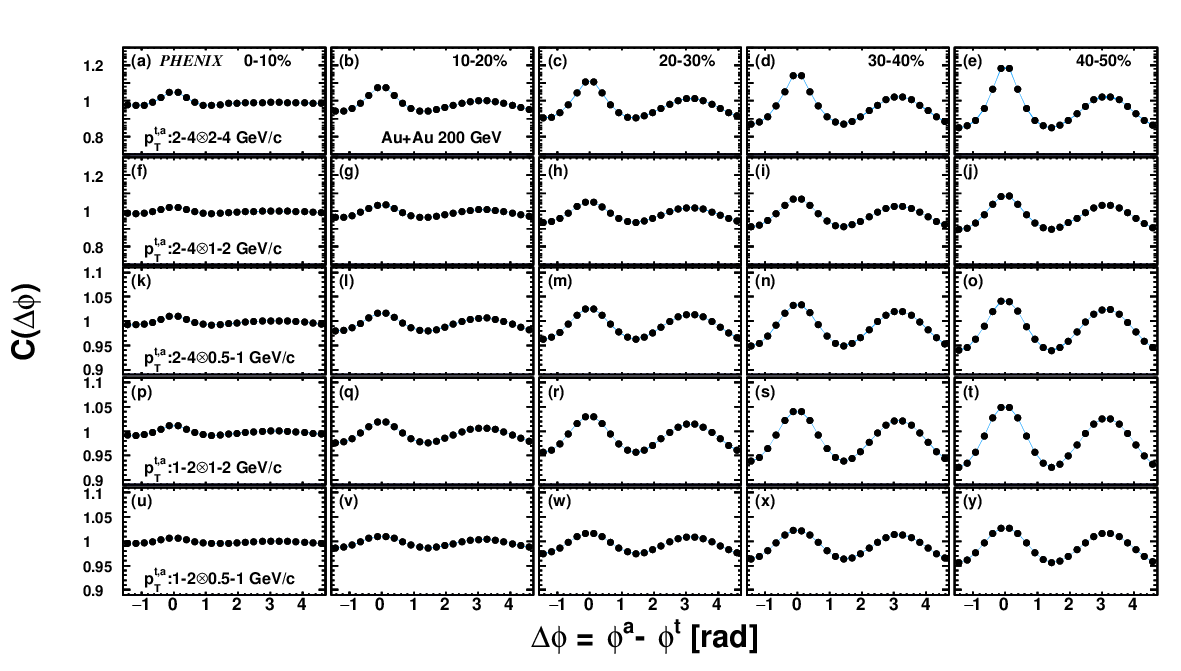} %%22
\caption{
Correlations $C(\Delta\phi)$ of dihadron pairs measured in Au$+$Au 
collisions before subtracting the underlying event-model with several 
$p_T$ selections of the trigger and associated particles ($p_T^{t,a}$):
((a)-(e)) ($2<\ptt<4$)~$\otimes$~($2<\pta<4$)~GeV/$c$,
((f)-(j)) ($2<\ptt<4$)~$\otimes$~($1<\pta<2$)~GeV/$c$,
((k)-(o)) ($2<\ptt<4$)~$\otimes$~($0.5<\pta<1$)~GeV/$c$,
((p)-(t)) ($1<\ptt<2$)~$\otimes$~($1<\pta<2$)~GeV/$c$, and
((u)-(y)) ($1<\ptt<2$)~$\otimes$~($0.5<\pta<1$)~GeV/$c$.
The columns represent centrality bins 0\%--10\% (a,f,k,p,u), 10\%--20\% 
(b,g,l,q,v), 20\%--30\% (c,h,m,r,w), 30\%--40\% (d,i,n,s,x), 40\%--50\% 
(e,j,o,t,y). 
Systematic uncertainties due to track matching are shown 
by blue bands around the points.
}
\label{fig:LowPtInclusiveCorWOVnSubtraction}
\end{figure*}

\begin{figure*}[htb]
\begin{minipage}{0.99\linewidth}
\includegraphics[width=0.89\linewidth]{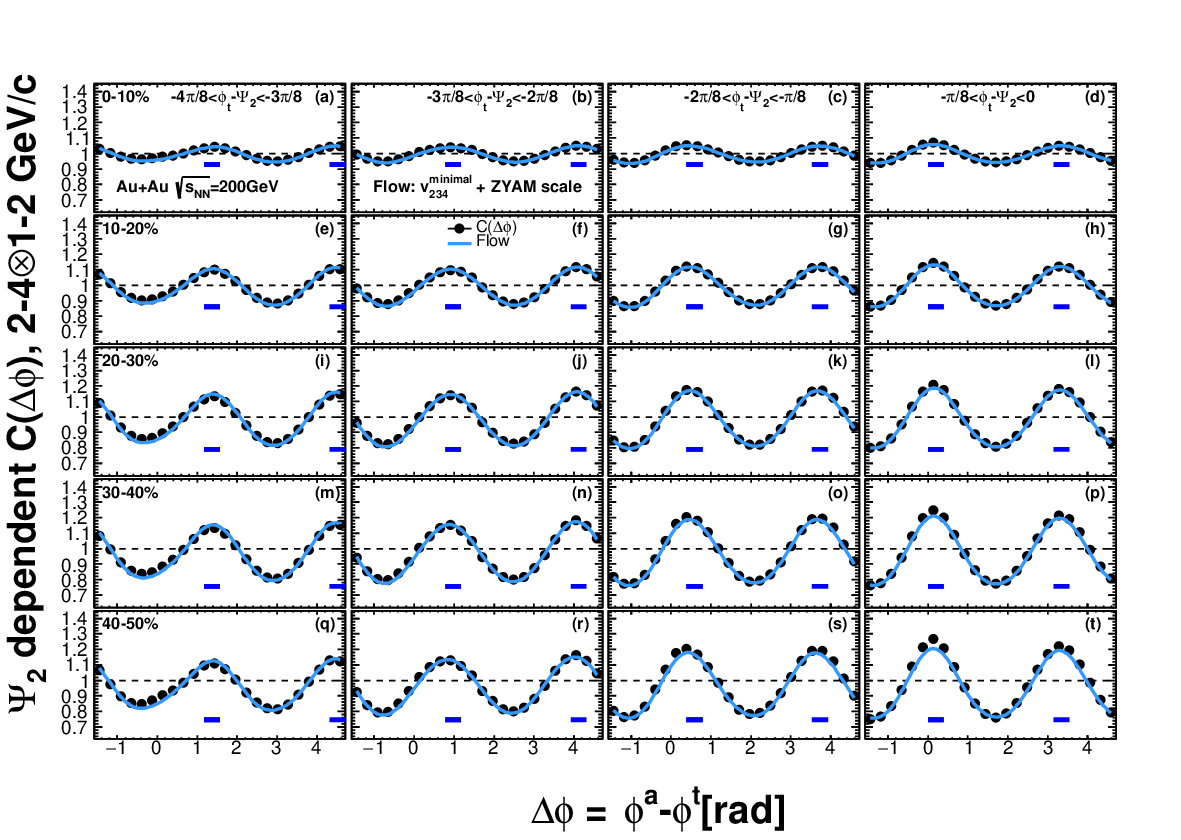}
\caption{
$\Psi_2$-dependent correlations $C(\Delta\phi,\phi_s)$ and flow 
backgrounds $F(\Delta\phi,\phi_s)$ for 
\mbox{($2<\ptt<4$)~$\otimes$~($1<\pta<2$)}~GeV/$c$. Trigger particle 
azimuthal angle relative to the event plane 
\mbox{$\phi_s=\phi^t-\Psi_2$} is selected out of plane (left) to in 
plane (right). Centrality is 0\%--10\% (top) to 40\%--50\% (bottom).
}
\label{fig:FlowCorFitNPFA5x4Pt100}
\end{minipage}
%\end{figure*}
%=================================================== Fig_24
%\begin{figure*}[htb]
\begin{minipage}{0.99\linewidth}
\includegraphics[width=0.89\linewidth]{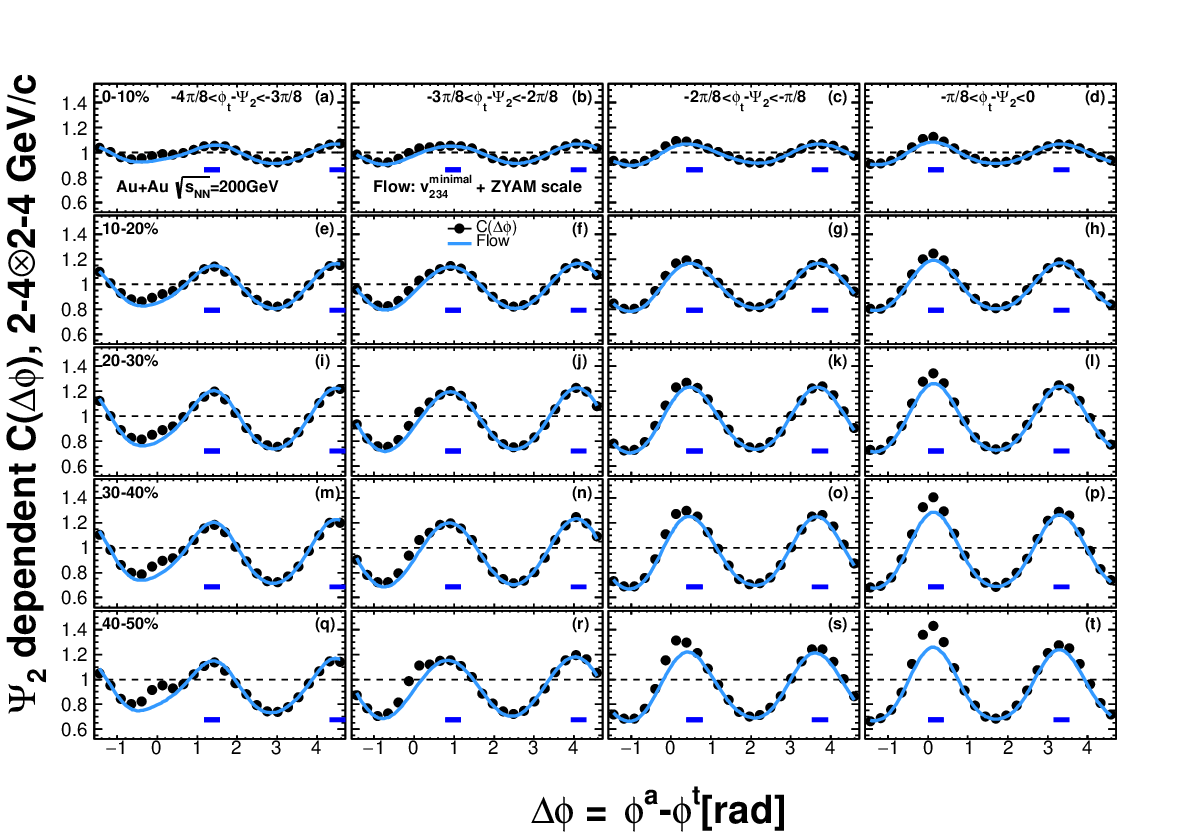}
\caption{
$\Psi_2$-dependent correlations $C(\Delta\phi,\phi_s)$ and flow backgrounds 
$F(\Delta\phi,\phi_s)$ for \mbox{($2<\ptt<4$)~$\otimes$~($2<\pta<4$)}~GeV/$c$. 
Trigger particle azimuthal angle relative to the event plane 
\mbox{$\phi_s=\phi^t-\Psi_2$} is selected out of plane (left) to in plane (right). 
Centrality is 0\%--10\% (top) to 40\%--50\% (bottom).
}
\label{fig:FlowCorFitNPFA5x4Pt110}
\end{minipage}
\end{figure*}

\begin{figure*}[htb]
\begin{minipage}{0.99\linewidth}
\includegraphics[width=0.89\linewidth]{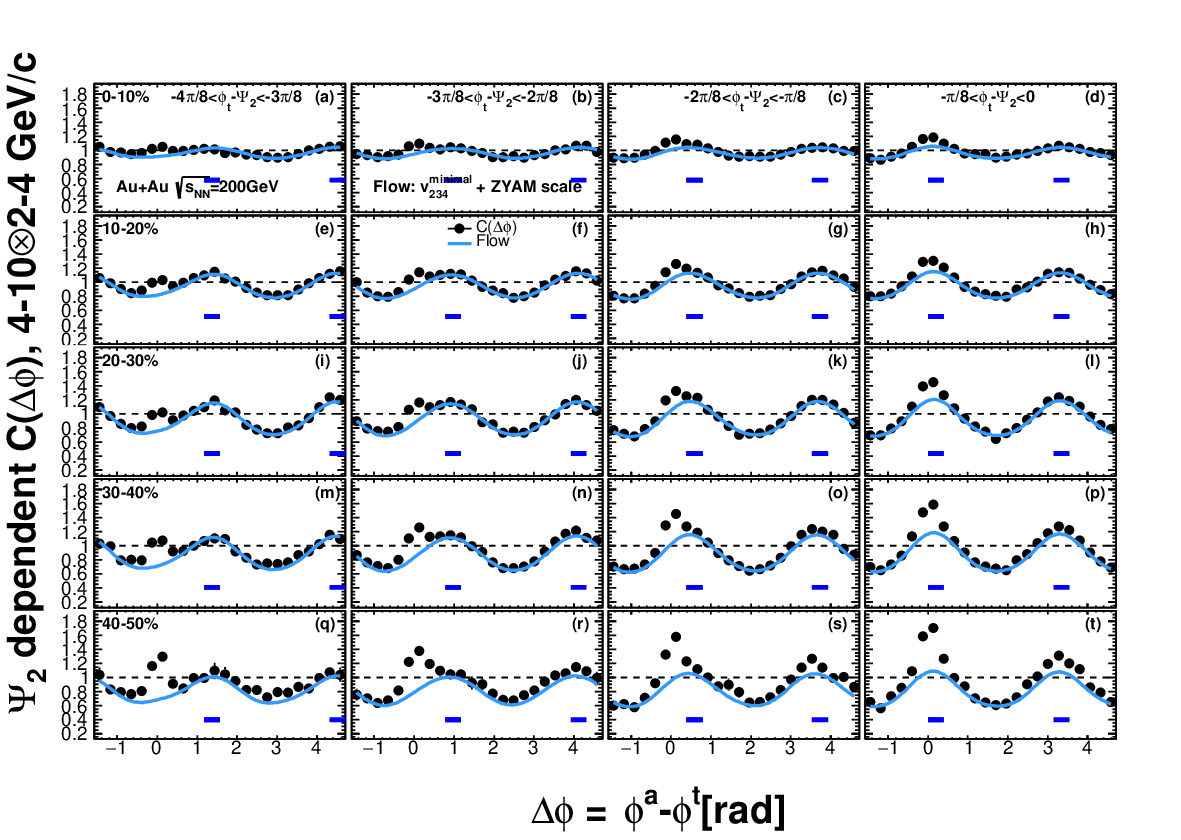}
\caption{
$\Psi_2$-dependent correlations $C(\Delta\phi,\phi_s)$ and flow backgrounds 
$F(\Delta\phi,\phi_s)$ for \mbox{($4<\ptt<10$)~$\otimes$~($2<\pta<4$)}~GeV/$c$. 
Trigger particle azimuthal angle relative to the event plane 
\mbox{$\phi_s=\phi^t-\Psi_2$} is selected out of plane (left) to in plane (right). 
Centrality is 0\%--10\% (top) to 40\%--50\% (bottom).
}
\label{fig:FlowCorFitNPFA5x4Pt210}
\end{minipage}
%\end{figure*}
%=================================================== Fig_26
%\begin{figure*}[htb]
\begin{minipage}{0.99\linewidth}
\includegraphics[width=0.89\linewidth]{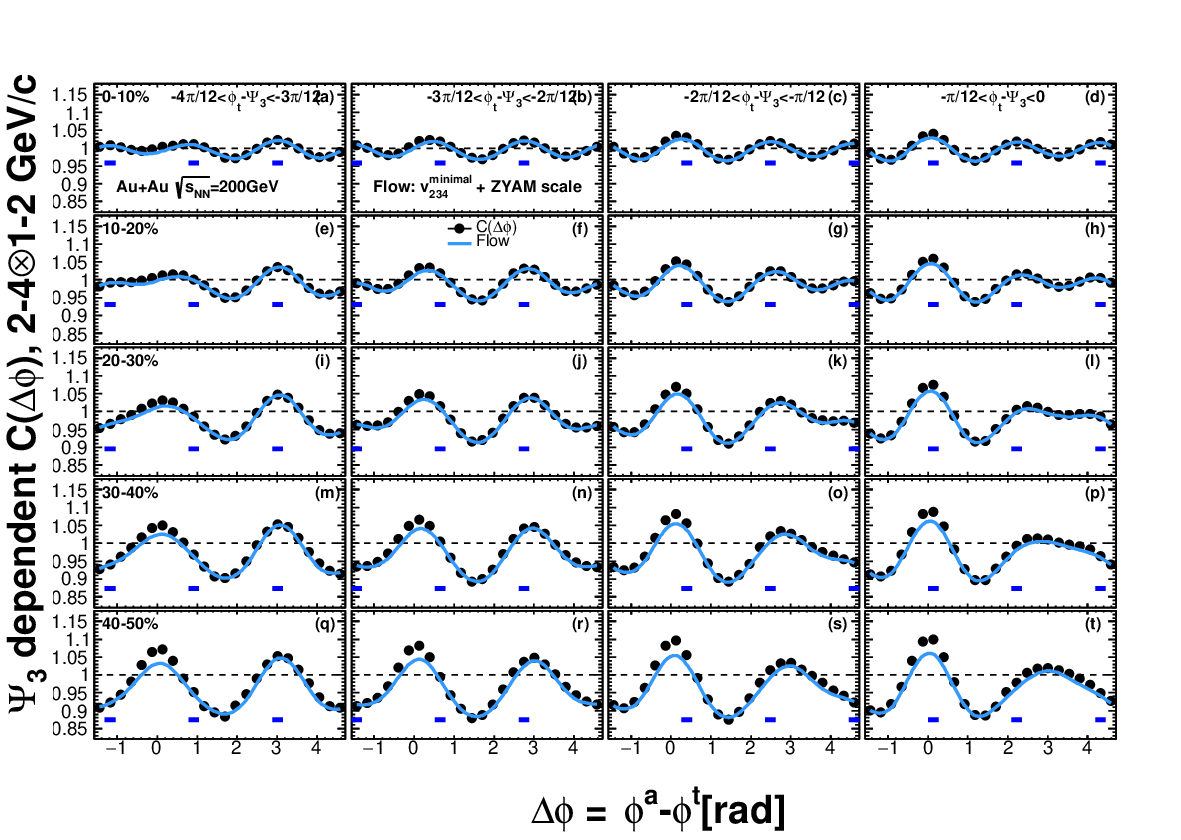}
\caption{
$\Psi_3$-dependent correlations $C(\Delta\phi,\phi_s)$ and flow backgrounds 
$F(\Delta\phi,\phi_s)$ for \mbox{($2<\ptt<4$)~$\otimes$~($1<\pta<2$)}~GeV/$c$. 
Trigger particle azimuthal angle relative to the event plane 
\mbox{$\phi_s=\phi^t-\Psi_3$} is selected out of plane (left) to in plane (right). 
Centrality is 0\%--10\% (top) to 40\%--50\% (bottom).
}
\label{fig:FlowCorFitNPFA5x4Pt101}
\end{minipage}
\end{figure*}

\begin{figure*}[htb]
\begin{minipage}{0.99\linewidth}
\includegraphics[width=0.89\linewidth]{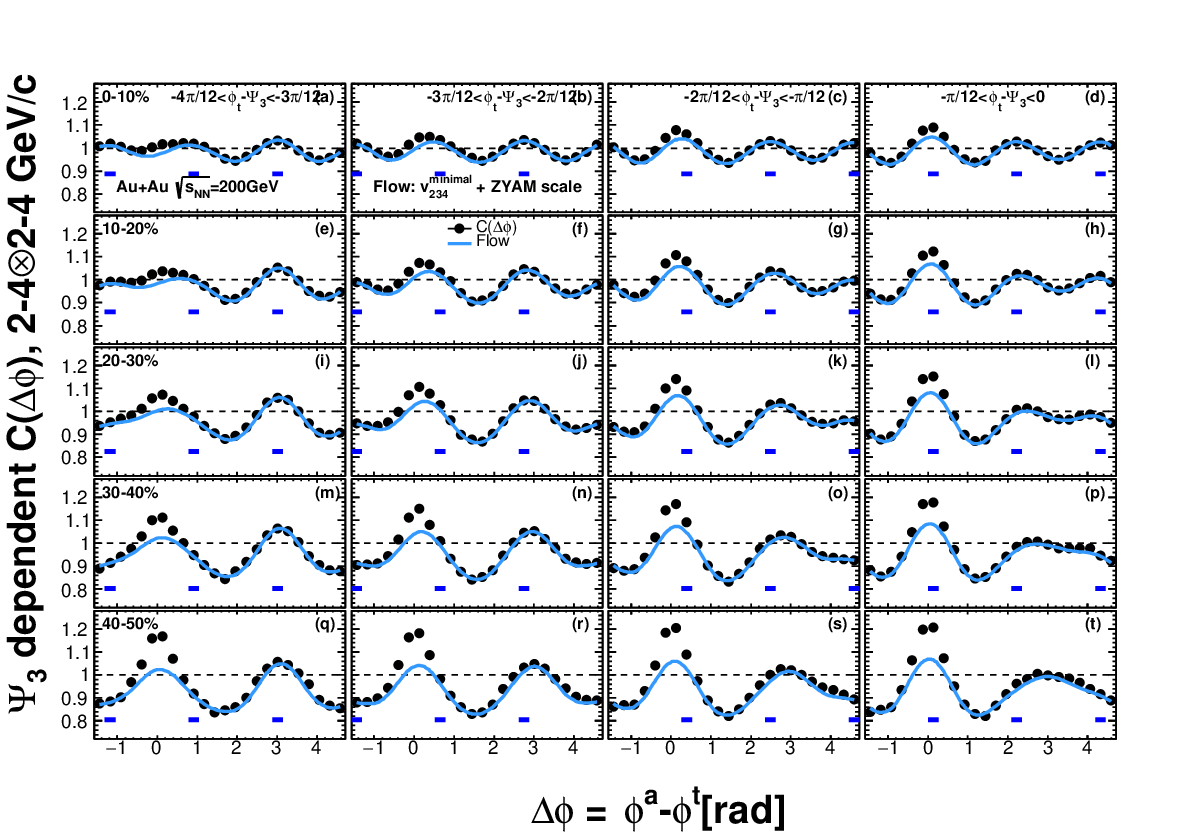}
\caption{
$\Psi_3$-dependent correlations $C(\Delta\phi,\phi_s)$ and flow backgrounds 
$F(\Delta\phi,\phi_s)$ \mbox{($2<\ptt<4$)~$\otimes$~($2<\pta<4$)}~GeV/$c$. Trigger 
particle azimuthal angle relative to the event plane \mbox{$\phi_s=\phi^t-\Psi_3$} 
is selected out of plane (left) to in plane (right). Centrality is 0\%--10\% (top) 
to 40\%--50\% (bottom).
}
\label{fig:FlowCorFitNPFA5x4Pt111}
\end{minipage}
%\end{figure*}
%=================================================== Fig_28
%\begin{figure*}[htb]
\begin{minipage}{0.99\linewidth}
\includegraphics[width=0.89\linewidth]{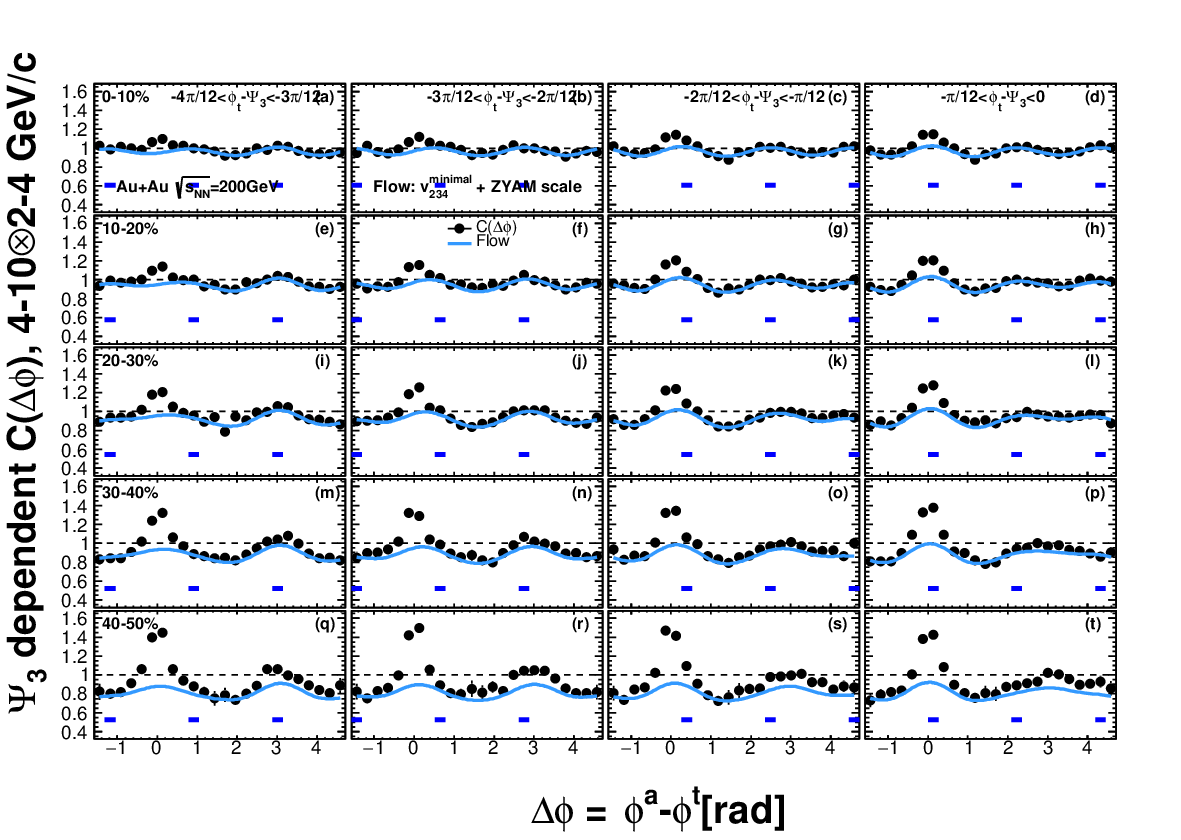}
\caption{
$\Psi_3$-dependent correlations $C(\Delta\phi,\phi_s)$ and flow backgrounds $F(\Delta\phi,\phi_s)$ for \mbox{($4<\ptt<10$)~$\otimes$~($2<\pta<4$)}~GeV/$c$. Trigger particle azimuthal angle relative to the event plane \mbox{$\phi_s=\phi^t-\Psi_3$} is selected out of plane (left) to in plane (right). Centrality is 0\%--10\% (top) to 40\%--50\% (bottom).
}
\label{fig:FlowCorFitNPFA5x4Pt211}
\end{minipage}
\end{figure*}

\begin{figure*}[htb]
\includegraphics[width=0.99\linewidth]{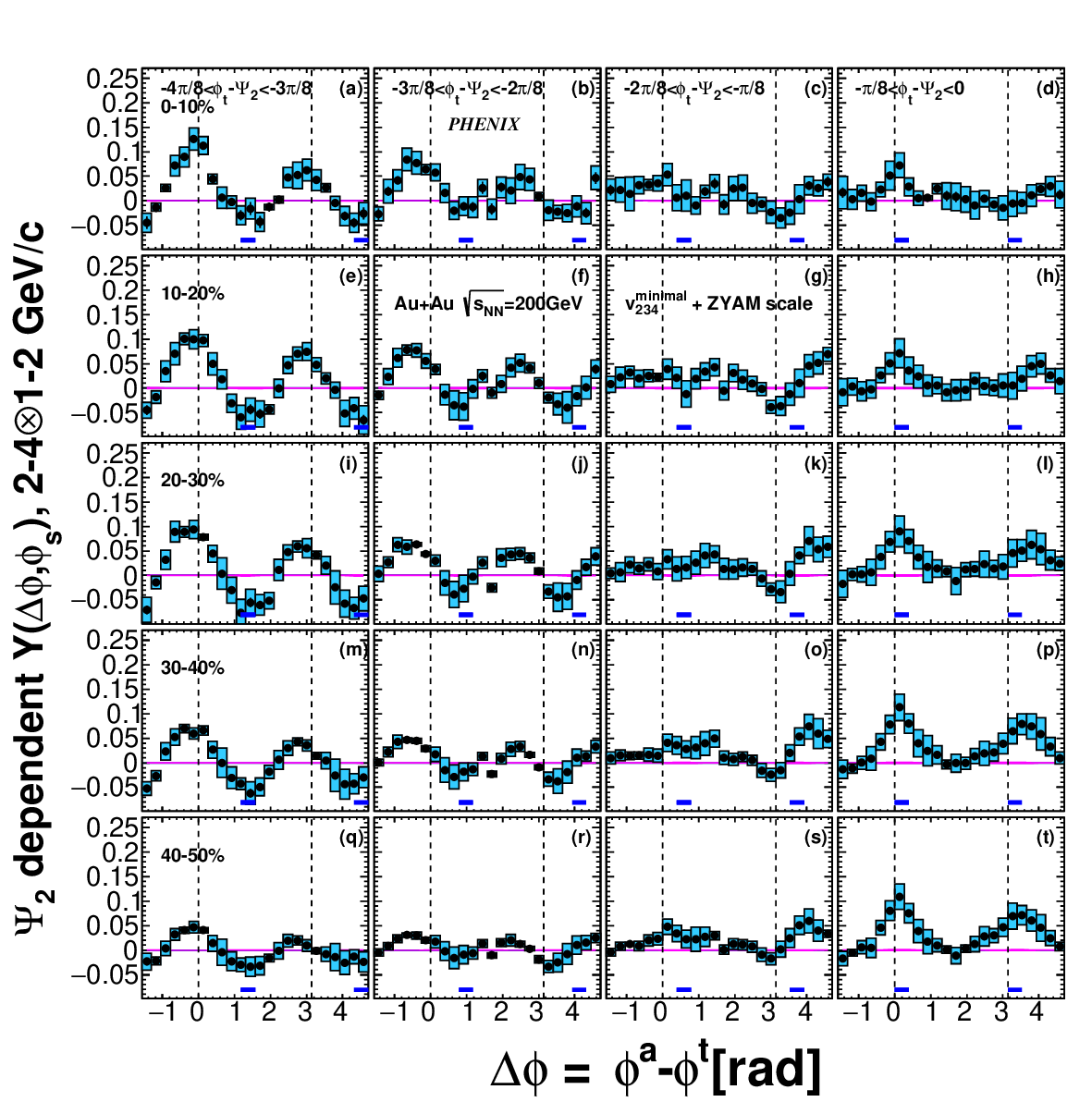}
\caption{
$\Psi_2$-dependent per-trigger yields $Y(\Delta\phi,\phi_s)$ for 
\mbox{($2<\ptt<4$)~$\otimes$~($1<\pta<2$)}~GeV/$c$. Trigger particle azimuthal 
angle relative to event-planes \mbox{$\phi_s=\phi^t-\Psi_2$} is selected out of 
plane (left) to in plane (right). Centrality is 0\%--10\% (top) to 40\%--50\% 
(bottom).
}
\label{fig:PTY4x5Pt10_0}
\end{figure*}

\begin{figure*}[htb]
\includegraphics[width=0.99\linewidth]{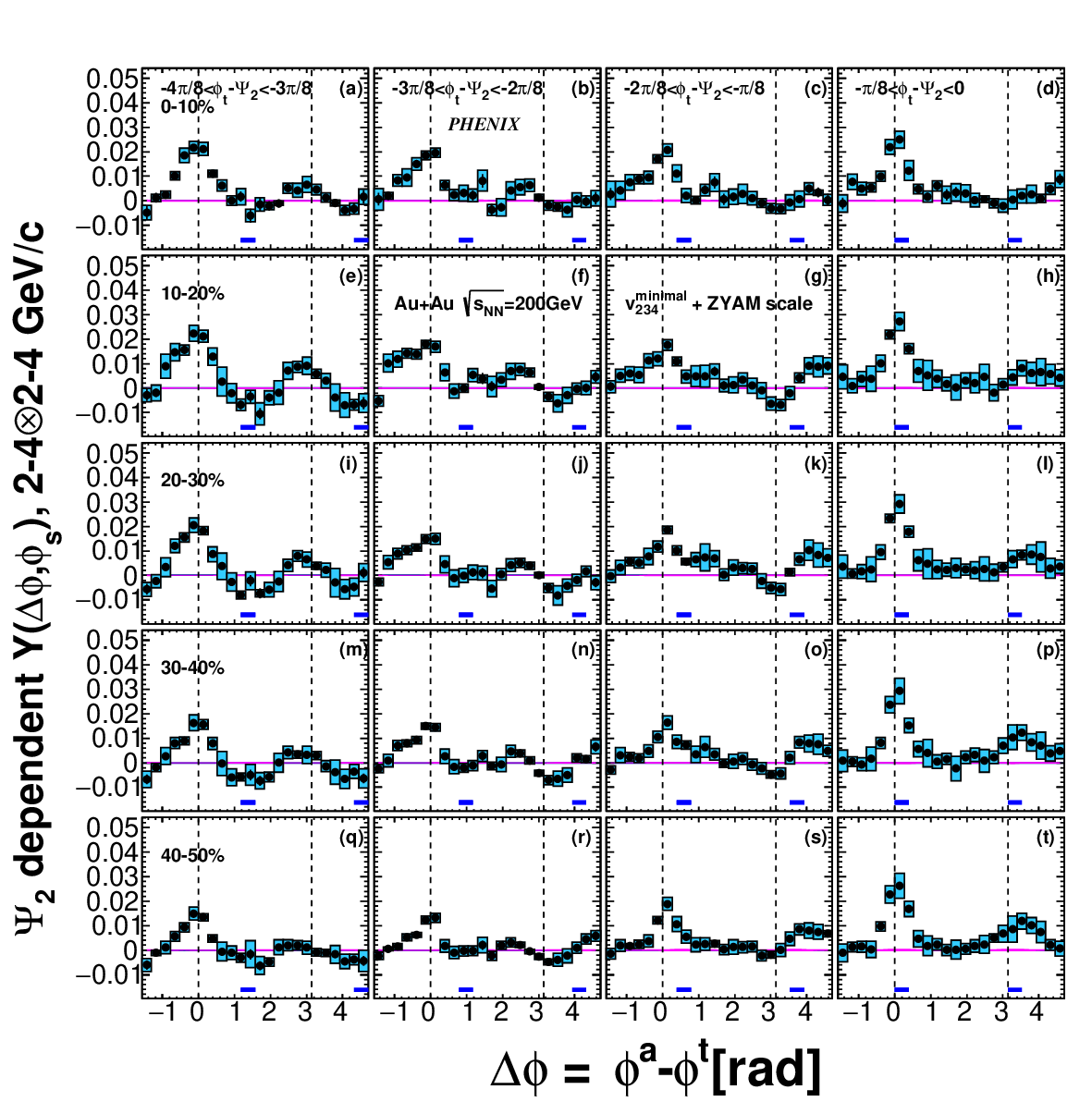}
\caption{
$\Psi_2$-dependent per-trigger yields $Y(\Delta\phi,\phi_s)$ for 
\mbox{($2<\ptt<4$)~$\otimes$~($2<\pta<4$)}~GeV/$c$. Trigger particle azimuthal 
angle relative to event-planes \mbox{$\phi_s=\phi^t-\Psi_2$} is selected out of 
plane (left) to in-plane (right). Centrality is 0\%--10\% (top) to 40\%--50\% 
(bottom).
}
\label{fig:PTY4x5Pt11_0}
\end{figure*}

\begin{figure*}[htb]
\includegraphics[width=0.99\linewidth]{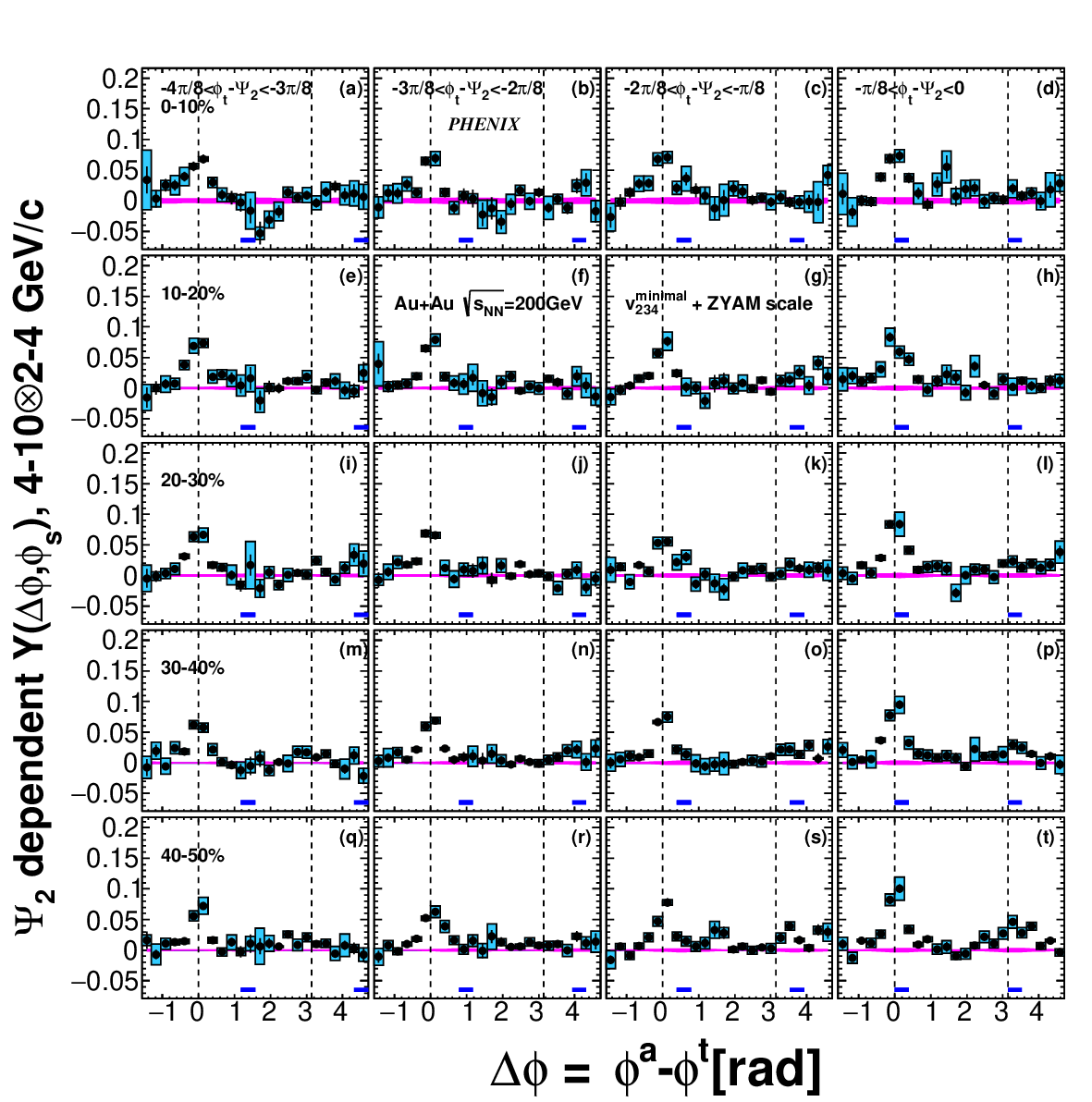}
\caption{
$\Psi_2$-dependent per-trigger yields $Y(\Delta\phi,\phi_s)$ for 
\mbox{($4<\ptt<10$)~$\otimes$~($2<\pta<4$)}~GeV/$c$. Trigger particle azimuthal 
angle relative to event-planes \mbox{$\phi_s=\phi^t-\Psi_2$} is selected out of 
plane (left) to in plane (right). Centrality is 0\%--10\% (top) to 40\%--50\% 
(bottom).
}
\label{fig:PTY4x5Pt21_0}
\end{figure*}

\begin{figure*}[htb]
\includegraphics[width=0.99\linewidth]{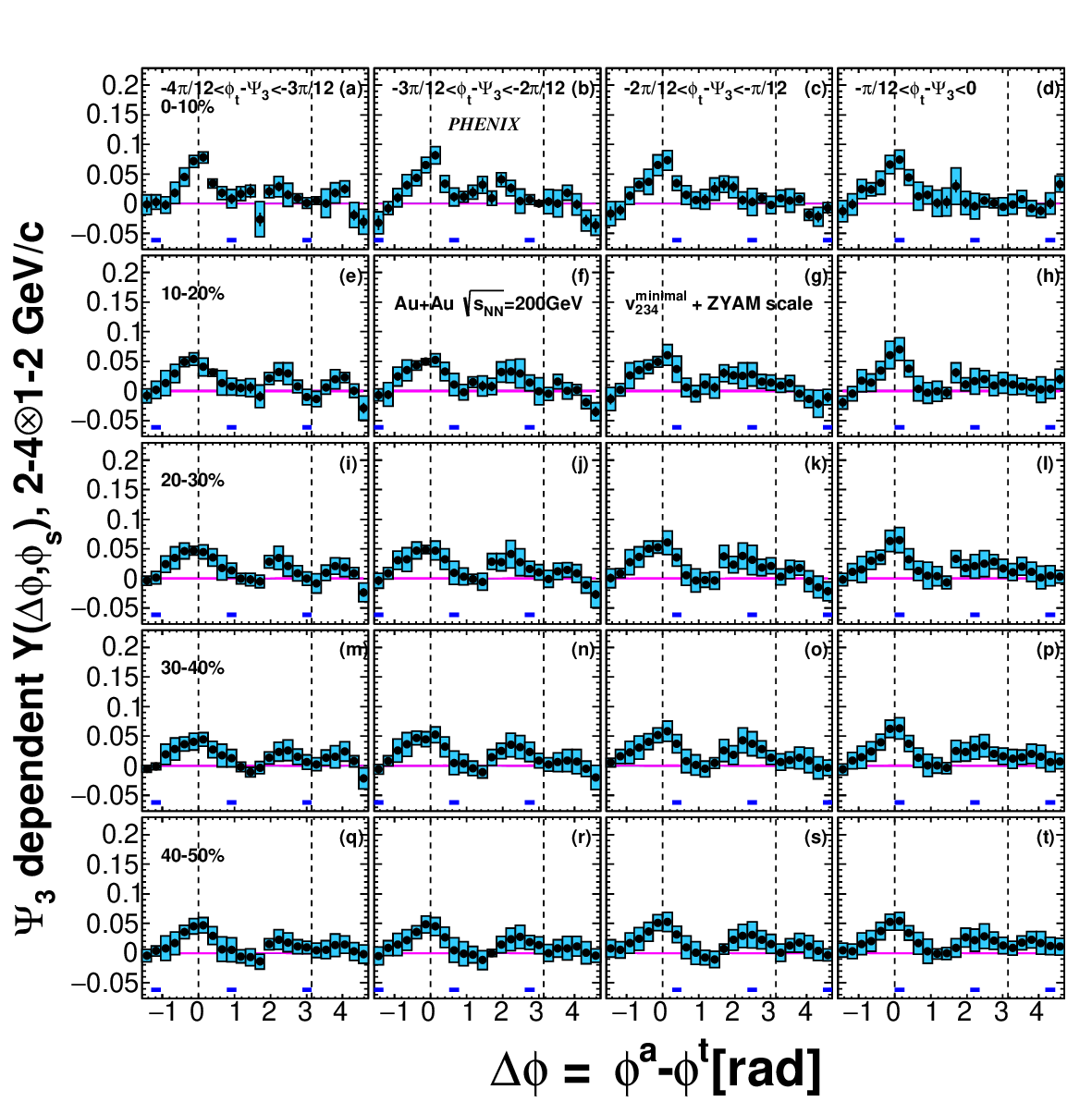}
\caption{
$\Psi_3$-dependent per-trigger yields $Y(\Delta\phi,\phi_s)$ for 
\mbox{($2<\ptt<4$)~$\otimes$~($1<\pta<2$)}~GeV/$c$. Trigger particle azimuthal 
angle relative to event-planes \mbox{$\phi_s=\phi^t-\Psi_3$} is selected out of 
plane (left) to in-plane (right). Centrality is 0\%--10\% (top) to 40\%--50\% 
(bottom).
}
\label{fig:PTY4x5Pt10_1}
\end{figure*}

\begin{figure*}[htb]
\includegraphics[width=0.99\linewidth]{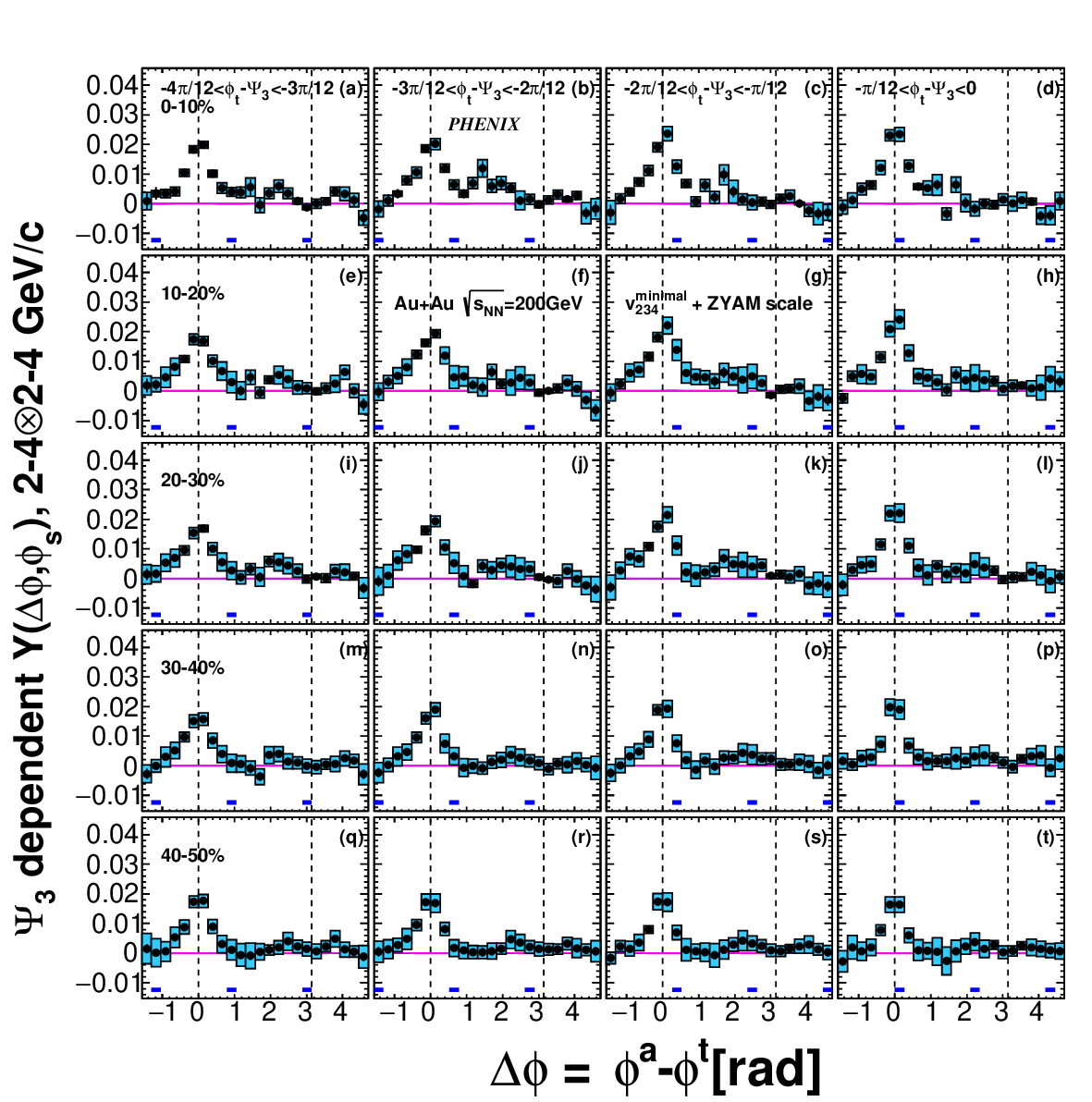}
\caption{
$\Psi_3$-dependent per-trigger yields $Y(\Delta\phi,\phi_s)$ for 
\mbox{($2<\ptt<4$)~$\otimes$~($2<\pta<4$)}~GeV/$c$. Trigger particle azimuthal 
angle relative to event-planes \mbox{$\phi_s=\phi^t-\Psi_3$} is out of plane (left) 
to in plane (right). Centrality is 0\%--10\% (top) to 40\%--50\% (bottom).
}
\label{fig:PTY4x5Pt11_1}
\end{figure*}

\begin{figure*}[htb]
\includegraphics[width=0.99\linewidth]{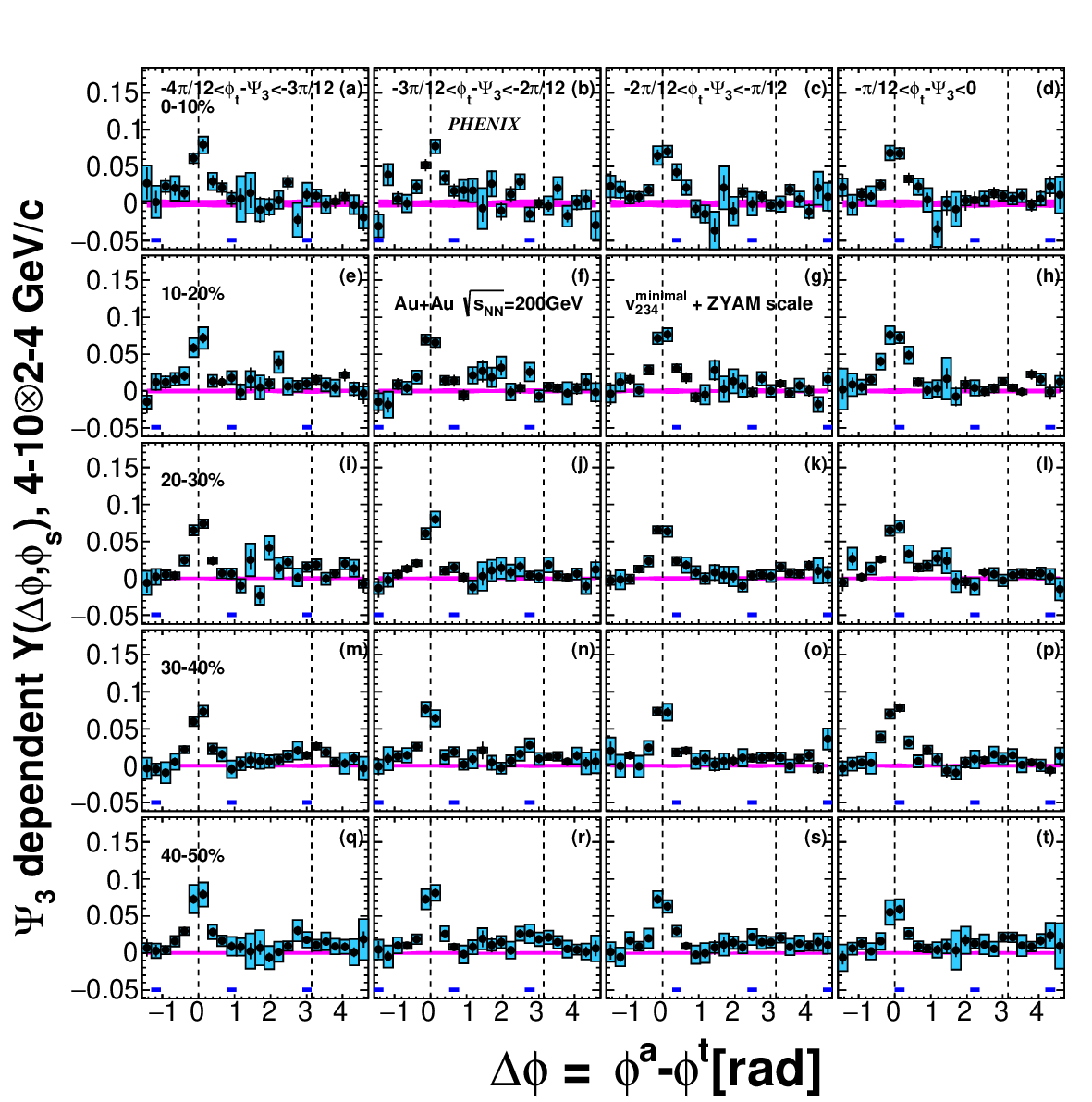}
\caption{
$\Psi_3$-dependent per-trigger yields $Y(\Delta\phi,\phi_s)$ for 
\mbox{($4<\ptt<10$)~$\otimes$~($2<\pta<4$)}~GeV/$c$. Trigger particle azimuthal 
angle relative to event-planes \mbox{$\phi_s=\phi^t-\Psi_3$} is selected out of 
plane (left) to in plane (right). Centrality is 0\%--10\% (left) to 40\%--50\% 
(right).
}
\label{fig:PTY4x5Pt21_1}
\end{figure*}

Inclusive correlations before subtracting the underlying event-model are shown 
in Figs.~\ref{fig:HighPtInclusiveCorWOVnSubtraction}--\ref{fig:LowPtInclusiveCorWOVnSubtraction}.
Event plane-dependent correlations and simulated-flow distributions are shown 
in Figs.~\ref{fig:FlowCorFitNPFA5x4Pt100}--\ref{fig:FlowCorFitNPFA5x4Pt211}.
Event-plane-dependent per-trigger Yields are shown 
in Figs.~\ref{fig:PTY4x5Pt10_0}--\ref{fig:PTY4x5Pt21_1}.

%=================================================== Fig_21

%=================================================== Fig_23
%\begin{turnpage}

%=================================================== Fig_25

%=================================================== Fig_27
%\end{turnpage}

%############################################################### Appendix (2nd plots)

%=================================================== Fig_29

%=================================================== Fig_30

%=================================================== Fig_31

%=================================================== Fig_32

%=================================================== Fig_33

%=================================================== Fig_34

\clearpage   %added by Takahito

%\bibliography{ppg173x2}

 
\end{document}